\documentclass[a4paper,11pt]{article}       
\usepackage{amsmath}
\usepackage{amssymb}
\usepackage{graphicx}
\usepackage{color}
\usepackage{bbm}
\usepackage{natbib}

\usepackage{tikz}
\usetikzlibrary{arrows,positioning}

\usepackage{algorithm}
\usepackage{algorithmic}
\usepackage{listings}
\usepackage{tabularx}
\usepackage{xspace}
\usepackage{multirow}
\usepackage{bm}  
\usepackage{verbatim}
\usepackage{colortbl}
\usepackage{mathtools}

\usepackage{setspace}

\usepackage[colorlinks,
			linkcolor=cyan,
			filecolor=blue,      
			urlcolor=red,
			citecolor=cyan]{hyperref}

\usepackage[left=2cm,right=2cm,top=2cm,bottom=2cm]{geometry}

\newcommand{\wu}[1]{#1}

\newcommand{\wuu}[1]{#1}

\newcommand{\wump}[1]{#1}

\numberwithin{equation}{section}

\newtheorem{theorem}{Theorem}[section]
\newtheorem{lemma}{Lemma}[section]
\newtheorem{definition}{Definition}[section]

\newtheorem{example}{Example}[section]

\newtheorem{remark}{Remark}[section]

\newtheorem{fact}{Fact}[section]

\newcommand{\G}{\Gamma}
\newcommand{\K}{\mathcal{K}}
\newcommand{\KK}{\mathcal{K}}
\renewcommand{\P}{\mathcal{P}}
\newcommand{\D}{\mathcal{U}}
\newcommand{\N}{\mathbb{N}}
\renewcommand{\S}{\mathcal{S}}
\newcommand{\E}{\mathbb{E}}
\newcommand{\1}{\mathbbm{1}}
\newcommand{\PP}{\mathbb{P}}
\newcommand{\var}{\mathbb{VAR}}

\newcommand{\WuSecondRevision}[1]{#1}

\newcommand{\WUU}[1]{#1}

\newcommand{\WuThird}[1]{#1}

\allowdisplaybreaks[4]

\title{A convergence analysis of the price of anarchy in atomic congestion games}

\author{Zijun Wu\footnote{Z Wu (\texttt{wuzj@hfuu.edu.cn}) is with Institute for Applied Optimization, Department of Artificial Intelligence and Bigdata, Hefei University, Jinxiu 99, Hefei, Anhui, China.}, 
	Rolf H. M{\"o}hring\footnote{R M{\"o}hring (\texttt{rolf.moehring@tu-berlin.de}) is with Kombinatorische Optimierung und Graphenalgorithmen (COGA), Fakult{\"a}t II--Mathematik und Naturwissenschaften,
		Institut f{\"u}r Mathematik, Sekr. MA 5--1,
		Technische Universit{\"a}t Berlin, Strasse des 17. Juni 136, Berlin, 10623, Germany.}, 
	Chunying Ren and Dachuan Xu
	\footnote{C Ren (\texttt{renchunying@emails.bjut.edu.cn}) and D Xu (\texttt{xudc@bjut.edu.cn})
	are with Department of Operations Research and Information Engineering, Beijing University of Technology, Pingleyuan 100, Beijing, 100124, China.}
}

\begin{document}

\maketitle






\begin{minipage}{0.9\textwidth}
	\begin{center}
		\textbf{Abstract}
	\end{center}

\WuSecondRevision{We analyze the convergence of the price of anarchy (PoA) of 
	Nash equilibria in atomic congestion games with growing total demand $T$. 
	When the cost functions are polynomials of the same degree, we obtain explicit rates
	for a rapid convergence of the PoAs of pure and mixed Nash equilibria 
	to $1$ in terms of $1/T$ and $d_{max}/T$,
	where $d_{max}$ is the \emph{maximum} demand controlled by an individual. 
	\WuSecondRevision{Similar convergence results} carry over  to the random inefficiency of the random flow induced by an arbitrary mixed Nash equilibrium.
	For arbitrary polynomial cost functions, we 
	derive a related convergence rate for the PoA of pure Nash equilibria (if they exist) when
	the demands fulfill certain regularity conditions and 
	$d_{max}$ is bounded as $T\to \infty.$ 
	In this \WUU{general} case, also the PoA of mixed Nash equilibria converges to $1$ as $T\to\infty$ when $d_{max}$ is bounded.  Our results constitute the ﬁrst convergence 
	analysis for the PoA in atomic congestion 
	games and show that selﬁsh behavior is well justiﬁed when the total demand is large.
}

\begin{flushleft}
\textbf{Keywords:}	atomic congestion games, pure
	and mixed Nash equilibria, \wu{price of anarchy and inefficiency of equilibria}
\end{flushleft}

\end{minipage}


\newpage

\tableofcontents

\newpage

\section{Introduction}
\label{sec:Introduction}

The price of anarchy (PoA, \citep{Papadimitriou2001Algorithms}) is an important notion in \emph{algorithmic game theory} (\citep{Nisan2007}) and has been investigated intensively during the last two decades in \emph{congestion games}
(\citep{Dafermos1969,Rosenthal1973A}), starting with the pioneering paper of
\citep{Roughgarden2002} on the PoA of pure Nash equilibria in \emph{non-atomic congestion games}
(\citep{Dafermos1969}) with affine linear cost functions. Much of this work has \WUU{then} been devoted to worst-case upper bounds of the PoA for different types of cost
functions $\tau_a(\cdot)$, and the influence of the network topology on these upper bounds, see, e.g., \citep{Nisan2007} for an overview.

Much less attention has been paid to the evolution of the PoA as a function of the growing total demand, although this is quite important for traffic and transportation networks in which the demands tend to be high. Only recently, it has been shown empirically (\citep{Youn2008Price,O2016Mechanisms,Monnot2017}) and analytically (\citep{Colini2016On,Colini2017WINE,Colini2020OR,Wu2019}) for \emph{non-atomic} congestion games that the PoA of pure Nash equilibria actually converges to 1 with growing total demand for a large class of cost functions that includes all polynomials.

Non-atomic congestion games have the special feature that
	every \emph{individual} user (player) is \emph{infinitesimal} and controls a \emph{negligible}
	amount of demand, and so has a \emph{negligible influence} on the performance of the whole
	game.
	This can be stated alternatively as that the demands
	are \emph{arbitrarily splittable}. 
	Prototypical non-atomic congestion games are traffic networks in which
		each (travel) origin-destination pair has an arbitrarily splittable traffic demand
		that need to be distributed on paths
		connecting the origin and the destination.
	A direct consequence is the \emph{essential
		uniqueness} (\citep{Roughgarden2002}) of pure Nash equilibria in  non-atomic congestion games,
	which plays a pivotal role in the convergence \WuSecondRevision{analysis} of the PoA 
	\WuSecondRevision{of pure Nash equilibria} by
	\citep{Colini2016On,Colini2017WINE,Colini2020OR}
	and \citep{Wu2019}.

In general, demands may not be arbitrarily splittable or even may not be split
at all. This is captured by \emph{atomic congestion games} (\citep{Rosenthal1973A}).
A prototypical such game is a transportation network in which
each user wants to transport a certain unsplittable demand of a good along a single path of that network. 
In this case, the congestion game is
\emph{finite} (\citep{Nash1950}), and each individual user is no longer infinitesimal and has
a \emph{non-negligible influence} on the whole game, and thus in particular on
the existence and \WuSecondRevision{other} properties of Nash equilibria. When 
\WuSecondRevision{the game is \emph{unweighted,}} i.e.,
\WuSecondRevision{users have equal demands}, then pure Nash equilibria exist, but may have \WuSecondRevision{\emph{different}} cost and so are \WuSecondRevision{\emph{not}} essentially
unique,
see, e.g., \citep{Rosenthal1973A,Roughgarden2007Introduction-routing}.
When \WuSecondRevision{the game is \emph{weighted},} 
\WuSecondRevision{i.e., users have unequal demands,} then pure Nash equilibria \WuSecondRevision{need} 
\WuSecondRevision{\emph{not exist}} and \WuSecondRevision{one has to resort to} \emph{mixed} Nash equilibria except for \emph{particular} cases, see, e.g., \citep{Nash1950,Fotakis2005,Harks2011,Harks2012}.

This raises an important question \WuSecondRevision{if and how much} the \emph{non-negligible 
	role} of individuals in \WuSecondRevision{atomic congestion games may} influence the total (transportation) inefficiency for growing
total (transportation) demand compared to \WuSecondRevision{their} \emph{negligible role} in \WuSecondRevision{non-atomic congestion games.}
This asks for a  convergence analysis of the PoA of both, \emph{pure} and \emph{mixed}, Nash equilibria for growing
total demands in atomic congestion games.


\subsection{Our contribution}

To address this question, we study the evolution  of the PoA for 
growing total demand $T$ in atomic congestion games with \emph{unsplittable} demands
and \emph{polynomial} cost functions. While our results hold for arbitrary atomic congestion games, we will mostly use the notation of transportation networks, since
they are more intuitive.

Our analysis covers the PoAs for both, pure and mixed, Nash equilibria. When pure Nash equilibria exist, then
	we \WuSecondRevision{call the ratio of their worst-case cost over 
	the social optimum cost  the \emph{atomic PoA},} see
	\eqref{eq:PoA-Atomic-def}. This distinguishes it \WuSecondRevision{from}
		the PoA of pure Nash equilibria in non-atomic congestion games, which
		\WuSecondRevision{is called} the \emph{non-atomic PoA} in this paper, see \eqref{eq:PoA-Non-Atomic-def}.
		\WuSecondRevision{Since mixed Nash equilibria are probability distributions, they
			induce random flows on the transportation network.}
	\WuSecondRevision{We then} call the ratio of the worst-case
	\emph{expected} cost \WuSecondRevision{of these random flows induced by} mixed Nash equilibria
	over the social optimum cost
	the \emph{mixed PoA},
	see \wu{\eqref{eq:MixedPoA}},
	\WuSecondRevision{and call} the ratio of the \emph{random} cost of \WuSecondRevision{the random flow induced by  a specific} mixed Nash equilibrium 
	over the social optimum cost
	the \emph{random PoA of that mixed Nash equilibrium}, see \eqref{eq:randomPoA}.
	
	\WuSecondRevision{
		The atomic PoA measures the inefficiency
		of selfish \emph{deterministic} choices, while the mixed and random PoAs quantify the 
		inefficiency of selfish \emph{random} choices in \emph{expectation} and as a \emph{stochastic
		variable}, respectively. 
		They are thus different. In particular, the random PoA is a random variable, and
		the atomic PoA is bounded by the mixed PoA, since
	    pure Nash equilibria in atomic congestion games can be considered as particular mixed Nash equilibria that result in deterministic choices of users.}

	We first derive upper bounds \WuSecondRevision{on} the atomic, mixed and random PoAs
	for polynomial cost functions
	of the \emph{same degree}, which \WuSecondRevision{cover \emph{BPR cost functions} (\citep{BPR}) that \WuSecondRevision{are
	of the form} $\xi_a\cdot x^{\beta}+\gamma_a$.} In this analysis, we apply the technique of \emph{scaling} \wu{that
		was used implicitly 
		in \citep{Colini2020OR} and formalized and extended
		in \citep{Wu2019} \WuSecondRevision{and \citep{Wu2019Sensitivity}}}.
	
	Using this technique, we show
		that the atomic PoA is \WuSecondRevision{$1+O(\frac{1}{T})+O(\sqrt{\frac{d_{max}}{T}})$}
		when pure Nash equilibria exist, 
		see Theorem~\ref{thm:MainResults}.
		Here, $T$
		is the \emph{total demand} and $d_{max}$
		is the \WuSecondRevision{\emph{maximum demand over all individuals} (simply, 
			\emph{maximum individual demand}),} which reflects
		to a certain extent the
		possible influence of an individual. Moreover, we show that the mixed PoA is \WuSecondRevision{$1+O(\frac{1}{T})+O(\frac{d_{max}^{1/6}}{T^{1/6}}),$} see 
		\WuSecondRevision{Theorem~\ref{thm:WeightedCase}b.}
		\WuSecondRevision{These upper bounds converge quickly} to $1$ as 
		$T\to \infty$ and $\frac{d_{max}}{T}\to 0$.
		We also explore the
		probability distribution of the random PoA of an arbitrary
		mixed Nash equilibrium \WuSecondRevision{and obtain with Chebyshev's inequalities} in
	\WuSecondRevision{Theorem~\ref{thm:WeightedCase}a} that the random PoA is \WuSecondRevision{bounded from above
	by} $1+O(\tfrac{1}{T})+O(\tfrac{d_{max}^{1/6}}{T^{1/6}})$ with 
\WuSecondRevision{an overwhelming} probability  of
		$1-O(\tfrac{d_{max}^{1/3}}{T^{1/3}})$. \WuSecondRevision{This shows} that
		an arbitrary mixed Nash equilibrium is also efficient \WuSecondRevision{as a random variable.}
		\WuSecondRevision{We further} illustrate that both conditions $T\to \infty$ and $\frac{d_{max}}{T}\to 0$ are necessary for these convergence results,
		see Example~\ref{example:Counterexample-Poly-Same-Degree} and Example~\ref{example:Counter2}.

	We then investigate 
		conditions   for  the convergence
		of the atomic PoA and the mixed PoA for arbitrary polynomial cost functions.
		We demonstrate first that the conditions $T\to \infty$ and $\frac{d_{max}}{T}\to 0$ 
		are \emph{no longer} sufficient for the convergence of the atomic PoA to $1$, since the cost functions may have different degrees and the (transportation)
		origin-destination pairs may have asynchronous demand growth rates. This may result in significantly discrepant influences of 
		different origin-destination pairs on the limits of the PoAs, 
		see Example~\ref{example:Divergence} \WuSecondRevision{or \citep{Wu2019}.}
		
			To capture these discrepant influences, we employ  the \emph{asymptotic decomposition} technique introduced by \citep{Wu2019}.
			We
		show \WuSecondRevision{for arbitrary polynomial cost functions} that
		both, the mixed PoA and \WuSecondRevision{the worst-case ratio of the total cost 
		of the \emph{expected} flow of a mixed Nash equilibrium 
		over the social optimum cost}, converge to $1$ as 
		$T\to \infty,$ when the maximum individual demand $d_{max}$
		is \emph{bounded} from above by
		a constant independent of \WuSecondRevision{the growth of} $T$, see \WuSecondRevision{Theorem~\ref{thm:GeneralPoly}a--b.}
		\WuSecondRevision{Note that 
			the total cost of the expected flow of a mixed Nash equilibrium
			need not coincide with the \emph{expected cost} of the random flow of that mixed Nash equilibrium, which is used in the definition of the mixed PoA,
			and that the condition ``$d_{max}$ is bounded from above''
			is necessary for these convergence results, see 
			Example~\ref{example:Divergence}.} \WuSecondRevision{To obtain these results, we
			have coupled the asymptotic decomposition technique with Chernorff-Hoeffding inequalities, see Appendices~\ref{proof:MainStep}--\ref{proof:Subgame-U-Sto-Exp}.}
		
		\WuSecondRevision{Hence, the atomic PoA converges also to $1$
		in this general case when pure Nash equilibria exist.}
		To analyze its convergence speed, we show, with a result
			by  \citep{Colini2020OR} for the convergence rate 
		of the non-atomic PoA \WuThird{and with a result by \citep{Wu2019Sensitivity}
		for the sensitivity of the non-atomic PoA,} that 
	   the atomic PoA (if pure Nash equilibria exist) converges to $1$ at a rate of 
      \WuThird{$O(T^{-\frac{1}{2\cdot \beta_{max}}}),$ when \WuThird{$\beta_{max}=\max_{a\in A}\beta_a> 0$ is the maximum of
      	the degrees $\beta_a$ of the polynomial cost functions $\tau_a$, the maximum individual demand} $d_{max}$ is bounded from above, and 
      the ratio $\frac{d_k}{T}$ of the total demand $d_k$ of each origin-destination pair $k$ over $T$ is bounded away from $0$,
  see Theorem~\ref{thm:GeneralPoly}c.}

\WuSecondRevision{In summary,} this paper \WuSecondRevision{presents for atomic congestion games
with growing total demands} the \WuSecondRevision{\emph{first}} convergence analysis of the atomic and mixed PoAs, and the \WuSecondRevision{\emph{first}} probabilistic analysis
	of the random PoA. 
	While individual users
		have a non-negligible role in atomic
		congestion games, our convergence results show that 
		this does not significantly increase the total transportation
		inefficiency for a large total demand $T$
		when the maximum individual demand $d_{max}$
		is very small compared to $T$.
	Our convergence results imply, \WuSecondRevision{in addition
	to \citep{Colini2016On,Colini2017WINE,Colini2020OR} and \citep{Wu2019},}
	that 
	\WuSecondRevision{pure Nash equilibria, mixed Nash equilibria and social optima 
	of an atomic congestion game with a large total demand} are almost equally efficient, and even 
	as efficient as the social optima of the corresponding
	non-atomic congestion games, \WuSecondRevision{see \eqref{eq:GeneralPoly-Obj}--\eqref{eq:Wu-AD} in Appendix~\ref{proof:MainStep}.}
	
	Thus, both pure Nash equilibria and mixed Nash equilibria
	in congestion games with
	a large total demand need not be bad.  
	The
	selfish choice of strategies leads then to an almost optimal behavior,  \WuSecondRevision{regardless whether} users employ mixed or pure strategies,
	and whether their transportation demands are splittable or not. Users may then \WuSecondRevision{restrict to} pure strategies and need not consider mixed strategies.
	Although that \WuSecondRevision{need} not lead to
	an equilibrium, it 
simplifies their decisions, and benefits both their own cost
and the total cost of the whole transportation network.


\subsection{Related work}
\label{subsec:RelatedWorks}

\subsubsection{Existence of equilibria}

The existence of equilibria in atomic congestion games was obtained in, e.g., \citep{Rosenthal1973A,Fotakis2005,Harks2011,Harks2012} and others.
\citep{Rosenthal1973A} showed that an arbitrary \emph{unweighted}
atomic congestion game has a pure Nash equilibrium.
\citep{Fotakis2005} showed \wu{that} an arbitrary \emph{weighted}
atomic congestion game $\G$ \wuu{with affine linear cost functions} \wu{is a \emph{potential game}
	(\citep{Shapley1996}) and thus has a pure Nash equilibrium}.
Moreover, \citep{Harks2011} proved
that if $\mathcal{C}$ is a class of cost functions such that every weighted atomic congestion game
$\G$ with cost functions in $\mathcal{C}$
is a potential game, then $\mathcal{C}$ contains only affine linear functions.
The existence of pure Nash equilibria in weighted atomic congestion games
was further studied by \citep{Harks2012}. Beyond these cases,
we have to consider mixed Nash equilibria
in atomic congestion games, as \citep{Nash1950} has shown that 
every finite game has a mixed Nash equilibrium.

\subsubsection{Worst-case upper bounds on the price of anarchy}

\citep{Papadimitriou1999} proposed to quantify the inefficiency of equilibria in arbitrary congestion games from a worst-case perspective. This resulted in the concept of the \emph{price of
	anarchy} (PoA) that is usually defined as the ratio of the worst-case cost of  (pure or mixed) Nash equilibria over
the social optimum cost, see \citep{Papadimitriou2001Algorithms}.

A wave of research has been started with the pioneering paper of
\citep{Roughgarden2002} on the PoA of pure Nash equilibria in non-atomic congestion games with affine linear cost functions.
Examples are \citep{Roughgarden2001,Roughgarden2003,Roughgarden2005,Roughgarden2015,Roughgarden2002,Roughgarden2004,Christodoulou2005,Correa2004a,Correa2005,Perakis2007}
and others. They investigated the worst-case upper bounds of the PoA \WuSecondRevision{of pure Nash equilibria}
in both atomic and non-atomic congestion games for different types of cost
functions $\tau_a(\cdot)$, and analyzed the influence of the network topology on these bounds. For
non-atomic congestion games, this upper bound is $\frac{4}{3}$
for affine linear cost functions (\citep{Roughgarden2002}), and 
$\Theta(\frac{\beta}{\ln\beta})$ for polynomial cost functions
of degree at most $\beta$ (\citep{Roughgarden2004}).
For \wuu{unweighted} atomic congestion games, \citep{Christodoulou2005} showed that this upper bound is $\frac{5}{2}$
for affine linear cost functions, and 
$\beta^{\Theta(\beta)}$ for polynomial cost functions
of \wu{degree} at most $\beta.$
\WuSecondRevision{Hence, the non-atomic PoA is not larger than the atomic PoA in general.}
Moreover, these upper bounds are independent
of the network topology, see, e.g., \citep{Roughgarden2003}.
\wump{\citep{Roughgarden2003,Roughgarden2015}} also developed a $(\lambda,\mu)$-smooth method by which one can obtain a \emph{tight} and \emph{robust} worst-case upper bound.
This method was then reproved by \citep{Correa2005} from a geometric 
perspective. Besides, \citep{Perakis2007} generalized the 
analysis to non-atomic congestion games with non-separable
and  
asymmetric cost maps.

\subsubsection{Convergence of the price of anarchy}

Recent papers have empirically studied the PoA
	of pure Nash equilibria in non-atomic congestion games with \emph{BPR cost functions}
	(\citep{BPR}) of the same degree $\beta>0$ and real traffic demands.   
	\citep{Youn2008Price}
	observed that the empirical PoA
	of pure Nash equilibria depends crucially on the total demand. Starting from~1, it grows with some oscillations, and
	ultimately becomes 1 again as the total demand increases. A similar observation was made by \citep{O2016Mechanisms}.
	They even conjectured that the PoA of pure Nash equilibria
	in non-atomic congestion games with BPR cost functions
of the same degree $\beta>0$ converges to $1$ \WuSecondRevision{at a rate of} $O\big(T^{-2\cdot\beta}\big)$ when the total
	demand $T$ becomes large. \citep{Monnot2017} showed that traffic choices of commuting students in Singapore are near-optimal and that the empirical PoA
	of pure Nash equilibria is much smaller than known worst-case upper bounds. Similar observations have been reported by \citep{Jahn2004System}.

These observations have been recently confirmed  by \citep{Colini2016On,Colini2017WINE,Colini2020OR} and \citep{Wu2019}.
	\citep{Colini2016On,Colini2017WINE,Colini2020OR} were
	the first to theoretically analyze the convergence of the PoA
	of pure Nash equilibria in non-atomic congestion games with growing total
	demand.

\citep{Colini2016On} showed \WuSecondRevision{that
	the PoA of pure Nash equilibria converges} to $1$ as the total demand 
	$T\to\infty$ \WuSecondRevision{when the} non-atomic congestion game \WuSecondRevision{has} a \emph{single}
	origin-destination pair
	and \emph{regularly varying} (\citep{Bingham1987Regular}) cost functions.
	This convergence result was then substantially extended by \citep{Colini2017WINE}
	to \emph{multiple} origin-destination pairs
	for both the case $T\to 0$ and the case $T\to\infty,$
	when the ratio of the demand of each origin-destination pair
	over the total demand $T$ remains \wump{a positive constant} as $T\to 0$ or $\infty$.
	\citep{Colini2020OR}  \WuSecondRevision{further}
	extended these results to the cases
	where the demands and the cost functions together fulfill certain \emph{tightness and salience conditions}
	that allow the ratios of demands to vary in a certain pattern
	as $T\to 0$ or $\infty.$ Moreover, \WuSecondRevision{\citep{Colini2020OR}} illustrated
	by an example that the PoA of pure Nash equilibria in non-atomic congestion games \WuSecondRevision{need} not converge to $1$ as $T\to \infty$
	when the cost functions are not regularly varying.
	\WuSecondRevision{In addition, they showed that} the PoA 
	of pure Nash equilibria
	in non-atomic congestion games with polynomial cost functions
	\WuSecondRevision{converges to $1$ at a rate of $O(\frac{1}{T})$} when the ratio
	of the demand of each origin-destination pair over the total demand $T$
	\WuSecondRevision{remains} a  positive constant
	as $T\to 0$ or $\infty.$

	\citep{Wu2019} generalized the work of \citep{Colini2016On,Colini2017WINE,Colini2020OR}
	for growing total demand. 
	They formalized the \emph{scaling technique}
		used implicitly
		in \citep{Colini2016On,Colini2017WINE,Colini2020OR}, proposed
	a \emph{limit notion} for \WuSecondRevision{a sequence of games} with 
growing total demand, and developed
a general technical framework, called \emph{asymptotic decomposition},
for \WuSecondRevision{the convergence} analysis of the PoA.
With this framework, they
	 showed for non-atomic congestion games
	with {\em arbitrary} regularly varying \WuSecondRevision{cost} functions that the PoA
	of pure Nash equilibria converges to $1$
	as the total demand tends to $\infty$ \emph{regardless of the growth pattern of the demands.}
	In particular, they proved a convergence rate of $o(T^{-\beta})$ for BPR cost functions
	\wu{of degree $\beta$}
	and illustrated by examples that the conjecture proposed by \citep{O2016Mechanisms} \WuSecondRevision{need} not hold.



\citep{Wu2019Sensitivity} extended the techniques of  \citep{Wu2019}
\WuSecondRevision{to a sensitivity analysis of the PoA.}
\WuSecondRevision{
	For an arbitrary non-atomic congestion game $\G$ with 
	Lipschitz continuous cost functions 
	on $[0,T],$
	they proved that the cost of an $\epsilon$-approximate
	equilibrium of $\G$ deviates at most by $O(\sqrt{\epsilon})$
from that of a pure Nash equilibrium of $\G,$ and that $O(\sqrt{\epsilon})$
is a tight upper bound of this deviation.
Moreover,}
they defined a metric $||\G_1,\G_2||$
	for two arbitrary \WuSecondRevision{games in}
	a set of non-atomic congestion games \WuSecondRevision{with} the same \emph{combinatorial structure}.
\WuSecondRevision{That metric induces} a topological
space of such games and \WuSecondRevision{permits to consider} \emph{continuous
	real-valued maps} and the \emph{limit of a
	sequence of non-atomic congestion games}. 
\WuSecondRevision{\citep{Wu2019Sensitivity} used these notions for a}
comprehensive analysis of the \emph{H{\"o}lder continuity} of the PoA map of pure Nash equilibria 
in that topological space.
 They
showed that the PoA map is \emph{point-wise} continuous, but neither
Lipschitz continuous, nor \emph{uniformly} H{\"o}lder continuous.
\WuSecondRevision{However, it} 
is \emph{point-wise} H{\"o}lder continuous
with H{\"o}lder exponent $\frac{1}{2}$ on a \emph{dense} subspace, i.e.,
$|\rho_{nat}(\G_1)-\rho_{nat}(\G_2)|\in O(\sqrt{||\G_1,\G_2||})$
for any two non-atomic congestion games 
$\G_1$ and $\G_2$ of \WuSecondRevision{that} subspace,
where $\rho_{nat}(\G_i)$ denotes the PoA value of pure Nash equilibria of the game $\G_i,$ $i=1,2.$
\WuSecondRevision{This results in an approximate computation of the PoA $\rho_{nat}(\cdot),$
meaning that one can approximate $\rho_{nat}(\G)$ for irregular cost functions with 
$\rho_{nat}(\G')$ for relatively simpler polynomial cost functions when the polynomial cost functions
of $\G'$ are sufficiently close to the irregular cost functions of $\G$.}

\WuThird{As a byproduct of the above H{\"o}lder continuity analysis, \citep{Wu2019Sensitivity} showed 
	that the total cost difference between Nash equilibria of two non-atomic congestion games $\G_{1}$ and 
	$\G_{2}$ is in $O(\sqrt{\|\G_{1}-\G_{2}\|})$ when $\G_{1}$ and $\G_{2}$ have the same Lipschitz
	continuous cost functions. Moreover, when the two non-atomic congestion games $\G_{1}$ and $\G_{2}$ have the same demands  but different Lipschitz continuous cost functions, they proved a similar upper bound on the total cost difference between their Nash equilibria.
These results together with the convergence rate of \citep{Colini2020OR} will help us to obtain an explicit convergence rate of the atomic PoA for polynomial cost functions of different degres, see Theorem~\ref{thm:GeneralPoly}c and its proof in Appendix~\ref{proof:MainStep}.}


Conditions implying the convergence of mixed Nash equilibria in atomic congestion games to
pure Nash equilibria in non-atomic congestion games
have also been studied in, e.g.,   \citep{Haurie1985,Milchtaich2000,Jacquot2018,Jacquot2019,Cominetti2020}, and others.

Among these papers, \citep{Cominetti2020} is the closest to our
work. They showed that  mixed Nash equilibria
of an atomic congestion game with \emph{strictly increasing} cost functions
converge \emph{in distribution} to pure Nash equilibria of a limit non-atomic congestion game,
when the total demand $T$ converges to a constant $T_0\in (0,\infty)$, the maximum individual demand $d_{max}$ converges to $0,$
and the number of users converges to $\infty$.
Moreover, 
they showed that this convergence happens at a rate of $O(\sqrt{d_{max}})$ when the cost functions have \emph{strictly positive first-order derivatives.}
Consequently, the PoA of mixed Nash equilibria (i.e., the mixed PoA) in \WuSecondRevision{such an} atomic congestion game converges also to that
of pure Nash equilibria in a ``limit non-atomic congestion game'' under these conditions.


\WuSecondRevision{The results of \citep{Cominetti2020} are inspiring and seminal. 
	They confirm the intuition that atomic congestion games can be thought of as non-atomic congestion games
	when $d_{max}$ is \emph{tiny}, the number of users is \emph{huge}, and $T$ is  \emph{moderate}, i.e., 
	neither too small nor too large.
	Our convergence result for the mixed PoA actually generalizes those of \citep{Cominetti2020} to 
	the case that $T\to \infty$.
	This is a non-trivial generalization, since it does not require
	the existence of the limit non-atomic congestion game, which
   is a premise in the analysis of \citep{Cominetti2020}.}

\WuSecondRevision{Our work also extends the convergence results
for the PoA of pure Nash equilibria in non-atomic congestion games
that were obtained recently
by \citep{Colini2016On,Colini2017WINE,Colini2020OR} and \citep{Wu2019} to
convergence results for pure and mixed Nash equilibria in atomic congestion games.
This implies that selfishness is also good in ``atomic congestion''. In particular, our results show for arbitrary congestion games
with a large total demand that selfish choice of users
is almost as efficient as social optima, regardless whether demands are splittable or not, and whether users use pure strategies or mixed strategies.}


%

\subsection{Outline of the  paper}
\label{subsec:Arrangement}

The paper is organized as follows. We develop our results for arbitrary atomic congestion games. These and their relevant
\wu{concepts} are introduced in Section~\ref{sec:Model_Preliminaries} . \WuSecondRevision{We analyze the convergence of the PoAs for atomic congestion games 
	in Section~\ref{sec:MainResults}.}
Section~\ref{sec:ConverganceAnalysis} then presents \WuSecondRevision{our convergence results}
for polynomial cost functions
with the same degree. \WuSecondRevision{Subsequently,}
Section~\ref{sec:GeneralPoly} presents \WuSecondRevision{our convergence results}
for arbitrary polynomial cost functions.
We conclude with a short summary and discussion in Section~\ref{sec:Summary}.
To improve  readability, all proofs have been moved to an Appendix.

\section{Model and Preliminaries}\label{sec:Model_Preliminaries}

Our study involves both atomic
and  non-atomic congestion games.
To facilitate the discussion,  
we \wu{introduce a unified notation} in Section~\ref{subsec:representation}, and distinguish
games implicitly by properties of their strategy profiles, see Section~\ref{subsec:RoutingChoices}.


\subsection{Atomic and non-atomic congestion games}
\label{subsec:representation} 

We define an \WuSecondRevision{arbitrary atomic} congestion game 
with the \WuSecondRevision{notation} of transportation games (see, e.g., \citep{Rosenthal1973A,Nisan2007}),
since this is more intuitive and closer to practice. \WuSecondRevision{An
\emph{atomic congestion game}} $\G$ is thus associated with a transportation network $G=(V,A)$, and
\WuSecondRevision{represented symbolically by} a tuple $(\K,\P,\tau,\D,d)$ 
\WuSecondRevision{with} components
\WuSecondRevision{defined in (G1)--(G5).}
\begin{itemize}
	\item \textbf{(G1)} $\K$ is a finite \emph{non-empty} set of 
	\WuSecondRevision{(transportation) \emph{origin-destination} (O/D) pairs $(o_k,t_k)\in V\times V$ with $o_k\ne t_k.$}  \WuSecondRevision{We will denote an O/D pair $(o_k,t_k)$
	simply by its index $k$ when this is not ambiguous.}
	
	\item \textbf{(G2)} $\P=\cup_{k\in \KK}\P_k$
	with each $\P_k\subseteq 
	2^A\setminus \{\emptyset\}$ 
	\WuSecondRevision{denotes the}  non-empty \WuSecondRevision{set of all} \emph{paths} from
	\WuSecondRevision{the origin} $o_k$ to \WuSecondRevision{the destination} $t_k.$ 
	Here, \WuSecondRevision{a} path 
	is a non-empty \WuSecondRevision{subset of the arc set $A.$}
	\WuSecondRevision{Then} $\P_k\cap\P_{k'}=\emptyset$
	for $k,k'\in\K$ with 
	$k\ne k'.$
	
	\item \textbf{(G3)} $\tau=(\tau_a)_{a\in A}$ is a \WuSecondRevision{cost function vector,}
	\WuSecondRevision{s.t.} $\tau_a: [0,\infty)\to [0,\infty)$ is  \emph{non-negative}, \emph{continuous} and \emph{non-decreasing} \WuSecondRevision{and denotes} the
	flow-dependent \emph{latency} or \emph{cost} of  arc 
	$a\in A.$ 
	\WuSecondRevision{We assume that no arc can be used for free, i.e.,
	$\tau_a(x)>0$ for all pairs $(a,x)\in A\times (0,\infty).$ }
	
	\item \textbf{(G4)} Associated with each O/D pair $k\in\K$ is a finite non-empty set $\D_{k}$ of
    \WuSecondRevision{\emph{agents} that are 
    	individual users or players.}
    Then
	$\D=\cup_{k\in \KK}\D_k$
	is the \WuSecondRevision{agent set} of $\G$. 
	We assume that $\D_k\cap\D_{k'}= \emptyset$
	for all $k,k'\in \KK$ with $k\ne k'.$
	\item \textbf{(G5)} $d=(d_{k,i})_{k\in \KK,i\in \D_k}$ is a \emph{\WuSecondRevision{demand} vector,} \WuSecondRevision{where}
	$d_{k,i}>0$ denotes \WuSecondRevision{an \emph{unsplittable}} demand to be transported 
	by \WuSecondRevision{agent} $i\in \D_k.$
	So $\G$
	has the \emph{total (\wump{transportation}) demand} $T=T(\D,d):=\sum_{k\in \KK}d_k,$
	where $d_k:=\sum_{i\in\D_k}d_{k,i}$ is the \emph{demand}
	of \WuSecondRevision{O/D pair} $k\in \K$.
	We call $d_{max}:=\max_{i\in \D_k,k\in\KK}$ $d_{k,i}$ the
	\emph{maximum individual demand}
	of $\G.$ \wu{Note that $\G$ is \emph{unweighted}
		if $d_{k,i}\equiv \upsilon$ for all 
		$k\in\K$ and all $i\in\D_{k},$
		for a constant $\upsilon>0.$ 
		Otherwise, $\G$ is \emph{weighted}.}
\end{itemize}
\WuSecondRevision{To unify notation, we view a
non-atomic congestion game as a \emph{variant} of an atomic congestion game,
in which each agent $i\in \D_{k}$ is no longer
an individual user, but a \emph{population of infinitesimal
users,} who together have the demand $d_{k,i}.$ Hence, the demands
$d_{k,i}$ can be split arbitrarily over paths in $\P_k$ when 
$\G$ is non-atomic. This differs from an atomic congestion game,
in which the demands $d_{k,i}$ cannot be split.
With a little abuse of notation, we denote a non-atomic congestion game  
again by the same tuple
$\G=(\K,\P,\tau,\D,d)$. 
We will simply call  a tuple $\G$ 
a congestion game, and distinguish atomic and non-atomic 
congestion games by their \emph{atomic} and \emph{non-atomic profiles}
in Section~\ref{subsec:RoutingChoices}.}

The tuple $(\K,\P)$ together
with the transportation network $G$  constitutes the \emph{combinatorial structure} of 
$\G$. For ease of notation, we \wu{may fix an arbitrary
	network $G$ and an arbitrary tuple
	$(\K,\P),$ and}
denote $\G$
simply by 
$(\tau,\D,d)$.
Viewed as a general congestion game, the
arcs $a\in A$
and \WuSecondRevision{the} paths $p\in\P$ correspond to
resources and (pure) strategies,
see, e.g., \citep{Dafermos1969} and \citep{Rosenthal1973A}.
Although we \WuSecondRevision{use} the nomenclature
of transportation networks, the analysis and results below
are independent of this view and carry over to arbitrary congestion games.

\subsection{Atomic, non-atomic and mixed profiles}\label{subsec:RoutingChoices}


Users distribute their demands
\emph{simultaneously} and \emph{independently}
on paths in $\P$.
\WuSecondRevision{This results in}
a \emph{strategy profile} or simply
\emph{profile} \WuSecondRevision{$\Pi=(\Pi_i)_{i\in \D}=(\Pi_i)_{i\in\D_k,k\in\K}
=(\Pi_{i,p})_{i\in\D_k,p\in\P_k,k\in\K}$} satisfying the condition \eqref{eq:RoutingChoice-user-i},
\begin{equation}\label{eq:RoutingChoice-user-i}
\sum_{p'\in\P_k}\Pi_{i,p'}=1\text{ and }\Pi_{i,p}\ge 0\quad \forall i\in\D_k\ \forall p\in\P_k\ \forall k\in\KK.
\end{equation} 
\WuSecondRevision{We put}  $\Pi_{i,p}=0$ when  $i\in\D_k$ and 
$p\in\P_{k'}$ for some $k,k'\in\K$ with $k\ne k'.$
This extends a profile $\Pi$ naturally to a vector $(\Pi_{i,p})_{i\in\D,p\in\P}$
with components $\Pi_{i,p}$ satisfying \WuSecondRevision{condition~\eqref{eq:RoutingChoice-user-i}.}

A profile $\Pi$ is called \emph{atomic}
if $\Pi$ is \emph{binary}. In this case,
\WuSecondRevision{$\Pi_{i,p}\in \{0,1\}, i\in\D_{k}, p\in\P_k, k\in\K,$ 
indicates whether path $p$ is
used by $i,$ i.e., 
$\Pi_{i,p}=1,$ or not, i.e., $\Pi_{i,p}=0.$}
Condition~\eqref{eq:RoutingChoice-user-i} 
then means that  each $i\in\D_k$
satisfies \WuSecondRevision{his} demand $d_{k,i}$ 
 by a \emph{single} path $p\in\P_k$ in an atomic profile $\Pi$. 
So a congestion game 
$\G$ with only atomic
profiles is \WuSecondRevision{indeed} an  atomic congestion game whose
demands $d_{k,i}$ cannot be split.

In a non-atomic congestion game, each  \WuSecondRevision{agent} $i\in\D_{k}$ is
a population of infinitesimal users and can split the demand $d_{k,i}$ arbitrarily,
\WuSecondRevision{i.e.,} \WuSecondRevision{agents}
$i\in \D_k$ can \WuSecondRevision{send}
their demands $d_{k,i}$ \WuSecondRevision{along} several paths $p\in \P_k.$
This is captured by \emph{non-atomic} profiles.
The components $\Pi_{i,p}$
are then
\emph{fractions} of the demands $d_{k,i}$ deposited by \WuSecondRevision{agents} $i\in\D_k$ on paths $p\in\P_k$, \WuSecondRevision{i.e., agents $i$ totally allocate
$d_{k,i}\cdot \Pi_{i,p}$ units of demands to paths $p$.}
Hence, these $\Pi_{i,p}$ can take arbitrary values in $[0,1]$ when $\Pi$ is non-atomic.
Condition~\eqref{eq:RoutingChoice-user-i}
is then a \emph{feasibility} constraint for
non-atomic profiles that ensures that \emph{all} demands 
are satisfied. \WuSecondRevision{Clearly, a congestion game 
	is non-atomic when it has only non-atomic profiles.}


\wu{In
	a \emph{mixed} profile $\Pi,$ each $\Pi_i=(\Pi_{i,p})_{p\in\P_k}$}
is a \emph{probability distribution} over \wu{the} set $\P_k$
for all $i\in\D_k$ and all $k\in\KK.$ 
\wu{Then} the decisions are
\WuSecondRevision{random,} and   every \WuSecondRevision{agent}
$i\in\D_k$
delivers \WuSecondRevision{his} demand \WuSecondRevision{$d_{k,i}$}
on a single \emph{random path} \WuSecondRevision{$p_{k,i}(\Pi_i)$} drawn 
\emph{independently} from
$\Pi_i=(\Pi_{i,p})_{p\in\P_k},$ \WuSecondRevision{where
$\Pi_{i,p}\in [0,1]$ is the probability of the random event ``$p_{k,i}(\Pi_i)=p$''.}
\WuSecondRevision{Note that we consider mixed profiles only for atomic congestion games, although we use a unified notation for both atomic and non-atomic congestion games.
Note also that an atomic profile is a particular mixed profile with $\{0,1\}$-probabilities.}

\subsection{Multi-commodity \wu{flows} and their cost}
\label{subsec:Flow}
Each profile 
$\Pi$ \WuSecondRevision{induces} a \emph{multi-commodity flow} \wu{$f=(f_p)_{p\in\P}=(f_p)_{p\in\P_k,k\in\K}.$} 
When $\Pi$ is atomic or non-atomic, then
$f$ is \emph{deterministic} with \emph{flow value} 
$f_p:=\sum_{i\in\D_{k}} d_{k,i}\cdot \Pi_{i,p}$  for all $p\in \P_k$
and all $k\in\KK.$
We then call $f$ \emph{atomic} and \emph{non-atomic}, respectively.
\wu{There are} only finitely many atomic flows, as
the number $|\D|=\sum_{k\in \KK}|\D_k|$ of \WuSecondRevision{agents} is finite and 
\WuSecondRevision{the} demands $d_{k,i}$
cannot be split in an atomic flow.

When $\Pi$ is mixed, \WuSecondRevision{then
the flow $f=(f_p)_{p\in\P}$ is a \emph{random} vector in which
each component $f_p$ is
a weighted sum $\sum_{i\in\D_k} d_{k,i}\cdot 
\mathbbm{1}_{\{p\}}\big(p_{k,i}(\Pi_i)\big)$ of 
mutually independent \emph{Bernoulli} random variables 
$\1_{\{p\}}\big(p_{k,i}(\Pi_i)\big),$ where 
$p_{k,i}(\Pi_i)$ is the \emph{random path} draw from the distribution $\Pi_{i}$ by agent $i$ of 
O/D pair $k,$ and $\1_{\{p\}}(\cdot)$ is the indicator function of the membership of
the singleton $\{p\}.$ Then}
\begin{equation}\label{eq:ExpectedPathFlow}
\begin{split}
&\mathbb{E}_{\Pi}(f_p)=\sum_{i\in\D_{k}} d_{k,i}\cdot \Pi_{i,p},\\
&\mathbb{VAR}_{\Pi}(f_p)=
\sum_{i\in\D_{k}} d^2_{k,i}\cdot \Pi_{i,p}\cdot (1-\Pi_{i,p})
\end{split}
\end{equation}
for all $p\in\P_k$ and all $k\in\K.$ 
Here, we \WuSecondRevision{used} that \WuSecondRevision{agents} choose their paths \emph{mutually independently}, 
\WuSecondRevision{that $\E_{\Pi}[\1_{\{p\}}(p_{k,i}(\Pi_i))]=\Pi_{i,p}$
and $\var_{\Pi}[\1_{\{p\}}(p_{k,i}(\Pi_i))]$ $=$ $\Pi_{i,p}\cdot (1-\Pi_{i,p}),$} and that every
\WuSecondRevision{agent} $i\in\D_k$ \wu{transports}
his demand $d_{k,i}$
entirely on the single random
	path $p_{k,i}(\Pi_i)\in\P_k.$ 
We will write 
	$\E_{\Pi}(f):=\big(\E_{\Pi}(f_p)\big)_{p\in\P}$, and \wu{call $\E_{\Pi}(f)$ and} $f=(f_p)_{p\in\P}$ the \emph{expected flow} and the \emph{random flow}
of the mixed profile $\Pi,$ respectively. 

\wu{The expected flow} $\E_{\Pi}(f)$ is 
a non-atomic flow, and an arbitrary non-atomic
flow is the expected flow of a mixed
profile. Moreover, an atomic flow $f$ is
a particular random flow, in which
the random flow values $f_p$
have a variance of zero.
\wu{Note that}  each state
of a random flow is an atomic flow, and
the \wu{finite} set of all
atomic flows
is the \emph{state space} of random flows, i.e., 
$\sum_{f'\text{ is an atomic flow}}\PP_{\Pi}[f=f']=1$
for \wu{a mixed profile $\Pi$
	with random flow $f$.}

\wu{An} arbitrary flow $f$ induces an \emph{arc flow} $(f_a)_{a\in A}$
\wu{in which} component $f_a:=\sum_{p\in\P: a\in p} f_p$ \WuSecondRevision{is}
the  flow value on arc $a\in A.$ 
%
When $\Pi$ is atomic or non-atomic, then 
$f_a$ is again deterministic for
all $a\in A.$
When $\Pi$ is mixed, 
then each $f_a$ is \emph{random}, and has 
\wu{the} \emph{expectation} and \emph{variance} in \eqref{eq:ExpectedArcFlow},
\begin{equation}\label{eq:ExpectedArcFlow}
\begin{split}
&\mathbb{E}_{\Pi}(f_a)=\sum_{k\in\K}\sum_{p\in\P_k: a\in p}
\mathbb{E}_{\Pi}(f_p)=\sum_{k\in\K}\sum_{i\in \D_k}
d_{k,i}\cdot\sum_{p\in\P_k: a\in p}\Pi_{i,p}\\
&\hspace{2cm}:=\sum_{k\in\K}\sum_{i\in \D_k}
d_{k,i}\cdot \Pi_{i,a},\\
&\mathbb{VAR}_{\Pi}(f_a)
=\sum_{k\in\K}\sum_{i\in \D_k}
d^2_{k,i}\cdot \Pi_{i,a}\cdot (1-\Pi_{i,a}).
\end{split}
\end{equation}
Here, $\Pi_{i,a}:=\sum_{p\in\P_k: a\in p}\Pi_{i,p}\in [0,1]$ \WuSecondRevision{is} the probability that \WuSecondRevision{agent} $i$ \WuSecondRevision{of O/D pair $k\in\K$}
	uses \WuSecondRevision{arc} $a\in A$.
 Then
\eqref{eq:ExpectedArcFlow} follows since
\WuSecondRevision{agents} use \WuSecondRevision{an arc} $a\in A$ \emph{mutually independently}, and only if 
\WuSecondRevision{the} arc $a$ belongs to one of
	their random paths $p_{k,i}(\Pi_i).$

For a non-atomic flow, we need only to specify the \emph{O/D pair demand vector} 
	$(d_k)_{k\in\K}$ with $d_k=\sum_{i\in\D_k}d_{k,i},$ since \WuSecondRevision{the} demands $d_{k,i}$ are arbitrarily
	splittable, and two congestion games have the same
	set of non-atomic flows if and only if
	they have the same $(d_k)_{k\in\K}.$
	Nonetheless, the demand vector $d=(d_{k,i})_{k\in\K,i\in\D_k}$ need to be specified for atomic and random flows, as 
	the demands $d_{k,i}$ can then be not split.


\WuSecondRevision{Given a flow $f,$
an arc $a\in A$ has the cost
$\tau_a(f_a),$ and a path $p\in\P$ 
has the cost
	$\tau_{p}(f):=\sum_{a\in p}\tau_a(f_a).$}
When $f$ is atomic or non-atomic, then these \wu{cost values}
are deterministic. Every $i\in \D_k$ \wu{then}
has \wu{the deterministic cost} 
\begin{displaymath}
	\WuSecondRevision{C_{k,i}(f,\G)=}C_{k,i}(f,\tau,\D,d):=\sum_{p\in\P_k}d_{k,i}\cdot \Pi_{i,p}\cdot \tau_p(f),
\end{displaymath}
and all \WuSecondRevision{agents} together have \wu{the} (deterministic) \emph{total cost}  
\begin{displaymath}
	\WuSecondRevision{C(f,\G)}:=\sum_{k\in \KK}\sum_{i\in\D_k}C_{k,i}(f,\tau,\D,d)
	=\sum_{p\in\P}f_p\cdot \tau_p(f)=\sum_{a\in A} f_a\cdot \tau_a(f_a).
\end{displaymath}
Note that \wu{the cost} \WuSecondRevision{$C_{k,i}(f,\G)$} \wu{can be expressed} equivalently \wu{as}
\WuSecondRevision{$C_{k,i}(f,\G)=d_{k,i}\cdot \tau_{p_{k,i}(f)}(f)$}
when $f$ is \emph{atomic} \wu{and} $p_{k,i}(f)\in \P_k$
\WuSecondRevision{is} the \WuSecondRevision{single} path used by \WuSecondRevision{agent} $i$ in $f.$ 

The \wu{cost values} $\tau_a(f_a)$ and \wu{$\tau_p(f)$} are \emph{random} when 
$f$ is the random \wu{flow of a mixed profile $\Pi.$} \wu{Then} each 
$i\in\D_k$ has \wu{the} \emph{random} cost
\begin{displaymath}
	\WuSecondRevision{C_{k,i}(f,\G)}:=d_{k,i}\cdot \tau_{p_{k,i}(\Pi_i)}(f)=\sum_{p\in\P_k}d_{k,i}\cdot \1_{\{p\}}(p_{k,i}(\Pi_i))\cdot \tau_{p}(f),
\end{displaymath}
where $p_{k,i}(\Pi_i)$ \WuSecondRevision{is} again the random path of \WuSecondRevision{agent} $i\in\D_k.$
\wu{The} \emph{random total cost}
is \wu{then}
	$\WuSecondRevision{C(f,\G):=\sum_{k\in \KK}\sum_{i\in\D_k} C_{k,i}(f,\G)}.$
Consequently, 
all \WuSecondRevision{agents} \wu{together} have the \emph{expected total cost} 
\begin{displaymath}
	\begin{split}
	\WuSecondRevision{\mathbb{E}_{\Pi}[C(f,\G)]}\!=\! \sum_{k\in\KK}\sum_{i\in\D_k}\!\WuSecondRevision{\mathbb{E}_{\Pi}[C_{k,i}(f,\G)]}
	\!=\!\sum_{a\in A}\!\E_{\Pi}\big[f_a\cdot \tau_a(f_a)\big]
	\!=\!\sum_{p\in \P}\!\E_{\Pi}\big[f_p\cdot \tau_p(f)\big].
	\end{split}
\end{displaymath}
\wu{The} expected total cost \WuSecondRevision{$\mathbb{E}_{\Pi}[C(f,\G)]$} of \wu{a} random
flow $f$ \WuSecondRevision{need not} \wu{equal} the total cost
\WuSecondRevision{$C(\E_{\Pi}(f),\G)$}
of its expected flow $\E_{\Pi}(f).$
\wu{But} they coincide \wu{when}  $\Pi$ is atomic.

\WuSecondRevision{We} denote 
\emph{atomic}, \emph{non-atomic} and \emph{random}
flows by $f_{at}=(f_{at,p})_{p\in \P},$ $f_{nat}=(f_{nat,p})_{p\in \P}$
and $f_{ran}$ $=$ $(f_{ran,p})_{p\in \P},$
respectively, \wu{and will} not refer explicitly to the corresponding profiles \wu{since they are clear from the context.}

\subsection{Social optima and equilibria}\label{subsec:Model-Essentials}

Consider an arbitrary congestion game $\G.$ An atomic flow $f^*_{at}$
is an \emph{atomic system optimum} \wump{(atomic SO),} if 
\WuSecondRevision{$C(f_{at}^*,\G)\le C(f_{at},\G)$}
for every atomic flow $f_{at}.$ 
Similarly, a non-atomic flow $f^*_{nat}$ 
is a \emph{non-atomic SO} if 
\WuSecondRevision{$C(f_{nat}^*,\G)$ $\le$  $C(f_{nat},\G)$}
for every non-atomic flow $f_{nat},$
and a random flow $f^*_{ran}$
is a \wump{\emph{mixed SO}} if 
\WuSecondRevision{$\mathbb{E}_{\Pi^*}[C(f_{ran}^*,\G)]$ $\le$ $ \mathbb{E}_{\Pi}[C(f_{ran},\G)]$ }
for each random flow $f_{ran},$ where 
$\Pi^*$ and $\Pi$ are the mixed profiles
\WuSecondRevision{of $f^*_{ran}$ and $f_{ran},$ respectively.}

\wu{The expected} total cost of
an arbitrary mixed SO flow
\wu{coincides} with that of \wu{an} arbitrary atomic SO flow, since
the set of atomic flows is the
state space of random flows and
every atomic flow is a 
random flow with zero variance.
Moreover, the total cost of an atomic SO flow
is not smaller than
that of a non-atomic SO flow, since
every atomic SO flow is also a 
non-atomic flow. 
We summarize \wu{this} in Lemma~\ref{lemma:SO-Cost-Comparision}.
\begin{lemma}\label{lemma:SO-Cost-Comparision}
	Consider an arbitrary congestion game $\G$
	with a mixed SO flow 
		$f_{ran}^*$ of a mixed profile 
		$\Pi^*,$ an atomic SO flow $f_{at}^*,$
		and a  non-atomic SO flow $f_{nat}^*.$
	Then  \WuSecondRevision{$\mathbb{E}_{\Pi^*}[C(f_{ran}^*,\G)]$ $=$ $ C(f_{at}^*,\G)$ $\ge$ $ C(f_{nat}^*,\G).$}
\end{lemma}

\WuThird{Similar to the different types of SO flows in Lemma~\ref{lemma:SO-Cost-Comparision}, congestion games admit also \emph{Nash equilibrium flows} of different types. In each of them an individual does \emph{not} 
	benefit from unilaterally changing his strategy. 
	Hence, a Nash equilibrium flow is essentially a steady-state of the network that is stable under unilateral selfish behavior.
	Since we consider 
three types of flows, i.e., atomic, non-atomic and random flows, we define their Nash equilibria separately.}

An atomic flow $\tilde{f}_{at}=(\tilde{f}_{at,p})_{p\in\P}$ is  an atomic \emph{(pure) Nash equilibrium} (NE),
if 
		\WuSecondRevision{$C_{k,i}(\tilde{f}_{at},\G)=d_{k,i}\cdot \tau_{p_{k,i}(\tilde{f}_{at})}(\tilde{f}_{at})\le C_{k,i}(f'_{at},\G)=d_{k,i}\cdot \tau_{p'}(f'_{at})$}
	for all $k\in\KK$, all $i\in\D_k$ and all $p'\in\P_k,$
where \wuu{$p_{k,i}(\tilde{f}_{at})\in\P_k$} is the 
path \wu{used} by \WuSecondRevision{agent} $i\in\D_k$ in
atomic flow $\tilde{f}_{at},$ and $f'_{at}=(f'_{at,p})_{p\in\P}$ 
is an atomic flow with
components $f'_{at,p}$ defined in \eqref{eq:user-Moving}.
\begin{equation}\label{eq:user-Moving}
f'_{at,p}=
\begin{cases}
\tilde{f}_{at,p}&\text{if }p\notin \{p_{k,i}(\tilde{f}_{at}),p'\},\\
\tilde{f}_{at,p}-d_{k,i}&\text{if }p=p_{k,i}(\tilde{f}_{at}),\\
\tilde{f}_{at,p}+d_{k,i}&\text{if }p=p',
\end{cases}
\quad \forall p\in\P.
\end{equation}
Clearly, $f'_{at}$ is \wu{the atomic flow}
obtained by \WuThird{only} moving $i$ from \wu{$p_{k,i}(\tilde{f}_{at})$} to
$p'$, and so differs slightly from $\tilde{f}_{at}$
when 
$d_{max}$ is tiny.
\WuSecondRevision{\citep{Rosenthal1973A} has shown 
the existence of atomic NE flows for unweighted atomic congestion games.
Weighted atomic congestion games usually do not have atomic NE flows, except
for particular cases, e.g., affine linear cost functions, see
\citep{Harks2011} and \citep{Harks2012}.}

\WuSecondRevision{Since the} cost functions $\tau_a(\cdot)$ are non-decreasing,
non-negative and continuous, and since each \WuSecondRevision{agent} in a non-atomic flow
is a \WuSecondRevision{population}	of infinitesimal users, non-atomic (pure) NE \WuSecondRevision{are identical to} \emph{Wardrop equilibria} (WE, \citep{Wardrop1952}), see, e.g., \citep{Beckmann1956,Roughgarden2002}. Thus 
a non-atomic flow $\tilde{f}_{nat}=(\tilde{f}_{nat,p})_{p\in\P}$ 
is a \emph{non-atomic NE}
if \WuThird{and only if} it fulfills
\emph{Wardrop's first principle},  i.e., $\tau_p\big(\tilde{f}_{nat}\big)
\le \tau_{p'}\big(\tilde{f}_{nat}\big)$
for any two paths $p,p'\in\P_k$ with
$\tilde{f}_{nat,p}>0$ for each 
$k\in \KK$. \WuThird{Here, we note that the cost of each path does not change when an infinitesimal
user unilaterally changes his path. Hence}
a path $p\in \P_k$ is used,
\wu{i.e., $\tilde{f}_{nat,p}>0,$} in a non-atomic NE flow 
$\tilde{f}_{nat}$
only if $\tau_p\big(\tilde{f}_{nat}\big)=\min_{p'\in\P_k}
\tau_{p'}\big(\tilde{f}_{nat}\big).$
\WuSecondRevision{\citep{Dafermos1980} has shown that 
non-atomic NE flows always exist, and can be characterized equivalently
by the \emph{variational inequality}~\eqref{eq:WE-VarIneq},
\begin{equation}\label{eq:WE-VarIneq}
\sum_{a\in A}\tau_a\big(\tilde{f}_{nat,a}\big)
\cdot \big(f_{nat,a}-\tilde{f}_{nat,a}\big)\ge 0,
\end{equation}
for all non-atomic flows
	$f_{nat}$.
Moreover, \citep{Roughgarden2002} have shown that 
non-atomic NE flows are \emph{essentially unique}, i.e., 
$\tau_a(\tilde{f}_{nat,a})=\tau_a(\tilde{f}'_{nat,a})$
for each $a\in A$ for two arbitrary non-atomic NE flows
$\tilde{f}_{nat}$ and $\tilde{f}'_{nat}.$
Clearly, atomic and non-atomic NE flows differ.
Nonetheless, both of them are pure Nash equilibria.}

\WuThird{Mixed NE flows directly generalize atomic NE flows by considering random flows of mixed profiles.
Formally,}
a random flow $\tilde{f}_{ran}$ is
a \emph{mixed NE flow} \WuSecondRevision{if, for each $i\in \D_k$ and
each $k\in\K,$ }
\begin{equation}\label{eq:MNE-def}
\E_{\tilde{\Pi}}[C_{k,i}(\tilde{f}_{ran},\!\G)]\!=\!\mathbb{E}_{\tilde{\Pi}_{-i}}\big[d_{k,i}\!\cdot\! 
\tau_p(\tilde{f}_{ran|i,p})\big]
\!\le\! \WuSecondRevision{\mathbb{E}_{\tilde{\Pi}_{-i}}\big[d_{k,i}\!\cdot\! 
\tau_{p'}(\tilde{f}_{ran|i,p'})\big]}
\end{equation}
\WuSecondRevision{when} $p,p'\in\P_k$
\WuSecondRevision{are two arbitrary paths} with $\tilde{\Pi}_{i,p}>0,$ $\tilde{\Pi}=(\tilde{\Pi}_j)_{j\in \D}$
is \wu{the} mixed profile of $\tilde{f}_{ran},$ \WuSecondRevision{and $\tilde{\Pi}_{-i}
	=(\tilde{\Pi}_j)_{j\in \D\setminus \{i\}}$ is the mixed profile
	of all agents other than $i$ in $\tilde{\Pi},$} \wu{see also \citep{Cominetti2020}.}
\WuSecondRevision{Herein, 
	$\tilde{f}_{ran|i,p}=(\tilde{f}_{ran,p''|i,p})_{p''\in\P}$
	is the random flow in which agent $i$ uses the fixed path $p$ and 
	the others still follow the mixed profile $\tilde{\Pi}_{-i},$ i.e.,
\begin{displaymath}
	   \tilde{f}_{ran,p''|i,p}=
		\begin{cases}
			\tilde{f}_{ran,p''} &\text{if } p''\in \cup_{k''\in \K\setminus \{k\}}\P_{k''},\\
			d_{k,i}+\sum_{j\in\D_k\setminus\{i\}}d_{k,j}
			\cdot \1_{\{p''\}}(p_{k,j}(\tilde{\Pi}_j)) &\text{if } p''=p,\\
			\sum_{j\in\D_k\setminus\{i\}}d_{k,j}
			\cdot \1_{\{p''\}}(p_{k,j}(\tilde{\Pi}_j)) &\text{if } \ p''\in \P_k\setminus\{p\},
		\end{cases}
\end{displaymath}
for all paths $p''\in \cup_{k''\in\K}\P_{k''}.$}
\WuSecondRevision{Inequality~\eqref{eq:MNE-def} then means that 
	each support of the mixed strategy $\tilde{\Pi}_{i}$ of an agent
	$i\in \D$ is the best response to the mixed profile $\tilde{\Pi}_{-i}$
	of his opponents when $\tilde{f}$ is a mixed NE flow with mixed profile $\tilde{\Pi}$.
	Hence, no agent can reduce his (expected) cost by unilaterally changing
	his mixed strategy when the random flow is a mixed NE.
Since atomic congestion games equipped with only atomic profiles
are finite games, mixed NE flows always exist, see \citep{Nash1950}.
Note that atomic NE flows are mixed NE flows
with zero variance, but mixed NE flows need not be atomic NE flows, see, e.g., \citep{Nisan2007}.}

\begin{remark}[The mixed Wardrop equilibria]\label{remark:Mixed-WE}\WuSecondRevision{
	Note that one may consider also random flows $f_{ran}$ in which all paths with \emph{positive}
	expected flow values have minimum expected cost, i.e., 
	\begin{equation}\label{eq:MWE}
		\E_{\Pi}[\tau_p(f_{ran})]
		\le \E_{\Pi}[\tau_{p'}(f_{ran})]
	\end{equation}
	for two arbitrary paths $p,p'\in\P_k$ with $\E_{\Pi}(f_{ran,p})>0$ for
	each $k\in\K,$
	where $\Pi$ is the mixed profile of $f_{ran}.$
	Such random flows then generalize WE flows of non-atomic congestion games
	in atomic congestion games. We thus call them 
	\emph{mixed WE flows}. Using \emph{Brouwer's fixed point theorem}
	(\citep{Brouwer1910}) and  an argument similar to that in \citep{Dafermos1980}
	for the existence of WE flows in non-atomic congestion games, we can show easily that
	mixed WE flows always exist in atomic congestion games, see
	Lemma~\ref{lemma:Existence-WE-flows} in Appendix~\ref{app:Existence_Of_WE_Flows}. 
	The convergence results presented in this paper  carry also over to the inefficiency of mixed WE flows.
	In fact, we can even view mixed NE flows as mixed WE flows in the convergence analysis of the PoA of
mixed NE, since mixed NE flows approximate mixed WE flows when 
$\frac{d_{max}}{T}$ is tiny, see, e.g., \eqref{eq:MNE-def}--\eqref{eq:MWE}, 
\eqref{eq:Epsilon-mixed-WE-flow} in Appendix~\ref{proof:ExpectedFlowIsEpWE}, and Appendix~\ref{proof:MainStep}. Nonetheless, we will not go deeper into the discussion of mixed WE flows,
so as to save space.}
\end{remark}
\begin{example}\label{example:NonMutualInclusive}
	Consider \wu{the} congestion game $\G$ with
	one O/D pair $(o,t)$ (i.e., $\K=\{1\}$) and two parallel paths (arcs) shown in Figure~\ref{fig:NE-Noninclusive}.
	We label the upper and lower arcs
	as $u$ and \WuSecondRevision{$\ell,$} respectively.
	\wu{$\G$ has cost functions} $\tau_u(x)=x^2$ and
	\WuSecondRevision{$\tau_\ell(x)\equiv 2,$}
	\wu{and} two \WuSecondRevision{agents} with \wu{O/D pair $(o,t)$ and}
	\wu{demand $2$ each.} 
	\begin{figure}[!htb]
		\centering
		\begin{tikzpicture}[
		>=latex]
		\node[scale=0.4,circle,fill=black,label=left:{$o$}](o){};
		\node[scale=0.4,circle,fill=black,right =2of o,label=right:{$t$}](t){};
		\draw[->,thick] (o) to [out=90,in=90] node[above]{$x^2$} (t);
		\draw[->,thick] (o) to [out=-90,in=-90] node[below]{$2$} (t);
		\end{tikzpicture}
		\caption{An example of atomic, non-atomic, and mixed NE flows}
		\label{fig:NE-Noninclusive}
	\end{figure}
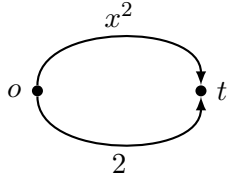
	\wu{Then} $\G$ has a unique atomic NE flow \WuSecondRevision{$\tilde{f}_{at}
	=(\tilde{f}_{at,u},\tilde{f}_{at,\ell})=(0,4),$}
	since \WuSecondRevision{an agent} \wu{using} the upper arc $u$ has a cost \wu{of} at least
	\WuSecondRevision{$4>\tau_\ell(x)\equiv 2$} and can always benefit \wu{by} moving to the lower arc $\ell.$
	Moreover, $\G$ has \wu{the} unique non-atomic NE flow 
	$\tilde{f}_{nat}=(\sqrt{2},4-\sqrt{2}),$ since demands
	 can be arbitrarily split \WuSecondRevision{in a non-atomic flow,}
	and a non-atomic NE flow fulfills Wardrop's first
	principle. So the \wu{sets} of atomic \wu{and} non-atomic NE flows
	\wu{of $\G$ do not}
	overlap.
	
	\WuSecondRevision{Clearly, $\tilde{f}_{at}$ is also the unique mixed NE flow, since
	the expected cost of the upper arc $u$ is always larger than the constant
cost of the lower arc $\ell$ when either of the two agents uses the upper
arc. Hence, neither the set of mixed NE flows nor the set of their expectations
need to intersect the set of non-atomic NE flows.
Moreover, by a little calculation, one can also see that neither
the set of mixed WE flows (Remark~\ref{remark:Mixed-WE})  nor the set
of their expectations 
intersects the sets of atomic and non-atomic NE flows in this Example.
This means that these equilibrium notions are mutually different, although
atomic and mixed NE flows coincide in this Example.}
\end{example}

\subsection{The price of anarchy}\label{subsec:PoA}


\wu{Since} we consider non-atomic, atomic and mixed NE flows,
we define four PoAs in \eqref{eq:PoA-Atomic-def}--\eqref{eq:randomPoA}, 
\wu{in which} $\tilde{f}_{nat}, f^*_{nat}$ and $f^*_{at}$ are
an arbitrary non-atomic NE flow, \WuSecondRevision{an arbitrary non-atomic  SO flow
	and an arbitrary atomic SO flow, respectively.}
We call \WuSecondRevision{$\rho_{at}(\G)$}
the \emph{atomic} PoA, 
\WuSecondRevision{$\rho_{nat}(\G)$} the \emph{non-atomic} PoA,
\WuSecondRevision{$\rho_{mix}(\G)$} the \emph{mixed} PoA,
and \WuSecondRevision{$\rho(\tilde{f}_{ran},\G)$}
the \emph{random} PoA of the mixed NE flow 
$\tilde{f}_{ran}.$ Here, 
\WuSecondRevision{we recall that non-atomic NE flows are essentially unique.}
\begin{align}
&\WuSecondRevision{\rho_{at}(\G)}\!:=\!\max\!\Big\{
\frac{\WuSecondRevision{C(\tilde{f}_{at},\G)}}{\WuSecondRevision{C(f^*_{at},\G)}}:
\tilde{f}_{at}\text{ is an atomic NE flow of }\G
\Big\}\label{eq:PoA-Atomic-def}\\
&\WuSecondRevision{\rho_{nat}(\G)}:=
\frac{\WuSecondRevision{C(\tilde{f}_{nat},\G)}}{\WuSecondRevision{C(f_{nat}^*,\G)}}\label{eq:PoA-Non-Atomic-def}\\
&\WuSecondRevision{\rho_{mix}(\G)}\!:=\!\max\!\Big\{
\frac{\WuSecondRevision{\E_{\tilde{\Pi}}[C(\tilde{f}_{ran},\G)]}}{
	\WuSecondRevision{\E_{\Pi^*}[C(f^*_{ran},\G)]}}:
\tilde{f}_{ran},f_{ran}^*\text{ are mixed NE and SO flows}\Big\}\notag\\
&\hspace{0.5cm}=\max\Big\{
\frac{\E_{\tilde{\Pi}}[\WuSecondRevision{C(\tilde{f}_{ran},\G)}]}{\WuSecondRevision{C(f^*_{at},\G)}}:
\tilde{f}_{ran} \text{ is a mixed NE flow of }\G\Big\}\label{eq:MixedPoA}\\
&\WuSecondRevision{\rho(\tilde{f}_{ran},\G)}:=\frac{\WuSecondRevision{C(\tilde{f}_{ran},\G)}}{\WuSecondRevision{C(f^*_{at},\G)}}\label{eq:randomPoA}
\end{align}


Note that \WuSecondRevision{$\rho(\tilde{f}_{ran},\G)$}
is a random variable and thus differs from 
the deterministic values
$\rho_{at}(\G),$ $\rho_{nat}(\G)$
and $\rho_{mix}(\G).$ 
Moreover, \WuSecondRevision{$\rho_{nat}(\G)$
differs from $\rho_{at}(\G)$ and $\rho_{mix}(\G),$
see Example~\ref{example:NonMutualInclusive}, 
in which $\rho_{nat}(\G)=\frac{18}{18-\sqrt{6}}
>\rho_{at}(\G)=\rho_{mix}(\G)=1.$
Although $\rho_{at}(\G)$ and $\rho_{mix}(\G)$
coincide in Example~\ref{example:NonMutualInclusive},
they differ in general, and $\rho_{mix}(\G)\ge \rho_{at}(\G).$
In particular, neither $\rho_{nat}(\G)\ge \rho_{at}(\G)$ nor 
$\rho_{nat}(\G)\ge \rho_{mix}(\G)$ holds in general, see, e.g., \citep{Christodoulou2005}. 
Thus the known convergence results of the non-atomic PoA 
in \citep{Colini2016On,Colini2017WINE,Colini2020OR}
and \citep{Wu2019} do not naturally carry over to
random, atomic and mixed PoAs.}


\WuSecondRevision{Due to the ``no free arc'' assumption in (G3), all PoAs are different from $\frac{0}{0},$ and take values in $[1,\infty)$. This follows from Lemma~\ref{lemma:SO-Cost-Comparision}, and the fact that the non-atomic SO cost
is strictly positive, see \citep{Wu2019Sensitivity}.}

\section{Convergence results of the PoAs in atomic congestion games}
\label{sec:MainResults}

\WuSecondRevision{We now analyze the convergence of the PoAs for
atomic congestion games with \emph{polynomial} cost functions,
i.e., all $\tau_a(\cdot)$ have the form
\begin{equation}\label{eq:GeneralPoly}
	\tau_a(x)=\sum_{l=0}^{\beta_a}\eta_{a,l}\cdot x^{\beta_a-l},
	\quad \forall x\in [0,\infty),
\end{equation}
where $\beta_a\ge 0$ is an \emph{integer} degree, and 
$\eta_{a,l},$ $l=0,\ldots,\beta_a, a\in A,$ are the coefficients.
Since all $\tau_a(\cdot)$ are nondecreasing and no arc can be used for free, see (G3), all leading coefficients 
$\eta_{a,0},$ $a\in A,$ are strictly  \emph{positive}. We 
assume, w.l.o.g., that all other coefficients $\eta_{a,l}$ are also \emph{non-negative}.
This will simplify our 
analysis. Note that this is not restrictive, and our results  carry over to arbitrary polynomial cost functions.
We will come back to this later in Section~\ref{subsec:AtomicPoA},
Section~\ref{subsec:MixedPoA}
and Section~\ref{sec:GeneralPoly}, respectively.}


\subsection{Convergence results for  polynomial cost functions of the same degree}
\label{sec:ConverganceAnalysis}

\WuSecondRevision{We consider first}  polynomial cost functions  $\tau_a(\cdot)$ of 
\WuSecondRevision{the same degree $\beta_a\equiv\beta\ge 0,$ i.e., they have}
\wu{the} form~\eqref{eq:PolynomialCost}
\begin{equation}\label{eq:PolynomialCost}
\tau_a(x)=\sum_{l=0}^{\beta} \eta_{a,l}\cdot x^{\beta-l}\quad \forall
x\ge 0\ \forall a\in A.
\end{equation}
 This covers  BPR cost functions,
\WuSecondRevision{which are of the simpler form $\eta_{a,0}\cdot x^{\beta}+\eta_{a,\beta}$ and}  frequently used in urban traffic
to \wu{model} travel latency, see \citep{BPR}.


\WuSecondRevision{With these cost functions,}
the total cost of a non-atomic SO flow
 \WuThird{is at least}
$\frac{T^{\beta+1}\cdot \eta_{0,\min}}{|\P|^{\beta+1}}>0$ when 
$T>0,$ where
$\eta_{0,\min}:=\min_{a\in A}\eta_{a,0}>0,$
see \citep{Wu2019Sensitivity}. \WuSecondRevision{Note that there is at least one path
with a flow value of at least $\frac{T}{|\P|}$ in 
	an arbitrary non-atomic SO flow. Note also, that $x\cdot \tau_a(x)\ge \eta_{0,\min}\cdot x^{\beta+1}$ for all $a\in A$ and all $x\ge 0.$}


\subsubsection{\wu{An upper bound for the atomic PoA}}
\label{subsec:AtomicPoA}

Theorem~\ref{thm:MainResults} presents
\wu{an} upper bound \wu{for} the atomic PoA
in congestion games
with polynomial cost functions of \wu{the} same degree, see \eqref{eq:PolynomialCost}. 
\WuSecondRevision{Here, $\eta_{\max}:=\max\{\eta_{a,l}:a\in A,l=0,\ldots,\beta\}
	\ge \eta_{0,\min}>0,$ and 
	$\kappa:=\beta\cdot \eta_{\max}\cdot \big(
	1+\sum_{l=1}^{\beta}\frac{1}{T^l}
	\big)>0,$ which is a \emph{Lipschitz bound} for the Lipschitz continuous functions 
$\frac{\tau_a(T\cdot x)}{T^\beta}$ on the compact interval $[0,1],$
i.e., $\kappa$ satisfies the condition that $|\frac{\tau_a(T\cdot x)}{T^\beta}-\frac{\tau_a(T\cdot y)}{T^\beta}|
\le \kappa\cdot |x-y|$ for all $x,y\in [0,1]$ and all $a\in A.$}
\begin{theorem}\label{thm:MainResults}
	Consider an arbitrary congestion game $\G=(\tau,\D,d)$ \wu{with cost functions} $\tau_a(\cdot)$ of the form~\eqref{eq:PolynomialCost}.
	\wu{If $\G$ has atomic NE flows, then} 
	\WuSecondRevision{\begin{displaymath}
		\begin{split}
		\rho_{at}(\G)
		&\!\le\! 1\!+\!\frac{\beta\cdot \eta_{\max}\cdot |\P|^{\beta\!+\!1}}{\eta_{0,\min}}
		\cdot \sum_{l=1}^{\beta}\!\frac{1}{T^l}\!+\!\frac{|A|\cdot \kappa\cdot |\P|^{\beta\!+\!1}}{\eta_{0,\min}}\cdot \sqrt{|\P|\cdot |A|\cdot \frac{d_{max}}{T}}\\
		&\hspace{3cm}+\frac{ |A|\cdot \kappa\cdot |\P|^{\beta+2}}{\eta_{0,\min}}\cdot\frac{d_{max}}{T}.
		\end{split}
	\end{displaymath}}
	Here, we use \WuSecondRevision{the} convention that 
	$\sum_{l=1}^{\beta}\frac{1}{T^l}=0$
	when $\beta=0.$
\end{theorem}

The upper bound  
holds for all $T$ and $d_{max},$
and converges to $1$ at a rate
of $O(\frac{1}{T})+O(\sqrt{\frac{d_{max}}{T}})$
as $T\to \infty$ and 
$\frac{d_{max}}{T}\to 0$.
So the atomic PoA 
\wu{decays}  to $1$ \wu{quickly when $\G$
	has atomic NE flows.}
\WuSecondRevision{Examples~\ref{example:Counterexample-Poly-Same-Degree}--\ref{example:Counter2} show that the conditions
``$T\to \infty$''
and 	``$\frac{d_{max}}{T}\to 0$'' are necessary
for this convergence.}

\begin{example}\label{example:Counterexample-Poly-Same-Degree}
	Consider \WuSecondRevision{an} unweighted congestion game 
	$\G$ with \WuSecondRevision{the} network 
	of Figure~\WuSecondRevision{\ref{fig:NE-Noninclusive}, but
	cost functions $x$ and $x+1$ for the upper and lower arc, respectively.}
	Assume that $\G$ has $|\D|=4\cdot n$ \WuSecondRevision{agents} \wu{with} $\frac{1}{4\cdot n}$ 
	\wu{demand each.}
	Then $T\equiv 1$
	and $d_{max}=\frac{1}{4\cdot n}.$
	Clearly, $\G$ has only one atomic NE flow
	$\tilde{f}_{at},$
	in which
	all \WuSecondRevision{agents} \wu{use} the upper arc.
	So 
	$C(\tilde{f}_{at},\G)= 1.$
	 $\G$
	has also a unique atomic SO flow 
	$f_{at}^*,$ in which $3\cdot n$ \WuSecondRevision{agents} \wu{use} the upper arc and
	the remaining $n$ \WuSecondRevision{agents} \wu{use} the lower arc.
	Then $C(f_{at}^*,\G)
	=\frac{7}{8}$,
	\wu{and} $\rho_{at}(\G)=\frac{8}{7}$ for all $n,$
	which does not converge to $1$ when only $\frac{d_{max}}{T}=d_{max}\to 0.$
\end{example}

\begin{example}\label{example:Counter2}
	Consider \WuSecondRevision{an} unweighted congestion game 
	$\G$ \WuSecondRevision{again} with \WuSecondRevision{the} network 
	\wu{of} Figure~\WuSecondRevision{\ref{fig:NE-Noninclusive}, but now
		\WuSecondRevision{with}
	cost functions $x$ and $2\cdot x$ for the upper and lower arc, respectively.}
	Assume that there are two \WuSecondRevision{agents}
	\wu{with demand} $n$ \wu{each.}
	\wu{Then} $T=2\cdot n,$
	which tends to $\infty$ as $n\to \infty.$
	However, $\frac{d_{max}}{T}\to \frac{1}{2}>0$ as $n\to \infty.$
	Obviously, $\G$
	has only one atomic SO flow $f^*_{at},$
	in which one \WuSecondRevision{agent} uses the upper 
	and the other  the lower arc.
	So $C(f^*_{at},\G)=3\cdot n^2$.
	However, $\G$
	has two atomic NE flows.
	One atomic NE flow is just the unique SO flow.
	\wu{In} the other atomic NE flow,
	both \WuSecondRevision{agents} use 
	the upper arc, \wu{and its} total cost \wu{is} $4\cdot n^2$.
	Consequently, $\rho_{at}(\G)=\frac{4}{3}\not\to 1$ \wu{as} $T=2\cdot n\to \infty.$
\end{example}

\WuSecondRevision{We now prove} Theorem~\ref{thm:MainResults} with the technique
of scaling
\wu{from} \WuSecondRevision{\citep{Colini2020OR} and \citep{Wu2019}}.


\begin{definition}[Scaled games, \citep{Wu2019}]\label{def:ScaledGames}
	Consider an arbitrary congestion game $\G=(\tau,\D,d)$
	with arbitrary cost functions,
	and an arbitrary constant $g>0.$
	\wu{The \emph{scaled game} of $\G$ w.r.t. \emph{scaling factor} $g$ is the} congestion game $\G^{[g]}=\big(\tau^{[g]},\D,\bar{d}\big)$
	\wu{whose} cost function vector $\tau^{[g]}:=(\tau_a^{[g]})_{a\in A}$ has
	a component
	$\tau^{[g]}_a(x):=\frac{\tau_a(x\cdot T)}{g}$
	for each pair $(a,x)\in A\times [0,1],$ and
	\wu{whose} demand vector
	$\bar{d}=(\bar{d}_{k,i})_{i\in\D_k,k\in \KK}$ has a component $\bar{d}_{k,i}:=\frac{d_{k,i}}{T}$ for  each $i\in\D_k$
	and each $k\in \KK.$
\end{definition}


Lemma~\ref{lemma:Scaling} shows that 
 scaling does not change the four PoAs. 
 \WuSecondRevision{We omit the straightforward proof.
 Note that a  flow
$f$ of $\G$ corresponds to a flow $f^{[g]}:=\frac{f}{T}$
of $\G^{[g]},$
and $C(f,\G)=C(f^{[g]},\G^{[g]})\cdot g\cdot T.$} 

\begin{lemma}\label{lemma:Scaling}
	Consider an arbitrary congestion game $\G,$
	an arbitrary
	mixed NE flow $\tilde{f}_{ran}$
	of $\G$,
	and an arbitrary scaling factor $g>0.$
	Let $\G^{[g]}$
	be \wu{the} scaled game with factor $g$.
	Then $\rho_{at}(\G^{[g]})=\rho_{at}(\G),$  
	$\rho_{nat}(\G^{[g]})=\rho_{nat}(\G),$ and $\rho_{mix}(\G^{[g]})=\rho_{mix}(\G).$
	Moreover, $\tilde{f}_{ran}^{[g]}:=\frac{\tilde{f}_{ran}}{T}$
	is a mixed NE flow of the scaled game $\G^{[g]},$ and $\rho(\tilde{f}_{ran}^{[g]},\G^{[g]})=\rho(\tilde{f}_{ran},\G).$
\end{lemma}

Lemma~\ref{lemma:Scaling}
enables us to
prove Theorem~\ref{thm:MainResults}
by bounding 
$\rho_{at}(\G^{[g]})$ instead
\WuSecondRevision{of $\rho_{at}(\G)$.}
\wu{We can thus} \wu{purely} concentrate 
on the influence of $\frac{d_{max}}{T}$ on the convergence, as 
\WuSecondRevision{the total demand of $\G^{[g]}$ is}
$\bar{T}=T(\D,\bar{d}):=\sum_{k\in\K,i\in\D_k}\bar{d}_{k,i}=1.$
\wu{However, the} scaling factor $g$ must be chosen carefully, 
so as to ensure that the total cost in $\G^{[g]}$
is \wu{moderate,} i.e., neither too large nor too small.
\WuSecondRevision{Following} \citep{Wu2019},
we use 
$g:=T^{\beta}$ \wu{for
polynomial cost functions of the same degree $\beta.$} 
Then $\G^{[g]}$
has the \emph{scaled} cost function
\begin{equation}\label{eq:ScaledCost}
\tau^{[g]}_a(x)=\frac{\sum_{l=0}^{\beta}\eta_{a,l}\cdot \big(T\cdot x\big)^{\beta-l}}{g}
=\eta_{a,0}\cdot x^{\beta}+\sum_{l=1}^{\beta} \frac{\eta_{a,l}}{T^{l}}\cdot x^{\beta-l}
\end{equation}
for arc $a\in A,$ \wu{the} \emph{bounded} demand  $\bar{d}_{k,i}
=\frac{d_{k,i}}{T}\in [0,1]$
for
$i\in \D_k,$ \WuSecondRevision{and} the \emph{bounded} demand 
$\bar{d}_k:=
\frac{d_k}{T}\in [0,1]$
for $k\in\K.$
%
\WuSecondRevision{Consequently, each flow 
$f^{[g]}$
of $\G^{[g]}$ has \emph{bounded } arc flow values $f^{[g]}_{a}\in [0,1]$, and
$C(f^{[g]},\G^{[g]})\ge \frac{\eta_{0,\min}}{|\P|^{\beta+1}}.$}

\WuSecondRevision{Definition~\eqref{eq:PoA-Atomic-def} of the atomic PoA and
Lemma~\ref{lemma:SO-Cost-Comparision} together imply that}
\begin{equation}\label{eq:Non-Atomic-Transformation}
\rho_{at}(\G)
=\!\rho_{at}(\G^{[g]})
\le\! \rho_{nat}(\G^{[g]})\!+\!
\frac{|\max_{\tilde{f}^{[g]}_{at}}\  C(\tilde{f}^{[g]}_{at},\G^{[g]})\!-\!C(\tilde{f}^{[g]}_{nat},\G^{[g]})|}{C(f_{nat}^{*[g]},\G^{[g]})},
\end{equation}
where $\tilde{f}_{nat}^{[g]}$ and
$f^{*[g]}_{nat}$ are arbitrary non-atomic NE and SO flows of
$\G^{[g]}$, respectively, and
the maximization is taken over all atomic NE flows 
$\tilde{f}_{at}^{[g]}$ of $\G^{[g]}.$
%
%
\WuSecondRevision{With~\eqref{eq:Non-Atomic-Transformation},} we can then prove Theorem~\ref{thm:MainResults}
by \WuSecondRevision{upper bounding} 
\begin{equation}\label{eq:ScaledGameObj}
|\max_{\tilde{f}^{[g]}_{at}}\  C(\tilde{f}^{[g]}_{at},\G^{[g]})-C(\tilde{f}^{[g]}_{nat},\G^{[g]})|
\end{equation}
\wu{and} 
$\rho_{nat}(\G^{[g]})$, \wu{respectively.}
Here, we \WuSecondRevision{observe} that  
	$C(f_{nat}^{*[g]},\G^{[g]})\ge \frac{\eta_{0,\min}}{|\P|^{\beta+1}}>0.$
%
To \WuSecondRevision{that} end, we need \wu{the} notion of \emph{$\epsilon$-approximate
	non-atomic NE flow} \WuSecondRevision{and a result} from \citep{Wu2019Sensitivity}.

\begin{definition}\label{def:Epsilon-Approximate-NE-Flow}
		We call an arbitrary non-atomic flow $f_{nat}$ of $\G$ an \emph{$\epsilon$-approximate non-atomic NE flow} for a constant $\epsilon>0$
	if 
	$\sum_{a\in A}\tau_a(f_{nat,a})\cdot (f_{nat,a}
	-f'_{nat,a})\le \epsilon$
	for an arbitrary non-atomic flow 
	$f'_{nat}$ of $\G.$
\end{definition}

\citep{Wu2019Sensitivity} \wu{have shown} 
that the total cost difference between 
$\epsilon$-approximate and accurate non-atomic NE flows 
is in $O(\sqrt{\epsilon}),$ see Lemma~\ref{lemma:Ep-Approxi}.
\begin{lemma}[\citep{Wu2019Sensitivity}]\label{lemma:Ep-Approxi}
	Consider an arbitrary congestion game 
	$\G=(\tau,\D,d)$ with a total demand
	of $1$ \wu{ and 
		an arbitrary $\epsilon$-approximate non-atomic NE flow $\tilde{f}^{\epsilon}_{nat}$.} If all cost functions 
	are Lipschitz continuous (or Lipschitz bounded) on $[0,1]$ with
	a \WuSecondRevision{Lipschitz} constant $\kappa>0,$ i.e.,
	$|\tau_a(x)-\tau_a(y)|\le \kappa\cdot |x-y|$
	for all $(a,x,y)\in A\times [0,1]^2,$
	then \WuSecondRevision{$|C(\tilde{f}_{nat},\G)-C(\tilde{f}_{nat}^\epsilon,\G)|\le |A|
		\cdot \sqrt{\kappa \cdot \epsilon}+\epsilon,$
	and $|\tau_a(\tilde{f}_{nat,a})-
	\tau_a(\tilde{f}_{nat,a}^\epsilon)|\le \sqrt{\kappa\cdot \epsilon}$}
	for all $a\in A$ and all  non-atomic NE flows $\tilde{f}_{nat}$.
\end{lemma} 

\WuSecondRevision{Lemma~\ref{lemma:ApproximateWE} below shows
that $\tilde{f}_{at}^{[g]}$ is an $O(\frac{d_{max}}{T})$-approximate
non-atomic NE flow of $\G^{[g]}$. Then Lemma~\ref{lemma:Ep-Approxi}
yields a desired upper bound for \eqref{eq:ScaledGameObj},
see Lemma~\ref{lemma:ApproximateWE}c. We move the proof of Lemma~\ref{lemma:ApproximateWE} to Appendix~\ref{proof:ApproximateWE}.}


\begin{lemma}\label{lemma:ApproximateWE}
	Consider \wu{an arbitrary} congestion game $\G$
	\wu{as} in Theorem~\ref{thm:MainResults}. 
	Let $\G^{[g]}$ be its scaled game
	with factor $g=T^{\beta},$
	and let $\tilde{f}^{[g]}_{at}$
	and $\tilde{f}^{[g]}_{nat}$
	be arbitrary
	atomic and non-atomic NE flows,
	 respectively. 
	\WuSecondRevision{Then:}
	\begin{itemize}
		\item[a)] 
		$
		\tau^{[g]}_{p}\big(\tilde{f}_{at}^{[g]}\big)\le \tau^{[g]}_{p'}\big(\tilde{f}^{[g]}_{at}\big)+\frac{|A|\cdot \kappa\cdot d_{max}}{T}
		$
		for all $k\in\KK$ and all $p,p'\in \P_k$ with $f^{[g]}_{at,p}>0.$ 
		\item[b)] $\tilde{f}_{at}^{[g]}$
		is a $\frac{|\P|\cdot |A|\cdot \kappa\cdot d_{max}}{T}$-approximate non-atomic NE flow
		of $\G^{[g]}.$ 
		
		\item[c)] 
		$
		|C(\tilde{f}_{at}^{[g]},\G^{[g]})\!-\!C(\tilde{f}_{nat}^{[g]},\G^{[g]})|\!\le\! |A|\!\cdot\! \kappa\!\cdot\! \sqrt{|\P|\!\cdot\! |A|\!\cdot\! \frac{d_{max}}{T}}\!+\!|\P|\!\cdot\! |A|\!\cdot\! \kappa\!\cdot\!\frac{d_{max}}{T}.
		$
	\end{itemize}
\end{lemma}

Lemma~\ref{lemma:Non-Atomic-PoA-Bound} \WuSecondRevision{yields} an upper bound
for $\rho_{nat}(\G^{[g]}),$
which results in a convergence rate
of $O(\frac{1}{T}).$   Note that \citep{Wu2019} have shown a
stronger convergence rate of
$o(\frac{1}{T^{\beta}})$ for
BPR cost functions, and that \citep{Colini2020OR} have  shown a similar
rate \WuSecondRevision{as in Lemma~\ref{lemma:Non-Atomic-PoA-Bound}} for arbitrary
polynomial cost functions under
the condition that $\frac{d_k}{T}\ge \xi_k>0$
for some constant $\xi_k$ independent of $T$ for each $k\in\K.$
We move the proof of Lemma~\ref{lemma:Non-Atomic-PoA-Bound}
to Appendix~\ref{proof:Non-Atomic-PoA-Bound}.
\begin{lemma}\label{lemma:Non-Atomic-PoA-Bound}
	Consider \wu{an arbitrary} congestion game $\G$
	as in Theorem~\ref{thm:MainResults}.
	Let $\G^{[g]}$
	be the scaled game
	with scaling factor 
	$g=T^{\beta}.$ 
	Then 
	$
	\rho_{nat}(\G)=\rho_{nat}(\G^{[g]})
	\le 1+\frac{\beta\cdot \eta_{\max}\cdot |\P|^{\beta+1}}{\eta_{0,\min}}
	\cdot \sum_{l=1}^{\beta}\frac{1}{T^l}.
	$
\end{lemma}

Theorem~\ref{thm:MainResults} \wu{then} follows from Lemma~\ref{lemma:Scaling},  \eqref{eq:Non-Atomic-Transformation}, 
Lemma~\ref{lemma:ApproximateWE}c and
Lemma~\ref{lemma:Non-Atomic-PoA-Bound}.

\WuSecondRevision{The above proofs build essentially on inequality~\eqref{eq:Non-Atomic-Transformation}, Lemma~\ref{lemma:Ep-Approxi}
	and the Lipschitz continuity of the scaled cost functions $\tau^{[g]}_a(\cdot)$ on 
	$[0,1],$ but not on the sign of the coefficients $\eta_{a,l},$ $l=1,\ldots,\beta, a\in A.$
	Thus Theorem~\ref{thm:MainResults} indeed carries over to 
	arbitrary polynomial cost functions of the same degree $\beta\ge 0.$
When $\eta_{a,l}<0$ for some terms $l=1,\ldots,\beta$ and some arcs $a\in A,$ then $\frac{\eta_{0,\min}}{|\P|^{\beta+1}}$ may be larger than $C(f_{nat}^{*[g]},\G^{[g]}).$  Instead, $C(f_{nat}^{*[g]},\G^{[g]})$
can be bounded from below by $\min_{a\in A} \frac{1}{|\P|}\cdot \tau^{[g]}_a(\frac{1}{|\P|})\in \Theta(1)$.
The Lipschitz bound for the scaled cost functions is still
$\kappa=\beta\cdot \eta_{\max}\cdot \big(
1+\sum_{l=1}^{\beta}\frac{1}{T^l}
\big)>0,$ but with $\eta_{\max}:=\{|\eta_{a,l}|: a\in A,l=0,1,\ldots,\beta\}.$
Lemma~\ref{lemma:ApproximateWE} then still holds, since its proof
in Appendix~\ref{proof:ApproximateWE} does not
involve the sign of coefficients $\eta_{a,l}$, but only the Lipschitz continuity
of the scaled cost functions on $[0,1]$.
Although the proof
of Lemma~\ref{lemma:Non-Atomic-PoA-Bound} in Appendix~\ref{proof:Non-Atomic-PoA-Bound}
does involve the sign of coefficients $\eta_{a,l}$, it can be adapted accordingly.}


\subsubsection{\wu{Upper bounds for the mixed PoA and the random PoA}}
\label{subsec:MixedPoA}

\WuSecondRevision{Theorem~\ref{thm:WeightedCase} below proves similar upper
bounds for $\rho(\tilde{f}_{ran},\G)$ and $\rho_{mix}(\G),$  respectively, in terms
of $T,$ $\frac{d_{max}}{T}$ and constants $M_i,$ $i=1,\ldots,5$.
\WuSecondRevision{We hide the detailed values of these constants
	$M_i$ in Theorem~\ref{thm:WeightedCase}, since
	they are complicated expressions. Interested readers
	may find their values in the proof.}
When $T\to\infty$
and $\frac{d_{max}}{T}\to 0,$ these upper bounds  converge
(with an overwhelming probability for $\rho(\tilde{f}_{ran},\G)$)
to $1$ at a rate of $O(\frac{1}{T})+O(\frac{d^{1/6}_{max}}{T^{1/6}}).$
Note that $\rho_{mix}(\G)$  converges more slowly than 
$\rho_{at}(\G)$ since
$\rho_{at}(\G)\le \rho_{mix}(\G).$}
\WuSecondRevision{\begin{theorem}\label{thm:WeightedCase}
	Consider the same congestion game $\G$ as in Theorem~\ref{thm:MainResults}.
	Let  $\tilde{f}_{ran}$ be an arbitrary mixed NE flow
	of $\G.$
	Then the following statements hold.
	\begin{itemize}
		\item[a)] The random
		event ``$\rho(\tilde{f}_{ran},\G)
		\le 1+M_1\cdot \frac{1}{T}$
		$+M_2\cdot \frac{d_{max}^{1/6}}{T^{1/6}}$''
		occurs with a probability
		\wu{of at least} $1-M_3\cdot \frac{d_{max}^{1/3}}{T^{1/3}}.$
		
		\item[b)] $\rho_{mix}(\G)\le 1+M_4\cdot \frac{1}{T}+M_5\cdot \frac{d_{max}^{1/6}}{T^{1/6}}$
	\end{itemize}
Herein, $M_i>0,$ $i=1,\ldots,5,$ are constants independent of $d_{max}$
and $T.$ 
\end{theorem}}





\wu{We also prove}
Theorem~\ref{thm:WeightedCase}
with the scaled game $\G^{[g]}$ \wu{and Lemma~\ref{lemma:Scaling}}.
\WuSecondRevision{Let $\tilde{f}^{[g]}_{nat}$ and 
$f^{*[g]}_{nat}$ be an arbitrary non-atomic NE flow and
an arbitrary non-atomic SO flow of $\G^{[g]},$ respectively.}
\wu{We obtain} by Lemma~\ref{lemma:SO-Cost-Comparision},
\eqref{eq:MixedPoA} and \eqref{eq:randomPoA} that
\begin{equation}\label{eq:Mixed-Trans}
\rho_{mix}(\G^{[g]})\!\le 
\rho_{nat}(\G^{[g]})+
\frac{|\max_{\tilde{f}^{[g]}_{ran}}\  \E_{\tilde{\Pi}}[C(\tilde{f}^{[g]}_{ran},\G^{[g]})]\!-\!C(\tilde{f}^{[g]}_{nat},\G^{[g]})|}{C(f_{nat}^{*[g]},\G^{[g]})},
\end{equation}
and that
\begin{equation}\label{eq:random-Trans}
\rho(\tilde{f}^{[g]}_{ran},\G^{[g]})\!\le\! \rho_{nat}(\G^{[g]})+
\frac{|C(\tilde{f}^{[g]}_{ran},\G^{[g]})\!-\!C(\tilde{f}^{[g]}_{nat},\G^{[g]})|}{
	C(f_{nat}^{*[g]},\G^{[g]})},
\end{equation}
\WuSecondRevision{where $\tilde{f}^{[g]}_{ran}$ is an arbitrary
mixed NE flow of $\G^{[g]}.$}
\wu{Using} Lemma~\ref{lemma:Non-Atomic-PoA-Bound},
we now only need to \wu{derive} upper bounds 
for \wu{the numerators} of the two fractions in \eqref{eq:Mixed-Trans} \wu{and} \eqref{eq:random-Trans}, respectively.


\WuSecondRevision{Lemma~\ref{lemma:ExpectedFlowIsEpWE}a below
shows that the expected flow $\E_{\tilde{\Pi}}(\tilde{f}^{[g]}_{ran})$
of a mixed NE $\tilde{f}^{[g]}_{ran}$
is
an $\epsilon$-approximate non-atomic NE flow
with $\epsilon\in O(\frac{d_{max}^{1/3}}{T^{1/3}}).$
Lemma~\ref{lemma:Ep-Approxi} then yields
$|C(\E_{\tilde{\Pi}}(\tilde{f}^{[g]}_{ran}),
\G^{[g]})-C(\tilde{f}^{[g]}_{nat},
\G^{[g]})|$  $
\in$ $O(\frac{d_{max}^{1/6}}{T^{1/6}}).$
Then Lemma~\ref{lemma:ExpectedFlowIsEpWE}b--c
upper bound the total cost difference between a mixed NE flow $\tilde{f}^{[g]}_{ran}$ and its expected flow
$\E_{\tilde{\Pi}}(\tilde{f}^{[g]}_{ran})$ both in expectation and as a random variable.
Moreover, Lemma~\ref{lemma:ExpectedFlowIsEpWE} together with
Lemma~\ref{lemma:Non-Atomic-PoA-Bound} and \eqref{eq:Mixed-Trans}--\eqref{eq:random-Trans}
prove Theorem~\ref{thm:WeightedCase}.}
%
We move the detailed proof
of Lemma~\ref{lemma:ExpectedFlowIsEpWE} to Appendix~\ref{proof:ExpectedFlowIsEpWE}.
\begin{lemma}
	\label{lemma:ExpectedFlowIsEpWE}
	Consider \wu{the} congestion game $\G$ in Theorem~\ref{thm:WeightedCase},
	and \wu{the} scaling factor $g=T^{\beta}.$
	Let $\G^{[g]}$ be \wu{the} scaled game with factor $g,$ and let 
	$\tilde{f}^{[g]}_{ran}$ be an arbitrary 
	mixed NE flow of $\G^{[g]}$ \wu{with mixed} profile
	$\tilde{\Pi}.$
	\begin{itemize}
		\item[a)] When $\beta>0,$ \wu{then} the expected flow $\E_{\tilde{\Pi}}(\tilde{f}^{[g]}_{ran})$ is an $\epsilon$-approximate non-atomic NE flow
		with 
		\WuSecondRevision{$
		\epsilon=3\cdot |\P|\cdot \kappa\cdot |A|\cdot \big(1+\frac{|A|}{4\cdot \beta}\big)
		\cdot \big(\frac{d_{max}}{T}\big)^{1/3},
		$}
		and 
		$
		|C(\tilde{f}^{[g]}_{nat},\G^{[g]})
		-C\big(\E_{\tilde{\Pi}}(\tilde{f}^{[g]}_{ran}),\G^{[g]}\big)|\le |A|
		\cdot \sqrt{\kappa\cdot\epsilon}+\epsilon\in O(\frac{
			d_{max}^{1/6}
		}{T^{1/6}})
		$
		for an arbitrary non-atomic NE flow
		$\tilde{f}^{[g]}_{nat}$ of 
		$\G^{[g]}.$
		When $\beta=0,$ \wu{then} $\E_{\tilde{\Pi}}(\tilde{f}^{[g]}_{ran})$
		is \wu{a} non-atomic NE flow of
		$\G^{[g]}.$
		\item[b)] 
		Consider an arbitrary 
		constant $\delta\in (0,1/2).$ 
		The event
		``$
		|C(\tilde{f}^{[g]}_{ran},\G^{[g]})-
		C\big(\E_{\tilde{\Pi}}(\tilde{f}^{[g]}_{ran}),\G^{[g]}\big)|\le 
		|A|\cdot \big(\kappa+\eta_{\max}\cdot \sum_{l=0}^{\beta}\frac{1}{T^l}\big)\cdot \big(\frac{d_{max}}{T}\big)^{\delta}
		$'' \wu{occurs with a probability of at least 
			$1- \frac{|A|}{4}\cdot \big(\frac{d_{max}}{T}\big)^{1-2\cdot\delta}.$}
		\item[c)] $
		\Big|\E_{\tilde{\Pi}}\big[C(\tilde{f}^{[g]}_{ran},\G^{[g]})\big]-C\big(\E_{\tilde{\Pi}}(\tilde{f}^{[g]}_{ran}),\G^{[g]}\big)\Big|
		\le |A|\cdot 
		\big(\kappa+\big(1+\frac{|A|}{4}\big)\cdot\eta_{\max}\cdot \sum_{l=0}^{\beta}\frac{1}{T^l}\big)\cdot \big(\frac{d_{max}}{T}\big)^{1/3}.
		$
	\end{itemize} 
\end{lemma}

\WuSecondRevision{Similar to the proof for Lemma~\ref{lemma:ApproximateWE}
	in Appendix~\ref{proof:ApproximateWE},
	the proof of Lemma~\ref{lemma:ExpectedFlowIsEpWE} 
	in Appendix~\ref{proof:ExpectedFlowIsEpWE} 
does neither involve the sign of the coefficients $\eta_{a,l}$, but 
only the Lipschitz continuity of the scaled cost functions on $[0,1]$ and the finite upper bound 
$\max_{a\in A}\tau_a^{[g]}(1)$ $\in$ $\Theta(1).$ Hence, 
Lemma~\ref{lemma:ExpectedFlowIsEpWE} carries also over to 
arbitrary polynomial cost functions of the same degree, and so does
Theorem~\ref{thm:WeightedCase}.}

\WuSecondRevision{Note that \citep{Cominetti2020} have shown that the mixed NE flow
$\tilde{f}_{ran}$
of an atomic congestion game $\G$ converges in distribution to a non-atomic NE flow $\tilde{f}_{nat}$
of a limit non-atomic congestion game $\G^{(\infty)}$
when the cost functions $\tau_a$
are strictly increasing, $T\to
T_0$ for a constant $T_0>0,$ $d_{max}\to 0,$ and 
the number $|\D|$ of agents tends to $\infty.$
Combined with the scaling technique, this 
may imply also  that the mixed PoA in the scaled game $\G^{[g]}$ converges to $1$
for polynomial cost functions of the same degree
when $\frac{d_{max}}{T}\to 0$ as $T\to \infty$,
%
although  the cost functions of the atomic congestion games in 
the analysis of \citep{Cominetti2020}
are fixed and equal those of the limit non-atomic congestion game, and although
the scaled cost functions $\tau_a^{[g]}$ here depend on $T$ and vary
with the growth of $T$.
%
While implying a
similar convergence, we  aim at upper bounding
the mixed and random PoAs, and so have results for arbitrary
demand vectors $d$, i.e.,
neither need $T\to \infty$ nor need
$\frac{d_{max}}{T}\to 0$ in the proofs.}
\WuSecondRevision{Moreover, the results of
\citep{Cominetti2020} do not imply the convergence of the mixed PoA in atomic congestion games
with arbitrary polynomial cost functions for growing total demand, since then the 
atomic congestion games cannot be scaled to have a unified limit non-atomic congestion game
for all O/D pairs, see \citep{Wu2019}.}



\subsection{Concergence results for polynomial cost functions with arbitrary degrees}
\label{sec:GeneralPoly}

\WuSecondRevision{We consider now polynomial
cost functions with arbitrary degrees,} \WuSecondRevision{i.e.,
$\beta_{a}\ne \beta_{a'}$ may hold for \wu{some} \wu{arcs} $a\ne a'.$}
%
\WuSecondRevision{Example~\ref{example:Divergence} below
shows that the conditions
``$\frac{d_{max}}{T}\to 0$''
and ``$T\to \infty$'' are no \WuSecondRevision{longer} sufficient for the convergence of $\rho_{mix}(\G)$
and $\rho_{at}(\G)$
in this case.}
\begin{example}\label{example:Divergence}
	Consider a congestion game $\G$ with \wu{the}
	network \wu{of} Figure~\ref{fig:DivergencePoA}. $\G$ has two non-overlapping
	O/D pairs $(o_1,t_1)$
	and $(o_2,t_2),$ and 
	\wu{both} of \wu{them have} two parallel arcs.
	Assume that $(o_1,t_1)$
	has $2\cdot \sqrt{n}$ agents with \wu{each a} demand of $\sqrt{n},$
	and \wu{that} $(o_2,t_2)$
	has $2$ agents with \wu{the same} demand of $\sqrt{n}$
	\wu{each}.
	\wu{So}  $d_{max}=\sqrt{n}.$
	\wu{Then, as $n\to\infty,$} $T=2\cdot n+2\cdot \sqrt{n}\to \infty$ and
	$\frac{d_{max}}{T}=\frac{\sqrt{n}}{2\cdot n+2\cdot \sqrt{n}}
	\to 0$.
	\begin{figure}[!htb]
		\centering
		\begin{tikzpicture}[
		>=latex
		]
		\node[scale=0.4,circle,fill=black,label=left:{$o_1$}](o1){};
		\node[scale=0.4,circle,fill=black,right =2of o1,label=right:{$t_1$}](t1){};
		\node[scale=0.4,circle,right =2of t1,fill=black,label=left:{$o_2$}](o2){};
		\node[scale=0.4,circle,fill=black,right =2of o2,label=right:{$t_2$}](t2){};
		\draw[->,thick] (o1) to [out=90,in=90] node[above]{$x$}(t1);
		\draw[->,thick] (o1) to [out=-90,in=-90] node[below]{$x$}(t1);
		\draw[->,thick] (o2) to [out=90,in=90] node[above]{$x^3$}(t2);
		\draw[->,thick] (o2) to [out=-90,in=-90] node[below]{$8\cdot x^3+1$}(t2);
		\end{tikzpicture}
		\caption{The PoA \WuSecondRevision{need not} converge to $1$}
		\label{fig:DivergencePoA}
	\end{figure}
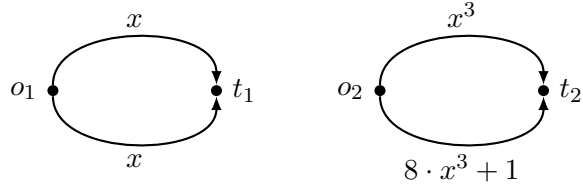
	However, 
	$\rho_{at}(\G)\to\frac{16}{9}>1$ 
	as $n\to \infty.$ \wu{This follows since} 
	$\G$ has the worst-case total cost
	of $2\cdot n+16\cdot n^{2}$
	for atomic NE flows, and 
	the total cost of $2\cdot n+9\cdot n^2+\sqrt{n}$
	for atomic SO flows \wu{when $n$ is large.}
\end{example}

\WuSecondRevision{While the game $\G$ in Example~\ref{example:Divergence}
is artificial, it shows that the convergence of the PoAs can be ruined by O/D pairs with small demands but polynomial cost functions of higher degrees, since they may dominate the PoAs completely when $T\to \infty$
and $d_{max}$ is unbounded.}
%
%
\WuSecondRevision{To ensure the convergence of the PoAs for polynomial cost functions of
	arbitrary degrees, we may thus need to impose a stronger condition that $d_{max}$
	is bounded when $T\to\infty.$
Theorem~\ref{thm:GeneralPoly} below confirms this.}
%
\begin{theorem}\label{thm:GeneralPoly}
	Consider an arbitrary congestion game $\G$ 
	with cost functions $\tau_a(\cdot)$ defined in
	\eqref{eq:GeneralPoly}. Assume that  $d_{max}$ is bounded from above by a constant
	$\upsilon>0$ independent of
	$T.$
	Then the following statements hold.
	\begin{itemize}
		\item[a)]  $
		\frac{\max_{\tilde{f}_{ran}}C\left(\E_{\tilde{\Pi}}\big(\tilde{f}_{ran}\big),\G\right)}{C(f^*_{nat},\G)}\to 1
		$
		as $T\to \infty,$
		where the maximization in the
		\wu{numerator} \wu{is taken over} all possible 
		mixed NE flows $\tilde{f}_{ran}$
		of $\G.$
		\item[b)] 
		$\rho_{mix}(\G)\to 1$ as
		$T\to \infty.$
		\WuThird{\item[c)] 
			If $\G$ has atomic flows for all demand
			vectors $d,$ if $\beta_{max}=\max_{a\in A}\beta_a>0,$ and if $\frac{d_k}{T}=\frac{\sum_{i\in\D_k}d_{k,i}}{T}\ge \xi_k >0$
			for all $k\in\K$ and some constants $\xi_k >0$ independent of $T,$ then $\rho_{at}(\G) = 1+O(T^{-\frac{1}{2\cdot \beta_{max}}}).$}
	\end{itemize}
\end{theorem}

\WuSecondRevision{Theorem~\ref{thm:GeneralPoly}a
	states  that the expected flow $\E_{\tilde{\Pi}}[\tilde{f}_{ran}]$ of a mixed NE flow 
	$\tilde{f}_{ran}$ is as efficient as 
	a non-atomic SO flow for large $T$ when the polynomial cost functions have arbitrary degrees and
	$d_{max}$ is bounded.
	Theorem~\ref{thm:GeneralPoly}b then shows that $\rho_{mix}(\G)$ converges to $1$ for growing $T$
	in this more general case.
	Hence, if the atomic NE flows exist, then $\rho_{at}(\G)\to 1$
	as $T\to \infty,$ since $\rho_{at}(\G)\le \rho_{mix}(\G).$  
	In addition to the pure convergence in Theorem~\ref{thm:GeneralPoly}a--b, Theorem~\ref{thm:GeneralPoly}c shows \WuThird{that $\rho_{at}(\G)$ converges at a rate of 
	$O(T^{-\frac{1}{2\cdot\beta_{max}}})$}. This
	demonstrates how fast the convergence of the PoAs can be in this more general case, \WuThird{when} each O/D pair demand $d_k$ has
	a positive ratio
	$\frac{d_k}{T}$ as $T\to \infty.$
	So far, we are unable to remove this restrictive condition, as we do not see a way to compute
	a concrete upper bound  in terms of $\frac{1}{T}$ for $\rho_{at}(\G)$ when the cost functions
	have \emph{different} degrees and the O/D pairs have significantly \emph{asynchronous} demand growth rates.
}
%
%

\WuSecondRevision{Theorem~\ref{thm:GeneralPoly}c can be proved by
a scaling technique similar to the proofs of Theorem~\ref{thm:MainResults} and Theorem~\ref{thm:WeightedCase}. 
However, due to the absence of a unified scaling factor, similar arguments will not be applicable in the proofs of Theorem~\ref{thm:GeneralPoly}a--b,
for which we need a more sophisticated technique called \emph{asymptotic decomposition} developed by \citep{Wu2019}.
%
%
%
%
%
%
%
%
%
%
%
In fact, Example~\ref{example:Divergence} \wu{has shown} that different 
O/D pairs $k\in\K$
may have significantly \emph{discrepant} influences 
on the limits of the PoAs for polynomial cost functions with arbitrary degrees.
These discrepant influences are caused by the different degrees of polynomial cost functions 
and the asynchronous growth
rates of the demands of the O/D pairs.
The asymptotic decomposition technique enables us to capture
these discrepant influences from different O/D pairs $k\in\K$.
It puts O/D pairs $k\in\K$
with a similar influence on the limits
of the PoAs
together to form a ``subgame'', then analyzes
the resulting subgames independently and combines
	the convergence results for these subgames to a convergence result for the whole game $\G$.
Interested readers may refer to \citep{Wu2019} for a detailed introduction of this
general technique.}
\WuSecondRevision{We move a description of the asymptotic decomposition and the very long proof of Theorem~\ref{thm:GeneralPoly} to Appendix~\ref{proof:MainStep} 
	in order to save space and improve readability.}

\WuSecondRevision{Although we have assumed at the beginning
of Section~\ref{sec:MainResults} that the polynomial cost functions have
only non-negative coefficients, the proof of Theorem~\ref{thm:GeneralPoly}a--b
in Appendix~\ref{proof:MainStep}
is essentially independent of this condition. 
The proof of Theorem~\ref{thm:GeneralPoly}c uses the nonnegativity of the coefficients
to obtain explicit lower and upper bounds of the scaled cost function values on the domain $[0,1]$, 
which carries also over to polynomials of arbitrary degrees when we slightly adapt
the constants in those bounds.
Hence, the convergence results in Theorem~\ref{thm:GeneralPoly} hold
for arbitrary polynomial cost functions, even with non-negative real-valued exponents.}

\WuSecondRevision{With the asymptotic decomposition, the
	convergence results for the non-atomic PoA in \citep{Wu2019},
	and Lemma~\ref{lemma:SO-Cost-Comparision}, we can actually show in the proof that
all the flows, $\tilde{f}_{ran},$ $\E_{\tilde{\Pi}}(\tilde{f}_{ran}),$ $\tilde{f}_{nat},$ $f^{*}_{at},$ 
$f^*_{nat},$ are equally efficient when $T\to\infty$ and $d_{max}$ is bounded,
see \eqref{eq:AD-Obj-Mixed} in Appendix~\ref{proof:MainStep}.
In particular, to obtain the convergence results in Theorem~\ref{thm:GeneralPoly}a--b, we 
have considered a mixed NE flow as an approximate mixed WE flow
(see Remark~\ref{remark:Mixed-WE}) in the proof, and 
so these convergence results carry also over to the ``PoA'' of mixed WE flows.
Hence, we need not distinguish between atomic and non-atomic congestion games for quantifying
the inefficiency of selfish choices of users, when the cost functions are polynomials, the total demand
$T$ is large, and the individual maximum demand $d_{max}$ is bounded.}

\section{Summary}
\label{sec:Summary}

\WuSecondRevision{We \wu{have} studied the inefficiency of 
both pure and mixed Nash equilibria in atomic congestion games with unsplittable demands.}

\WuSecondRevision{When \wu{the} cost functions are polynomials of
\wu{the} same degree,
we \wu{derive} upper bounds for the atomic, mixed and
random PoAs, respectively. These upper bounds \wu{tend} to $1$ 
quickly as $T\to \infty$
and  $\frac{d_{max}}{T}\to 0$.}

\WuSecondRevision{When \wu{the} cost functions are polynomials of arbitrary degrees
and $d_{max}$ is bounded, we show that
the mixed PoA converges again to $1$
as $T\to\infty.$ 
Moreover, we \wu{illustrate} that \wu{this} need not
hold \wu{when} $d_{max}$ is unbounded.
To demonstrate the convergence rates in this more general case, we show in addition that the atomic PoA 
converges to $1$ at a rate of \WuThird{$O(T^{-\frac{1}{2\cdot \beta_{max}}})$}
under the relatively restrictive condition that all O/D pairs have demand proportions $\frac{d_k}{T}$ 
that do not vanish when $T\to \infty.$
 \wu{However, it is still open and challenging to obtain concrete convergence rates
 	without this condition.}}
 
 \WuSecondRevision{Nevertheless,
our results already imply, under rather mild conditions, that \wu{pure and mixed Nash equilibria}
in atomic congestion games with
large unsplittable demands need not be bad.
This, together with
studies of
\citep{Colini2016On,Colini2017WINE,Colini2020OR} and \citep{Wu2019}, \wu{indicates} that 
\wu{the} selfish \wu{choice of strategies leads to a near-optimal behavior
in arbitrary congestion games with large total demands}, regardless whether users \wu{choose} mixed or pure strategies,
and whether the \wu{demands} are splittable or not.}

 \WuSecondRevision{The convergence rate} of the PoAs for arbitrary polynomial cost functions \WuSecondRevision{under
arbitrary demand growth pattern remains an important future research topic. It} is a crucial step for further bounding the PoAs in a congestion game with a high demand and arbitrary analytic cost functions. Note that
	analytic cost functions can be approximated with polynomials, and that the 
	\wump{H{\"o}lder continuity results} in \citep{Wu2019Sensitivity} seem to indicate that this approximation \WuSecondRevision{of} analytic cost functions may also be used for the PoAs.

While pure Nash equilibria 
	need not exist in \WuSecondRevision{arbitrary finite games,} \citep{Nash1950} has shown that every finite game
	has a mixed Nash equilibrium. 
	Since the user choices in a mixed Nash equilibrium are \WuSecondRevision{random,} the
	probability distribution of the random PoA might be a more
	\WuSecondRevision{suitable} measure for the inefficiency of mixed Nash equilibria.
	Our analysis of the random PoA for atomic congestion games
	with polynomial cost functions of the same degree
	has already provided the first positive evidence \WuSecondRevision{in that direction,} which may 
	apply also to finite games of other types. Thus another important
	future research topic is  to generalize \WuSecondRevision{the}
	probabilistic analysis of the random PoA to finite games of other types.

In our study, we have assumed that the cost functions
are \emph{separable}, i.e., each arc $a\in A$ has a cost function
depending only on its own flow value $f_a$. 
However, it may happen also that the cost of some arc
	$a\in A$
	depends not only on $f_a$, but also on
	flow values $f_b$ of other arcs $b\in A$.
	Then the cost functions are called \emph{non-separable},
	see, e.g., \citep{Perakis2007}.
\WuSecondRevision{A convergence analysis of atomic,} mixed and non-atomic PoAs for congestion games
with \emph{non-separable} cost functions would also be 
	an interesting future research topic,
\wu{as worst-case upper bounds of the non-atomic PoA in
	such games have already been obtained by \citep{Chau2003} and \citep{Perakis2007}.
\WuSecondRevision{In fact, the expected flow $\E_{\Pi}[f_{ran}]$ of a mixed WE flow 
	$f_{ran}$ introduced in Remark~\ref{remark:Mixed-WE} 
is essentially a non-atomic NE flow of a congestion game with
 the expected cost
$\E_{\Pi}[\tau_a(f_{ran,a})]$ as non-separable cost when viewed as a non-atomic flow of that congestion game.
Hence, the proof of Theorem~\ref{thm:GeneralPoly} has already provided a first 
positive example for a convergence analysis of the PoAs for non-separable
cost functions, although the expected cost is still rather simple compared with general non-separable cost functions.}}

\section*{Acknowledgement}
The first author acknowledges support from the National Natural Science Foundation of China with grant No.~61906062, support from the Natural Science Foundation of Anhui Province of China with grant No.~1908085QF262, support from the Talent Foundation
of Hefei University with grant No.~1819RC29, and support
from the Natural Science Foundation of the Education Department of Anhui Province 
of China with grant No.~KJ2019A0834.
The second author acknowledges support
from the Natural Science Foundation of the Education Department
of Anhui Province of China with grant No.~KJ2019A0834, and support from the National Natural Science Foundation of China with grant No.~12131003. The fourth author acknowledges support from the National Natural Science Foundation of China with grants No.~12131003 and No.~11531014.


\bibliographystyle{plainnat}
\bibliography{game}

\appendix

\section{\WuSecondRevision{Detailed Proofs}}

\subsection{The existence of mixed WE flows}
\label{app:Existence_Of_WE_Flows}

\WuSecondRevision{
\begin{lemma}\label{lemma:Existence-WE-flows}
	Every congestion game $\G=(\tau,\D,d)$ has a mixed WE flow.
\end{lemma}

\textbf{Proof of Lemma~\ref{lemma:Existence-WE-flows}:}

We use Brouwer's fixed point theorem and 
an argument similar to that in \citep{Dafermos1980}.
	Inequality~\eqref{eq:MWE} is equivalent to the variational inequality that
	\begin{displaymath}
		\sum_{k\in\K}\sum_{p\in\P_k}	\E_{\Pi}[\tau_p(f_{ran})]\cdot (\E_{\Pi}[f_{ran,p}]-\E_{\Pi'}[f'_{ran,p}])\le 0
	\end{displaymath}
	for an arbitrary mixed profile $\Pi'$ with random flow $f'_{ran}.$
	Brouwer's fixed theorem implies  that there is a fixed point 
	$\Pi^\alpha$ of the continuous map
	\begin{displaymath}
		D_\alpha(\Pi'):=\arg\min_{\Pi''}\ \sum_{k\in\K}\sum_{p\in\P_k}\left|\E_{\Pi''}\left[f''_{ran,p}\right
		]-\E_{\Pi'}\left[f'_{ran,p}\right]
		+\alpha\cdot \E_{\Pi'}\left[\tau_p(f'_{ran})\right]\right|^2
	\end{displaymath}
for an arbitrary
$\alpha>0.$
This follows since $D_{\alpha}(\cdot)$ maps the space of all mixed profiles
continuously into a subspace, and since
the space of all mixed profiles is convex and compact.
This fixed point $\Pi^\alpha$ fulfills the condition that
	\begin{equation}\label{eq:Existence_Of_MWE}
		\begin{split}
			\sum_{k\in\K}\sum_{p\in\P_k}
			&\E_{\Pi^\alpha}[\tau_p(f^\alpha_{ran})]\cdot 
			\left(\E_{\Pi^\alpha}[f^{\alpha}_{ran,p}]-
			\E_{\Pi''}[f''_{ran,p}]\right)\\
			&\le \frac{1}{2\cdot \alpha}\cdot \sum_{k\in\K}\sum_{p\in\P_k}
			\left(\E_{\Pi^\alpha}[f^{\alpha}_{ran,p}]-
			\E_{\Pi''}[f''_{ran,p}]\right)^2
		\end{split}
	\end{equation}
	for an arbitrary mixed profile $\Pi''$ with random flow $f''_{ran},$
	where $f^{\alpha}_{ran}$ is the random flow of $\Pi^\alpha.$
	Since the mixed profile sequence $(\Pi^\alpha)_{\alpha\in (0,\infty)}$ is
	bounded, there is an infinite subsequence $(\alpha_n)_{n\in \N}$
	such that $\alpha_n\to \infty$ and that
	$(\Pi^{\alpha_n})_{n\in\N}$ converges to 
	a limit mixed profile $\Pi,$  as $n\to \infty.$
	This limit mixed profile $\Pi$ has a mixed WE flow $f_{ran}$, since 
	inequality~\eqref{eq:Existence_Of_MWE} holds for an arbitrary 
	$\alpha>0$ and an arbitrary mixed profile $\Pi''.$ Here, we used that
	$\lim_{n\to \infty} \E_{\Pi^{\alpha_n}}[f^{\alpha_n}_{ran,p}]
	=\E_{\Pi}[f_{ran,p}]$ and $\lim_{n\to \infty}
	\E_{\Pi^{\alpha_n}}[\tau_p(f^{\alpha_n}_{ran})]=
	\E_{\Pi}[\tau_p(f_{ran})]$ as $n\to\infty,$ when
	$\Pi^{\alpha_n}\to\Pi$ as $n\to\infty.$
	This proves the existence of mixed WE flows.
\hfill $\square$}

%

\subsection{Stochastic inequalities}

Our proofs \WuSecondRevision{will} use Markov's inequality, 
	Chebyshev's inequality and Jensen's inequality.
	We summarize them in Lemma~\ref{lemma:Markov} below.
	\WuSecondRevision{\begin{lemma}
		\label{lemma:Markov}
		Let $X$ be a \wuu{non-negative} random variable  
		\wuu{whose expectation $\E(X)$ exists,} \wu{and let $\Delta> 0$ be an arbitrary constant.} Then 
		\begin{itemize}
			\item[a)] (Markov's inequality, see, e.g., \citep{Neuts1973}) \wuu{$\PP(X\ge \Delta)\le 
			\frac{\E(X)}{\Delta}.$}  
			\item[b)] (Chebyshev's inequality, see, e.g., \citep{Budny2014}) \wuu{$\PP(|X-\E(X)|\ge  \Delta)\le \frac{\var(X)}{\Delta^2}.$}
			\item[c)] (Jensen's inequality, see, e.g., \citep{McShane1937}) 
			$\E(h(X)|\mathcal{E})\ge h(\E(X|\mathcal{E}))$ for every convex function 
			$h:\mathbb{R}\to \mathbb{R}$ and an arbitrary random event $\mathcal{E}.$
		\end{itemize}
\end{lemma}}

\subsection{Proof of Lemma~\ref{lemma:ApproximateWE}}
\label{proof:ApproximateWE}

Note that Lemma~\ref{lemma:ApproximateWE} holds
trivially if \WuSecondRevision{the integer degree} $\beta=0,$ \wu{since
all cost functions $\tau^{[g]}_a(\cdot)$ are then}
positive constants, and so the total \wu{cost} of
atomic and non-atomic NE flows
coincide. We thus assume \wu{that} $\beta\ge 1.$

\wu{\textbf{Proof of  Lemma~\ref{lemma:ApproximateWE}a:}}

Consider now  an 
arbitrary $k\in \K$ and an arbitrary 
$i\in \D_k$.
Lemma~\ref{lemma:ApproximateWE}a
follows if
\begin{equation}\label{eq:ANE-WE-Approximate-Standard}
\tau^{[g]}_{p_{k,i}(\tilde{f}^{[g]}_{at})}\big(\tilde{f}_{at}^{[g]}\big)\le \tau^{[g]}_{p'}\big(\tilde{f}^{[g]}_{at}\big)+|A|\cdot \kappa\cdot\frac{d_{max}}{T}
\end{equation}
for all paths $p'\in \P_{k},$
where we recall that $p_{k,i}(\tilde{f}^{[g]}_{at})$
is the path of agent $i$ and \WuSecondRevision{that} $\kappa=\beta\cdot \eta_{\max}\cdot \big(
1+\sum_{l=1}^{\beta}\frac{1}{T^l}
\big)>0$ \WuSecondRevision{is the Lipschitz constant of scaled cost functions $\tau_a^{[g]}$
on $[0,1].$}

To prove \eqref{eq:ANE-WE-Approximate-Standard}, we consider an arbitrary \wu{path} $p'\in\P_k.$
\wu{Since $\tilde{f}^{[g]}_{at}$ is an} atomic NE flow, 
we obtain
\begin{equation}\label{eq:ANE-def}
\begin{split}
C_{k,i}(\tilde{f}^{[g]}_{at},\G^{[g]})=\!\frac{d_{k,i}}{T}
\cdot \tau^{[g]}_{p_{k,i}(\tilde{f}^{[g]}_{at})}(\tilde{f}^{[g]}_{at})
\le\! C_{k,i}(f^{[g]'}_{at},\G^{[g]})
\!=\!\frac{d_{k,i}}{T}
\cdot \tau^{[g]}_{p'}(f^{[g]'}_{at}),
\end{split}
\end{equation}
where $f^{[g]'}_{at}$
is an atomic flow of $\G^{[g]}$
 \wu{as defined in} \eqref{eq:user-Moving},
i.e., $f^{[g]'}_{at}$
is the resulting flow obtained by moving
$i$ from 
$p_{k,i}(\tilde{f}^{[g]}_{at})$
to $p'$ in the atomic NE flow
$\tilde{f}^{[g]}_{at}.$
\eqref{eq:ANE-def} \wu{implies}
further that 
\begin{equation}\label{eq:ANE-WE-near}
\tau^{[g]}_{p_{k,i}(\tilde{f}^{[g]}_{at})}\big(\tilde{f}^{[g]}_{at}\big)
=\sum_{a\in p_{k,i}(\tilde{f}^{[g]}_{at})}
\tau^{[g]}_{a}\big(\tilde{f}^{[g]}_{at,a}\big)
\le
\tau^{[g]}_{p'}\big(f^{[g]'}_{at}\big)
=\sum_{ a\in p'}
\tau^{[g]}_{a}\big(f^{[g]'}_{at,a}\big).
\end{equation}

Note that \wu{the} atomic
flows $\tilde{f}^{[g]}_{at}$
and $\tilde{f}^{[g]'}_{at}$
differ only \wu{in} the choice of $i.$
Note also that $i$ controls 
an amount $\bar{d}_{k,i}\le \frac{d_{max}}{T}$ of \wu{demand}
in $\G^{[g]}$.
So we obtain
for all $a\in A$
that
$|\tilde{f}^{[g]}_{at,a}-f^{[g]'}_{at,a}| 
\le \bar{d}_{k,i}\le  \frac{d_{max}}{T},$
\wu{where we} recall that $i$ uses only a single path
in \wu{an}y atomic flow.
This \wu{and}~\eqref{eq:ScaledCost} imply that
\begin{equation}
\label{eq:Arc-Cost-Diff}
\big|\tau_a^{[g]}\big(\tilde{f}^{[g]}_{at,a}\big)-
\tau_a^{[g]}\big(f^{[g]'}_{at,a}\big)\big|
\le \kappa\cdot |\tilde{f}^{[g]}_{at,a}-f^{[g]'}_{at,a}|
=\kappa\cdot\bar{d}_{k,i}\le  \kappa\cdot\frac{d_{max}}{T}\quad \forall a\in A.
\end{equation}
Here, we used that $\tau_a^{[g]}(x)$
is Lipschitz \wu{bounded} on $[0,1]$ with the
constant $\kappa$,
and \wu{that} all arc flow values of $\G^{[g]}$ are 
in $[0,1].$ Then \eqref{eq:Arc-Cost-Diff} and \eqref{eq:ANE-WE-near} \wu{imply}
that
$\tau^{[g]}_{p_{k,i}(\tilde{f}^{[g]}_{at})}\big(\tilde{f}^{[g]}_{at}\big) \le  \tau^{[g]}_{p'}\big(\tilde{f}^{[g]}_{at}\big)+|A|\cdot \kappa\cdot\frac{d_{max}}{T},$
which proves \eqref{eq:ANE-WE-Approximate-Standard}
due to arbitrary choice of $p'\in\P_k.$
This completes the proof of Lemma~\ref{lemma:ApproximateWE}a.

\wu{\textbf{Proof of  Lemma~\ref{lemma:ApproximateWE}b:}}

\wuu{Lemma~\ref{lemma:ApproximateWE}a yields
that}
$\max_{p\in \P_k:\ \tilde{f}^{[g]}_{at,p}>0}$ $
\tau^{[g]}_p(\tilde{f}^{[g]}_{at})$ $
\le $ $\min_{p\in \P_k}$ $
\tau^{[g]}_p(\tilde{f}^{[g]}_{at})
$ $+$ $\frac{|A|\cdot \kappa\cdot d_{max}}{T}$
for each $k\in \KK.$ This in turn implies
for an arbitrary non-atomic 
flow $f^{[g]}_{nat}$ \wu{that}
\WuSecondRevision{\begin{equation}\label{eq:NE-VarInequality-Approx}
\begin{split}
\sum_{a\in A}\tau^{[g]}_a
&\big(\tilde{f}^{[g]}_{at,a}\big)\cdot 
\big(f^{[g]}_{nat,a}\!-\tilde{f}^{[g]}_{at,a}\big)
=\!\sum_{p\in \P}\tau^{[g]}_p
\big(\tilde{f}^{[g]}_{at}\big)\cdot 
\big(f^{[g]}_{nat,p}\!-\!\tilde{f}^{[g]}_{at,p}\big)\\
&=\sum_{k\in\K}\sum_{p\in \P_k}\big(\tau^{[g]}_p
(\tilde{f}^{[g]}_{at})-\tau_{p^*_k}^{[g]}(\tilde{f}^{[g]}_{at})\big)\cdot 
\big(f^{[g]}_{nat,p}\!-\!\tilde{f}^{[g]}_{at,p}\big)\\
&\ge \sum_{k\in\K}\sum_{p\in \P_k: \tilde{f}_{at,p}^{[g]}>0}\big(\tau^{[g]}_p
(\tilde{f}^{[g]}_{at})-\tau_{p^*_k}^{[g]}(\tilde{f}^{[g]}_{at})\big)\cdot 
\big(f^{[g]}_{nat,p}\!-\!\tilde{f}^{[g]}_{at,p}\big)\\
&\ge\! -|\P|\cdot |A|\cdot \kappa\cdot\frac{d_{max}}{T}.
\end{split}
\end{equation}}
Here, we used that  \wu{the} total demand of
$\G^{[g]}$ \wu{is} $\bar{T}=1,$ and that 
\wu{
\begin{displaymath}
	\begin{split}
	\sum_{k\in \KK}\sum_{p\in \P_k}\tau_{p^*_k}^{[g]}
	\big(\tilde{f}^{[g]}_{at}\big)\cdot 
	\big(f^{[g]}_{nat,p}-\tilde{f}^{[g]}_{at,p}\big)= \sum_{k\in \KK}\tau_{p^*_k}^{[g]}
	\big(\tilde{f}^{[g]}_{at}\big)\cdot\sum_{p\in \P_k} 
	\big(f^{[g]}_{nat,p}-\tilde{f}^{[g]}_{at,p}\big)=0,
	\end{split}
\end{displaymath}}
where $p_k^*$ is
the least costly path
in $\P_k$ w.r.t. \wu{the}
atomic NE flow $\tilde{f}^{[g]}_{at}.$
By \wu{Definition}~\ref{def:Epsilon-Approximate-NE-Flow},
$\tilde{f}^{[g]}_{at}$
is an $\epsilon$-approximate
non-atomic NE flow of 
$\G^{[g]}$ with 
$\epsilon:=\frac{|\P|\cdot |A|\cdot \kappa\cdot d_{max}}{T}.$

\WuSecondRevision{In the sequel, we will use without further proof that a flow $f$ is a $|\P|\cdot \epsilon$-approximate non-atomic NE flow
when it satisfies the condition that 
\begin{displaymath}
	\max_{p\in\P_{k}: f_p>0} \tau^{[g]}_{p}(f)\le \min_{p\in\P_k}\ \tau^{[g]}_p(f)+\epsilon\quad 
	\forall k\in\K.
\end{displaymath}
This can be justified by an argument similar to that in \eqref{eq:NE-VarInequality-Approx}.}

\wu{\textbf{Proof of Lemma~\ref{lemma:ApproximateWE}c:}}
It follows immediately from Lemma~\ref{lemma:ApproximateWE}b and
Lemma~\ref{lemma:Ep-Approxi}.


\hfill$\square$

\subsection{Proof of Lemma~\ref{lemma:Non-Atomic-PoA-Bound}}
\label{proof:Non-Atomic-PoA-Bound}

\wu{Let} \WuSecondRevision{$\tilde{f}^{[g]}$} and
\WuSecondRevision{$f^{*[g]}$} \wu{be} non-atomic NE and SO flows
of \wu{the scaled game} $\G^{[g]},$
respectively.
Then
\WuSecondRevision{$\rho_{nat}
(\G^{[g]})
=\frac{\sum_{a\in A}\tau_a^{[g]}(\tilde{f}^{[g]}_a)\cdot \tilde{f}^{[g]}_a}{\sum_{a\in A}\tau_a^{[g]}(f^{*[g]}_a)\cdot f^{*[g]}_a}.$}
Note that \WuSecondRevision{$\tilde{f}^{[g]}$} is an optimal
solution of the \emph{non-linear program}
(NLP)~\eqref{eq:PotentialFunction},
\WuSecondRevision{\begin{equation}\label{eq:PotentialFunction}
\begin{split}
&\min\quad \Phi(y):=\sum_{a\in A}\int_{0}^{y_a}\tau^{[g]}_a(x)dx\\
&\text{s.t.}\quad \sum_{p\in \P_k}y_p=\bar{d}_k=
\sum_{i\in \D_k}\frac{d_{k,i}}{T}
=\frac{d_k}{T}\quad \forall k\in \KK,\\
&\qquad\ y_p\ge 0\quad \forall p\in \P,
\end{split}
\end{equation}}
see, e.g., \citep{Roughgarden2002,Roughgarden2007Introduction-routing}.
So \WuSecondRevision{$\Phi(\tilde{f}^{[g]})\le \Phi(f^{*[g]}).$}

As \wu{the} scaled cost functions 
$\tau_a^{[g]}(\cdot)$
\wu{have the} form \eqref{eq:ScaledCost}, we obtain that 
\WuSecondRevision{\begin{equation}\label{eq:Integral}
\begin{split}
\int_{0}^{y_a}\tau^{[g]}_a(x)dx
&=\frac{1}{\beta+1}\cdot \eta_{a,0}\cdot
y_a^{\beta+1}+\sum_{l=1}^{\beta}\frac{\eta_{a,l}}{(\beta-l+1)\cdot T^l}\cdot y_a^{\beta-l+1}\\
&=\frac{1}{\beta+1}\cdot \tau^{[g]}_a(y_a)\cdot y_a
+\sum_{l=1}^{\beta}\frac{l\cdot \eta_{a,l}}{(\beta-l+1)\cdot (\beta+1)\cdot T^l}\cdot y_a^{\beta-l+1}
\end{split}
\end{equation}
for all $a\in A$ and 
all $y_a\in [0,1].$} So 
\WuSecondRevision{\begin{displaymath}
	0\le \int_{0}^{y_a}\tau^{[g]}_a(x)dx-\frac{1}{\beta+1}\cdot \tau^{[g]}_a(y_a)\cdot y_a\le 
	\frac{\beta\cdot \eta_{\max}}{\beta+1}
	\cdot \sum_{l=1}^{\beta}\frac{1}{T^l}
\end{displaymath}
for all $a\in A$ and all $y_a\in [0,1].$} Here, we employ the
convention that 
$\sum_{l=1}^{\beta}\frac{1}{T^l}=0$
when $\beta=0.$
We thus obtain that 
\WuSecondRevision{\begin{displaymath}
	\sum_{a\in A}\tau_a^{[g]}(\tilde{f}^{[g]}_a)\cdot
	\tilde{f}^{[g]}_a
	\le (\beta+1)\cdot \Phi(\tilde{f}^{[g]})
	\le (\beta+1)\cdot \Phi(f^{*[g]})
	\le \sum_{a\in A}\tau_a^{[g]}(f^{*[g]}_a)\cdot f_a^{*[g]}
	+\beta\cdot \eta_{\max}
	\cdot \sum_{l=1}^{\beta}\frac{1}{T^l},
\end{displaymath}
which in turn implies that}
$
\rho_{nat}(\G^{[g]})
\le 1+\frac{\beta\cdot \eta_{\max}\cdot |\P|^{\beta+1}}{\eta_{0,\min}}
\cdot \sum_{l=1}^{\beta}\frac{1}{T^l}.
$
Here, we recall that $\eta_{0,\min}=\min_{a\in A}\eta_{a,0}>0,$ and that
the total cost \WuSecondRevision{$\sum_{a\in A}\tau_a^{[g]}(f_a^{*[g]})\cdot f_a^{*[g]}$}
is bounded from below by $\frac{\eta_{0,\min}}{|\P|^{\beta+1}}.$
This completes the proof of Lemma~\ref{lemma:Non-Atomic-PoA-Bound}.
\hfill$\square$

\subsection{Proof of Lemma~\ref{lemma:ExpectedFlowIsEpWE}}
\label{proof:ExpectedFlowIsEpWE}

\WuSecondRevision{Recall that $\beta$
	is the common degree of the polynomial cost functions, and is thus a non-negative integer.}
When $\beta=0,$ then \wu{the} scaled cost functions
$\tau_a^{[g]}(\cdot)$ are positive constants,
and  Lemma~\ref{lemma:ExpectedFlowIsEpWE}
\wu{holds} trivially. 
\WuThird{We thus assume  $\beta\ge 1.$} 

\WuSecondRevision{Consider now an arbitrary mixed NE flow
$f^{[g]}_{ran}$ of $\G^{[g]}.$} \wu{Chebyshev's inequality,} \wu{see} 
Lemma~\ref{lemma:Markov}b, 
implies  that
\begin{equation}\label{eq:DivergenceOfrandomFlow}
\begin{split}
	&\PP_{\tilde{\Pi}}\Big[\big|\tilde{f}^{[g]}_{ran,a}-\E_{\tilde{\Pi}}(\tilde{f}^{[g]}_{ran,a})\big|>\big(\frac{d_{max}}{T}\big)^{\delta}\Big]\le
	\big(\frac{T}{d_{max}}\big)^{2\cdot \delta}\cdot \var_{\tilde{\Pi}}(\tilde{f}^{[g]}_{ran,a})\\
	&\hspace{2cm}\le \big(\frac{T}{d_{max}}\big)^{2\cdot \delta}\cdot \sum_{i\in\D_k,k\in\KK}\frac{d^2_{k,i}}{4\cdot T^2}\\
	&\hspace{2cm}\le \frac{1}{4}\cdot \big(\frac{d_{max}}{T}\big)^{1-2\cdot\delta}\quad \forall a\in A\ \forall \delta\in (0,\frac{1}{2}).
\end{split}
\end{equation}
Here, we used 
\WuThird{
$\var_{\tilde{\Pi}}(\tilde{f}^{[g]}_{ran,a})$ $=$ $\sum_{i\in \D_k,k\in \KK}\frac{
	d_{k,i}^2}{T^2}\cdot \tilde{\Pi}_{i,a}\cdot (1-\tilde{\Pi}_{i,a})$.}
\wu{This follows since $\tilde{\Pi}_{i,a}
	=\sum_{p\in \P_k:a\in p}\tilde{\Pi}_{i,p}$}
is the probability \wu{that agent $i\in \D_k$ uses
	arc $a,$} \wu{since the} demand of
agent $i\in\D_k$  is $\frac{d_{k,i}}{T}$ in the scaled game $\G^{[g]}$
and \WuSecondRevision{since} $\G^{[g]}$ has total demand $\bar{T}=\sum_{k\in\K,i\in\D_k}\frac{d_{k,i}}{T}=1.$

\WuSecondRevision{We now show that the mixed NE flow $\tilde{f}^{[g]}_{ran}$
	is an \emph{approximate} mixed WE flow (see  Remark~\ref{remark:Mixed-WE}).
	Consider an arbitrary $k\in\K$ and an arbitrary
$p\in\P_k$ with 
$\tilde{\Pi}_{i,p}>0$ for some $i\in\D_k.$}

\WuSecondRevision{Note that 
$
	|\tilde{f}^{[g]}_{ran,a|i,p''}-\tilde{f}^{[g]}_{ran,a|i,p'}|
	\le \frac{d_{max}}{T}.
$ for all $a\in A$ and all $p',p''\in\P_{k}.$
Here, we recall that $\tilde{f}^{[g]}_{ran,a|i,p'}$ 
is the random flow of arc $a$ when agent $i$ uses the fixed path
$p'$ and the other agents $j\in\D\setminus\{i\}$ still follow their random paths 
drawn from $\tilde{\Pi}_{j}.$
Then
	\begin{align}
	|	\E_{\tilde{\Pi}}[\tau^{[g]}_{p'}(\tilde{f}^{[g]}_{ran})]-&\E_{\tilde{\Pi}_{-i}}[\tau^{[g]}_{p'}(\tilde{f}^{[g]}_{ran|i,p''})]|\notag\\
	&=\left|\sum_{p'''\in \P_k}\left[
		\E_{\tilde{\Pi}_{-i}}[\tau^{[g]}_{p'}(\tilde{f}^{[g]}_{ran| i,p'''})]-\E_{\tilde{\Pi}_{-i}}[\tau^{[g]}_{p'}(\tilde{f}^{[g]}_{ran|i,p''})]\right]\cdot \tilde{\Pi}_{i,p'''}\right|\notag\\
	&\le \sum_{p'''\in\P_k}\tilde{\Pi}_{i,p'''}\cdot \sum_{a\in p'}
		\E_{\tilde{\Pi}_{-i}}\left|\tau^{[g]}_a(\tilde{f}^{[g]}_{ran,a|i,p'''})-
		\tau^{[g]}_a(\tilde{f}^{[g]}_{ran,a|i,p''})\right|\notag\\
		&\le \kappa\cdot \sum_{p'''\in\P_k}\tilde{\Pi}_{i,p'''}\cdot \sum_{a\in p'}
		\E_{\tilde{\Pi}_{-i}}\big|\tilde{f}^{[g]}_{ran,a|i,p'''}-
		\tilde{f}^{[g]}_{ran,a|i,p''}\big|\notag\\
		&\le |A|\cdot \kappa\cdot \frac{
			d_{max}
		}{T},\qquad \forall p',p''\in \P_k,\label{eq:MixedNE-Expected-flow}
	\end{align}
since each $\tau_a^{[g]}$ is Lipschitz bounded on $[0,1]$ with Lipschitz
constant $\kappa$.
\eqref{eq:MixedNE-Expected-flow} implies that 
$\tilde{f}^{[g]}_{ran}$ is an approximate mixed WE flow,
i.e., 
\begin{equation}\label{eq:Epsilon-mixed-WE-flow}
	\begin{split}
		\E_{\tilde{\Pi}}[\tau^{[g]}_p(\tilde{f}^{[g]}_{ran})]&= \E_{\tilde{\Pi}_{-i}}[\tau^{[g]}_p(\tilde{f}^{[g]}_{ran|i,p})]
	\le \E_{\tilde{\Pi}_{-i}}[\tau^{[g]}_{p'}(\tilde{f}^{[g]}_{ran|i,p'})]\\
&\le \E_{\tilde{\Pi}}[\tau^{[g]}_{p'}(\tilde{f}^{[g]}_{ran})]
+|A|\cdot \kappa\cdot \frac{
	d_{max}
}{T}
	\end{split}
\end{equation}
for an arbitrary $p'\in\P_k.$ This follows since $\tilde{f}^{[g]}_{ran}$ is a mixed NE flow
and $\tilde{\Pi}_{i,p}>0.$
}

\WuSecondRevision{We now show with \eqref{eq:DivergenceOfrandomFlow} and
	\eqref{eq:Epsilon-mixed-WE-flow} that the expected
flow $\E_{\tilde{\Pi}}(\tilde{f}^{[g]}_{ran})$ is an $\epsilon$-approximate
non-atomic NE flow with $\epsilon$ tending to $0$ as $\frac{d_{max}}{T}\to 0.$}

\eqref{eq:DivergenceOfrandomFlow} implies
\begin{equation}\label{eq:DivergenceOfrandomFlowAll}
	\begin{split}
		\PP_{\tilde{\Pi}}&\Big[\forall a\in A: \big|\tilde{f}^{[g]}_{ran,a}-\E_{\tilde{\Pi}}(\tilde{f}^{[g]}_{ran,a})\big|\le \big(\frac{d_{max}}{T}\big)^{\delta}\Big]\\
		&\hspace{2cm}\ge 1- \frac{|A|}{4}\cdot \Big(\frac{d_{max}}{T}\Big)^{ 1-2\cdot\delta}=1-\mathbf{P}_{\delta},
	\end{split}
\end{equation}
where
$
\mathbf{P}_{\delta}:=\frac{|A|}{4}\cdot \big(\frac{d_{max}}{T}\big)^{1-2\cdot\delta}.
$
Consequently,
\begin{equation}\label{eq:Stoc-Expect-Arc-Cost-diff}
	\PP_{\tilde{\Pi}}\Big[\forall a\in A:\ |\tau_a^{[g]}(\tilde{f}^{[g]}_{ran,a})
	-\tau_a^{[g]}\big(\E_{\tilde{\Pi}}(\tilde{f}^{[g]}_{ran,a})\big)|
	\le \kappa\cdot \big(\frac{d_{max}}{T}\big)^{\delta}\Big]
	\ge 1-\mathbf{P}_{\delta},
\end{equation}
\WuSecondRevision{again since the scaled cost functions are Lipschitz continuous
	on $[0,1]$
with the Lipschitz constant $\kappa$.}

Note that \WuSecondRevision{$|\tau_a^{[g]}(\tilde{f}^{[g]}_{ran,a})
\!-\!\tau_a^{[g]}\big(\E_{\tilde{\Pi}}(\tilde{f}^{[g]}_{ran,a})\big)|\!\le\!  \tau^{[g]}_a(1)
\!\le\! \frac{\kappa}{\beta}\!=\!\sum_{l\!=\!0}^{\beta}\frac{\eta_{\max}}{T^l}$}
\wu{with probability $1$} for all 
$a\in A.$
This, together with \eqref{eq:Stoc-Expect-Arc-Cost-diff},
\wu{implies} that 
\wuu{\begin{align}
		|\E_{\tilde{\Pi}}&\big(\tau_a^{[g]}(\tilde{f}^{[g]}_{ran,a})\big)
		\!-\!\tau_a^{[g]}\big(\E_{\tilde{\Pi}}(\tilde{f}^{[g]}_{ran,a})\big)|
		\!\le\! \E_{\tilde{\Pi}}\big(|\tau_a^{[g]}(\tilde{f}^{[g]}_{ran,a})\!-\!
		\tau_a^{[g]}(\E_{\tilde{\Pi}}(\tilde{f}^{[g]}_{ran,a}))|\big)\notag\\
		&\le\! (1-\mathbf{P}_{\delta}) \cdot\kappa\cdot \big(\frac{d_{max}}{T}\big)^{\delta}+
		\mathbf{P}_{\delta}\cdot 
		\frac{\kappa}{\beta}\notag\\
		&=\kappa\cdot \big(\frac{d_{max}}{T}\big)^{\delta}+
		\frac{|A|}{4}\cdot \big(\frac{d_{max}}{T}\big)^{ 1-2\cdot\delta}\cdot 
		\frac{\kappa}{\beta}\quad \forall a\in A\ \forall 
		\delta\in (0,\frac{1}{2}).\label{eq:Expect-Expect-Arc-Cost-diff}
\end{align}}\wu{\eqref{eq:Expect-Expect-Arc-Cost-diff} uses} that \wu{the} random event \wuu{``$\kappa\cdot \big(\frac{d_{max}}{T}\big)^{\delta}<|\tau_a^{[g]}(\tilde{f}^{[g]}_{ran,a})-
	\tau_a^{[g]}(\E_{\tilde{\Pi}}(\tilde{f}^{[g]}_{ran,a}))|\le \frac{\kappa}{\beta}$''}
occurs with a probability  of \emph{at most} $\mathbf{P}_{\delta},$ 
since \wu{the} random event of \eqref{eq:Stoc-Expect-Arc-Cost-diff}
occurs with a probability of at least  $1-\mathbf{P}_{\delta}.$



\wu{Putting $\delta=\frac{1}{3}$
	in \eqref{eq:Expect-Expect-Arc-Cost-diff},} we obtain  that
\begin{equation}\label{eq:Expect-Expect-Arc-Cost-diff-Constant}
\begin{split}
|\E_{\tilde{\Pi}}\big(\tau_a^{[g]}(\tilde{f}^{[g]}_{ran,a})\big)
-\tau_a^{[g]}\big(\E_{\tilde{\Pi}}(\tilde{f}^{[g]}_{ran,a})\big)|
&\le \kappa\cdot \big(\frac{d_{max}}{T}\big)^{1/3}+
\frac{|A|}{4}\cdot \big(\frac{d_{max}}{T}\big)^{1/3}\cdot 
\frac{\kappa}{\beta}\\
&=\kappa\cdot \big(1+\frac{|A|}{4\cdot \beta}\big)
\cdot \big(\frac{d_{max}}{T}\big)^{1/3}\quad \forall a\in A.
\end{split}
\end{equation}
\eqref{eq:Expect-Expect-Arc-Cost-diff-Constant} in turn
implies that 
\begin{equation}\label{eq:Expect-Expect-Path-Cost-diff-Constant}
\begin{split}
|\E_{\tilde{\Pi}}\big(\tau_p^{[g]}(\tilde{f}^{[g]}_{ran})\big)
-\tau_p^{[g]}\big(\E_{\tilde{\Pi}}(\tilde{f}^{[g]}_{ran})\big)|
\le \kappa\cdot |A|\cdot \big(1+\frac{|A|}{4\cdot \beta}\big)
\cdot \big(\frac{d_{max}}{T}\big)^{1/3}\quad \forall p\in\P.
\end{split}
\end{equation}
\eqref{eq:Expect-Expect-Path-Cost-diff-Constant}  and \eqref{eq:Epsilon-mixed-WE-flow}
\wu{together then yield}
\WuSecondRevision{\begin{equation}\label{eq:Epsil-Expect-WE}
	\begin{split}
		\tau_p^{[g]}&\big(\E_{\tilde{\Pi}}(\tilde{f}^{[g]}_{ran})\big)\le
		\E_{\tilde{\Pi}}\big(\tau_p^{[g]}(\tilde{f}^{[g]}_{ran})\big)+\kappa\cdot |A|\cdot \big(1+\frac{|A|}{4\cdot \beta}\big)
		\cdot \big(\frac{d_{max}}{T}\big)^{1/3}\\
		&\le \E_{\tilde{\Pi}}\big(\tau_{p'}^{[g]}(\tilde{f}^{[g]}_{ran})\big)+\kappa\cdot |A|\cdot \big(1+\frac{|A|}{4\cdot \beta}\big)
		\cdot \big(\frac{d_{max}}{T}\big)^{1/3}+|A|\cdot \kappa\cdot \frac{
			d_{max}
		}{T}\\
		&\le \tau_{p'}^{[g]}\big(\E_{\tilde{\Pi}}(\tilde{f}^{[g]}_{ran})\big)+2\cdot\kappa\cdot |A|\cdot \big(1+\frac{|A|}{4\cdot \beta}\big)
		\cdot \big(\frac{d_{max}}{T}\big)^{1/3}
		+|A|\cdot \kappa\cdot \frac{
			d_{max}
		}{T}\\
	 &\le \tau_{p'}^{[g]}\big(\E_{\tilde{\Pi}}(\tilde{f}^{[g]}_{ran})\big)+3\cdot \kappa\cdot |A|\cdot \big(1+\frac{|A|}{4\cdot \beta}\big)
	 \cdot \big(\frac{d_{max}}{T}\big)^{1/3}
	\end{split}
\end{equation}
for all $k\in\KK,$ and any two paths 
	$p,p'\in\KK$ with expected flow value $\E_{\tilde{\Pi}}(\tilde{f}^{[g]}_{ran,p})>0.$}

\eqref{eq:Epsil-Expect-WE} \wu{yields} \WuSecondRevision{with a similar argument as in the proof of} Lemma~\ref{lemma:ApproximateWE}b
that \wu{the} expected non-atomic flow
$\E_{\tilde{\Pi}}[\tilde{f}^{[g]}_{ran}]$
is an $\epsilon$-approximate non-atomic NE flow
with 
\WuSecondRevision{$\epsilon:=3\cdot |\P|\cdot \kappa\cdot |A|\cdot \big(1+\frac{|A|}{4\cdot \beta}\big)
\cdot \big(\frac{d_{max}}{T}\big)^{1/3}.$} Lemma~\ref{lemma:ExpectedFlowIsEpWE}a then follows immediately
from Lemma~\ref{lemma:Ep-Approxi}.




Lemma~\ref{lemma:ExpectedFlowIsEpWE}b then \WuSecondRevision{follows}
from 
\eqref{eq:DivergenceOfrandomFlowAll}
and \eqref{eq:Stoc-Expect-Arc-Cost-diff}, since they together imply
that \wu{the} random event  
\begin{equation}\label{eq:Expected-Arc-ToCo-Diff}
\begin{split}
\forall a\in A: |\tilde{f}^{[g]}_{ran,a}
\cdot \tau_a^{[g]}(\tilde{f}^{[g]}_{ran,a})
-&\E_{\tilde{\Pi}}(\tilde{f}^{[g]}_{ran,a})
	\cdot \tau_a^{[g]}\big(\E_{\tilde{\Pi}}(\tilde{f}^{[g]}_{ran,a})\big)|\\
	&\le\! \big(\kappa\!+\!\eta_{\max}\!\cdot\! \sum_{l=0}^{\beta}\frac{1}{T^l}\big)\cdot \big(\frac{d_{max}}{T}\big)^{\delta}
\end{split}
\end{equation}
occurs with a probability  of \emph{at least}
$1-\mathbf{P}_{\delta}.$
Here, we \wu{use} that \WuSecondRevision{$\tau_a^{[g]}(\cdot)$
	is Lipschitz bounded in $[0,1]$ with Lipschitz constant $\kappa,$ that $|x\cdot \tau_a^{[g]}(x)-y\cdot
	\tau_a^{[g]}(y)|\le x\cdot |\tau_a^{[g]}(x)-\tau_a^{[g]}(y)|
	+\tau_a^{[g]}(y)\cdot |x-y|\le 
	(\kappa+\tau_a^{[g]}(y))\cdot |x-y|
	\le (\kappa+\tau_a^{[g]}(1))\cdot |x-y|$ for
	all $x,y\in [0,1],$ that}
$\G^{[g]}$
has arc flow values
in $[0,1],$ and \wu{that}
$\max_{a\in A}\max_{[0,1]}\tau_a^{[g]}(x)
\le \sum_{l=0}^{\beta}\frac{\eta_{\max}}{T^l}.$


\eqref{eq:DivergenceOfrandomFlowAll} and \eqref{eq:Expected-Arc-ToCo-Diff} 
yield
\begin{align}
|\E_{\tilde{\Pi}}&\big(\tilde{f}^{[g]}_{ran,a}\cdot \tau^{[g]}_a(\tilde{f}^{[g]}_{ran,a})\big)\!-\!
\E_{\tilde{\Pi}}(\tilde{f}^{[g]}_{ran,a})\cdot
\tau_a^{[g]}\big(\E_{\tilde{\Pi}}(\tilde{f}^{[g]}_{ran,a})\big)|\notag\\
&\le\! \E_{\tilde{\Pi}}\Big(\big|\tilde{f}^{[g]}_{ran,a}\cdot \tau^{[g]}_a(\tilde{f}^{[g]}_{ran,a})\!-\!\E_{\tilde{\Pi}}(\tilde{f}^{[g]}_{ran,a})\cdot
\tau_a^{[g]}\big(\E_{\tilde{\Pi}}(\tilde{f}^{[g]}_{ran,a})\big)\big|\Big)\notag\\
&\le \big(\kappa+\eta_{\max}\cdot \sum_{l=0}^{\beta}\frac{1}{T^l}\big)\cdot \big(\frac{d_{max}}{T}\big)^{\delta}
+\mathbf{P}_{\delta}\cdot \eta_{\max}\cdot \sum_{l=0}^{\beta}\frac{1}{T^l}\label{eq:Total-Expec-Diff}\\
&=\big(\kappa+\eta_{\max}\cdot \sum_{l=0}^{\beta}\frac{1}{T^l}\big)\cdot \big(\frac{d_{max}}{T}\big)^{\delta}
+\frac{|A|}{4}\cdot \big(\frac{d_{max}}{T}\big)^{ 1-2\cdot\delta}\cdot \eta_{\max}\cdot \sum_{l=0}^{\beta}\frac{1}{T^l}\notag
\end{align}
for all $a\in A$ \wu{and all} $\delta\in (0,\frac{1}{2}).$
Here, we \wu{use} that $\max_{a\in A}\max_{x\in [0,1]}x\cdot \tau_a^{[g]}(x)\le \sum_{l=0}^{\beta}
\frac{\eta_{\max}}{T^l}$, and that \wu{the} random event \eqref{eq:Expected-Arc-ToCo-Diff}
\wu{occurs} with a probability  of at least  $1-\mathbf{P}_{\delta},$
and so \wu{the complement} event
of \eqref{eq:Expected-Arc-ToCo-Diff} \wu{occurs} with
a probability  of at most $\mathbf{P}_{\delta}.$
Lemma~\ref{lemma:ExpectedFlowIsEpWE}c
then follows immediately from \eqref{eq:Total-Expec-Diff} when we put
$\delta=\frac{1}{3}$.
%
\hfill$\square$

\subsection{Proof of Theorem~\ref{thm:GeneralPoly}}
\label{proof:MainStep}

\WuThird{We first show Theorem~\ref{thm:GeneralPoly}c, and then prove
Theorem~\ref{thm:GeneralPoly}a--b with the technique of asymptotic decomposition
proposed by \citep{Wu2019}.}

\textbf{Proof of Theorem~\ref{thm:GeneralPoly}c:} 

\WuThird{We define $\beta=\max_{k\in\K}\min_{p\in\P_k}\max_{a\in p}\beta_a,$
	and put the scaling factor $g:=T^\beta.$
Here, we recall that the degree $\beta_a\ge 0$ of arc $a\in A$ is an integer.
We call a path $p\in \P=\cup_{k\in \K}\P_k$ with $\max_{a\in p} \beta_a\le \beta$ 
a \emph{tight} path, and an arc $a\in A$ with $\beta_a\le \beta$ a \emph{tight} arc.
Clearly, each O/D pair $k\in\K$ has at least one tight path $p\in\P_k.$
We denote by $\Gamma^{[g]}$ the resulting scaled game with scaling factor $g.$
This has a total demand of $1.$
}

\WuThird{Let $\tilde{f}^{[g]}_{nat}$ be an abitrary non-atomic NE flow of $\Gamma^{[g]},$ and 
	let $\tilde{f}_{at}^{[g]}$ be an arbitrary atomic NE flow of $\Gamma^{[g]}.$}

\WuThird{\citep{Colini2020OR} have shown that $\rho_{nat}(\Gamma)=\rho_{nat}(\Gamma^{[g]})
=1+O(\frac{1}{T})$ under the condition of Theorem~\ref{thm:GeneralPoly}c, i.e., 
$\frac{d_k}{T}\ge \xi_k$ for each $k\in \K$ for constants $\xi_k>0$ independent of $T.$
To obtain the convergence rate of the atomic PoA $\rho_{at}(\G)=\rho_{at}(\G^{[g]})$, we again need to upper bound only the cost difference 
$|C(\tilde{f}_{at}^{[g]},\Gamma^{[g]})-C(\tilde{f}_{nat}^{[g]},\G^{[g]})|$
because of inequality \eqref{eq:Non-Atomic-Transformation}. Here, we observe that non-atomic SO flows of $\Gamma^{[g]}$ have a cost of $\Omega(1),$ since every O/D pair $k\in\K$ has a total demand of $\frac{d_k}{T}\in \Theta(1)$ in $\G^{[g]},$ and since 
there is at least one O/D pair $k\in\K$ with $\min_{p\in\P_k}\max_{a\in p}\beta_a=\beta.$}

\WuThird{When all arcs are tight, i.e., $\beta_a\le \beta$ for all $a\in A,$ then all the scaled polynomial  cost functions
$\tau_a^{[g]}(x)$ of $\Gamma^{[g]}$ have bounded coefficients and degrees smaller than $\beta,$ and are 
thus Lipschitz continuous on $[0,1]$ with a Lipschitz constant independent of $T.$
Moreover, with arguments similar to those for Theorem~\ref{thm:MainResults}, 
we obtain immediately that $|C(\tilde{f}_{at}^{[g]},\Gamma^{[g]})-C(\tilde{f}_{nat}^{[g]},\G^{[g]})|\in O(\sqrt{\frac{1}{T}})$ and so  $\rho_{at}(\Gamma^{[g]})=1+O(\sqrt{\frac{1}{T}})$
by inequality~\eqref{eq:Non-Atomic-Transformation}.
Here, note that the maximum individual demand in $\G$ is $d_{\max}\le \upsilon$ for a constant 
$\upsilon>0$ independent of $T.$} 



\WuThird{Now assume that there are non-tight arcs $a\in A,$ i.e., arcs $a\in A$ with $\beta_a>\beta.$ Then
	the scaled cost functions $\tau_a^{[g]}(\cdot)$ of these non-tight arcs $a\in A$ need not
	 be Lipschitz continuous on $[0,1],$ since their coefficients may tend to $\infty$ with growing $T.$ A natural idea here is to remove the influence of these non-tight arcs in the analysis.}

\WuThird{Since each O/D pair $k\in\K$ has at least one tight path $p\in\P_k,$  we obtain that 
\begin{equation}\label{eq:Nontight_arc_cost_upperbound}
\eta_{0,\min}\cdot T^{\beta_a-\beta}\cdot (\tilde{f}_{at,a}^{[g]})^{\beta_a}\le \tau_a^{[g]}(\tilde{f}_{at,a}^{[g]})\le 
\eta_{\max}\cdot (\beta+1)\cdot |A|,\quad \forall a\in A.
\end{equation}
Here, we used that a tight path $p$ has 
a scaled cost of at most $\eta_{\max}\cdot (\beta+1)\cdot |A|$ in an arbitrary flow, as it contains at most 
$|A|$ many arcs, and has a flow value of at most $1$ in an arbitrary flow of the
scaled game $\Gamma^{[g]}.$ Moreover, by the definition of atomic NE flows, the 
scaled cost $\frac{d_{k,i}}{T}\cdot \tau_{p'}^{[g]}(\tilde{f}_{at}^{[g]})$ of an arbitrary individual $i\in\D_k$ with an arbitrary ``pure strategy" $p'\in \P_k$
will not decrease, even that individual unilaterally moves from path $p'$ to a 
tight path $p\in \P_k.$
}

 \WuThird{Hence, we obtain for each non-tight arc $a\in A$ that
	\begin{equation}\label{eq:Nontight_Arc_Flow_Upbound}
		\tilde{f}_{at,a}^{[g]}
		\le \theta_a(T):=
		\frac{\eta_{\max}\cdot |A|\cdot (\beta+1)}{\eta_{0,\min}}\cdot 
		T^{-\frac{\beta_a-\beta}{\beta_a}} \in o(1).
\end{equation}
Similarly, $\tilde{f}_{nat,a}^{[g]}\le \theta_a(T)$ for each non-tight arc $a\in A.$ Moreover, inequality~\eqref{eq:Nontight_Arc_Flow_Upbound}
implies for each $k\in\K$ and each non-tight path $p\in\P_k,$ i.e., 
$\max_{a\in p} \beta_a>\beta,$
that 
\begin{equation}\label{eq:Nontight_Path_Flow_Upbound}
	\tilde{f}^{[g]}_{at,p}\le\theta_p(T)\!:=\! \min_{a\in p:\beta_a\!>\!\beta}\theta_a(T)
	\!\in\!\Theta(
	T^{-\!\max_{a\in p:\beta_a\!>\!\beta}\frac{\beta_a\!-\!\beta}{\beta_a}}
	)\quad \text{and}
	\quad \tilde{f}^{[g]}_{nat,p}\!\le\! \theta_p(T),
\end{equation}
since the flow value of a path is not larger than the minimum flow value of arcs contained in that path.
}

\WuThird{Inequalities~\eqref{eq:Nontight_arc_cost_upperbound}--\eqref{eq:Nontight_Path_Flow_Upbound} actually indicate that we can ignore all non-tight arcs $a\in A$ 
	and all non-tight paths $p\in \P$
	in the analysis. In particular, we have
\begin{align}
		|C(\tilde{f}_{at}^{[g]},\G^{[g]})-C(\tilde{f}_{nat}^{[g]},\G^{[g]})|
		&\le |
		\sum_{a\in A: \beta_a\le \beta}
		\tilde{f}_{at,a}^{[g]}\cdot \tau_a^{[g]}(\tilde{f}_{at,a}^{[g]})
		-\sum_{b\in A:\beta_b\le \beta}
		\tilde{f}_{nat,b}^{[g]}\cdot \tau_b^{[g]}(\tilde{f}_{nat,b}^{[g]})
		|\notag\\
		&\hspace{1cm}+2\cdot \eta_{\max}\cdot (\beta+1)\cdot |A|\cdot \sum_{a\in A:\beta_a>\beta}
		\theta_a(T).\label{eq:Atomic-Nonatomic-NE-cost-difference-Import}
\end{align}
This provides a very good basis for further upper bounding the cost difference in this general case.}

\WuThird{For each O/D pair $k\in \P_k,$ we denote by $\P'_k=
	\{p\in \P_k:\ \max_{a\in p} \beta_a\le \beta\}$ the subset of all tight paths $p\in \P_k,$ and put $\P':=\cup_{k\in \K}\P'_k.$ 
	Moreover, 
%
%
%
we denote by $A'=\{a\in A:\ \beta_a\le \beta\}$ the subset of all tight arcs $a\in A.$}
	
		\WuThird{For each tight arc $a\in A',$ we define an auxiliary cost function
	\begin{displaymath}
		\sigma_{1,a}(x)=
		\begin{cases}
			\tau_a^{[g]}(x)&\text{if }a\notin p\text{ for all } p\in \P\setminus \P',\\
		\tau_a^{[g]}(x+\sum_{k\in\K}\sum_{p\in \P\setminus\P'_k}\tilde{f}_{at,p}^{[g]})	&\text{otherwise.}
		\end{cases}
\end{displaymath}
Then the restricted flow $\tilde{f}_{at|\P'}^{[g]}=
(\tilde{f}_{at,p}^{[g]})_{p\in\P'_k,k\in\K}$ is an atomic NE flow w.r.t. these tight paths 
$p\in\P'$ and w.r.t. the arc cost functions $\sigma_{1,a}(\cdot)$ of these tight arcs $a\in A'.$
This follows since $\tau_p^{[g]}(\tilde{f}_{at}^{[g]})=
\sigma_{1,p}(\tilde{f}_{at|\P'}^{[g]})=$ for all $p\in\P'.$
Here, we note that $\sigma_{1,p}(\tilde{f}_{at|\P'}^{[g]})
=\sum_{a\in p} \sigma_{1,a}(\tilde{f}_{at,a|\P'}^{[g]})=\sum_{a\in p}\tau_a^{[g]}(\tilde{f}_{at,a}^{[g]})$ for each
$p\in \P',$ and $\tilde{f}_{at,a|\P'}^{[g]}=\sum_{k\in\K}\sum_{p\in\P'_k} \tilde{f}_{at,p}^{[g]}$ for 
each $a\in A'.$
In particular, the total cost $C(\tilde{f}_{at|\P'}^{[g]},\Gamma_1^{[g]})$ of $\tilde{f}_{at|\P'}^{[g]}$ satisfies the condition that 
\begin{align}
		\sum_{a\in A'} &\tilde{f}_{at,a}^{[g]}\cdot \tau_a^{[g]}(\tilde{f}_{at,a}^{[g]})
		\ge 
		C(\tilde{f}_{at|\P'}^{[g]},\Gamma_1^{[g]})= 
		\sum_{a\in A'} \tilde{f}_{at,a|\P'}^{[g]}\cdot \sigma_{1,a}(\tilde{f}_{at,a|\P'}^{[g]})\notag\\
		&\ge \sum_{a\in A'} \tilde{f}_{at,a}^{[g]}\cdot \tau_a^{[g]}(\tilde{f}_{at,a}^{[g]})
		-\eta_{\max}\cdot |A|^2\cdot (\beta+1)\cdot \sum_{p\in \P\setminus \P'} \theta_p(T),\label{eq:Restricted-Atomic-NE-Total-Cost-In-Reduced-Game}
\end{align}
where the quantity $\theta_p(T)$ defined in  inequality~\eqref{eq:Nontight_Path_Flow_Upbound} is an upper bound of the atomic flow value $\tilde{f}_{at,p}^{[g]}$
on a non-tight path $p\in \P\setminus\P'$,
and $\G_{1}^{[g]}$ is the resulting ``reduced" scaled game, which ignores all non-tight paths 
$p\in \P\setminus \P'$ together with their ``demands" $\tilde{f}_{at,p}^{[g]}$, and, moreover,
has the auxiliary functions $\sigma_{1,a}(\cdot)$ as the cost functions of the  tight arcs $a\in A'.$
Here, we used inequalities~\eqref{eq:Nontight_arc_cost_upperbound}, \eqref{eq:Nontight_Path_Flow_Upbound}, $|A'|\le |A|,$ and the fact that
\begin{displaymath}
0\le 	\tilde{f}_{at,a}^{[g]}-\tilde{f}_{at,a|\P'}^{[g]}
\le \sum_{p\in \P\setminus\P'} \tilde{f}_{at,p}^{[g]}
\end{displaymath}
for each $a\in A'.$
}

 \WuThird{Let $\tilde{f}_{1,nat}^{[g]}$ be a non-atomic NE flow of $\Gamma_1^{[g]}.$ 
	Since  $\Gamma_1^{[g]}$ ignores all non-tight arcs $a\in A\setminus A',$
	all its cost functions $\sigma_{1,a}(\cdot)$ have coefficients bounded from above by a constant independent of 
	$T,$ and are thus Lipschitz continuous on $[0,1].$ While $\G_1^{[g]}$ ignores all demands
	$\tilde{f}_{at,p}^{[g]}$ of non-tight paths $p\in\P\setminus\P',$ inequality \eqref{eq:Nontight_Path_Flow_Upbound} implies that 
	$\G_{1}^{[g]}$ has a total demand tending to $1$ as $T\to \infty.$
	Hence,
	we obtain again by arguments similar to those for Theorem~\ref{thm:MainResults} that
	\begin{equation}\label{eq:Atomic-NonAtomic-Cost-Differ-In-Restricted-Game}
|C(\tilde{f}_{1,nat}^{[g]},\G_{1}^{[g]})-C(\tilde{f}^{[g]}_{at|\P'},\G_{1}^{[g]})|
\in O(\sqrt{\frac{1}{T}}).
\end{equation}
Here, we note that $\tilde{f}_{at|\P'}^{[g]}$ is an atomic NE flow of $\G_1^{[g]}.$}

\WuThird{We proceed similarly with the non-atomic NE flow $\tilde{f}_{nat}^{[g]},$ and consider its restriction $\tilde{f}_{nat|\P'}^{[g]}=(\tilde{f}_{nat,p}^{[g]})_{p\in\P'_k,k\in\K}$ to tight paths $p\in \P'.$ 
We define the auxiliary cost functions 
$\sigma_{2,a}(\cdot)$ for each tight arc $a\in A'$ and the resulting reduced scaled game 
$\G_2^{[g]}$ by using non-atomic flow values $\tilde{f}_{nat,p}^{[g]}$ instead of atomic flow values $\tilde{f}_{at,p}^{[g]}$ in the above definitions. Then we obtain also that $\tilde{f}^{[g]}_{nat|\P'}$ is a non-atomic NE flow of $\G_2^{[g]},$ and, moreover, 
\begin{align}
	\sum_{a\in A'} &\tilde{f}_{nat,a}^{[g]}\cdot \tau_a^{[g]}(\tilde{f}_{nat,a}^{[g]})
	\ge 
	C(\tilde{f}_{nat|\P'}^{[g]},\Gamma_2^{[g]})= 
	\sum_{a\in A'} \tilde{f}_{nat,a|\P'}^{[g]}\cdot \sigma_{2,a}(\tilde{f}_{nat,a|\P'}^{[g]})\notag\\
	&\ge \sum_{a\in A'} \tilde{f}_{nat,a}^{[g]}\cdot \tau_a^{[g]}(\tilde{f}_{nat,a}^{[g]})
	-\eta_{\max}\cdot |A|^2\cdot (\beta+1)\cdot \sum_{p\in \P\setminus \P'} \theta_p(T).\label{eq:Restricted-Nonatomic-NE-Total-Cost-In-Reduced-Game}
\end{align}
Inequalities~\eqref{eq:Atomic-Nonatomic-NE-cost-difference-Import}--\eqref{eq:Restricted-Nonatomic-NE-Total-Cost-In-Reduced-Game} yield that
\begin{equation}\label{eq:Atomic-Nonatomic-NE-cost-dfference-Import2}
	\begin{split}
		|C(\tilde{f}_{at}^{[g]},\G^{[g]})&-C(\tilde{f}_{nat}^{[g]},
		\G^{[g]})|
		\le |C(\tilde{f}_{1,nat}^{[g]},\G^{[g]}_1)-
		C(\tilde{f}_{nat|\P'}^{[g]},\G_{2}^{[g]})|\\
		&+O(\sqrt{\frac{1}{T}})+O(\sum_{a\in A\setminus A'} \theta_a(T) )
		+O(\sum_{p\in\P\setminus\P'} \theta_p(T)).
	\end{split}
\end{equation}
}

\WuThird{Note that $\G_1^{[g]}$ and $\G_{2}^{[g]}$ share the same path set $\P'$ and 
	the same arc set $A'.$ In particular, inequality \eqref{eq:Nontight_Path_Flow_Upbound} 
yields that the respective total demands of an arbitrary O/D pair $k\in\K$ in $\G_1^{[g]}$ and  $\G_2^{[g]}$ deviate from each other by at most 
$O(\sum_{p\in\P\setminus\P'}\theta_p(T)),$ and that
$|\sigma_{1,a}(x)-\sigma_{2,a}(x)|\in 
O(\sum_{p\in\P\setminus\P'}\theta_p(T))$ for all $x\in [0,1]$ and all 
$a\in A'.$
Hence, viewed as non-atomic congestion games,
the distance $\|\G_{1}^{[g]}-\G_{2}^{[g]}\|$ between $\G_1^{[g]}$ and $\G_2^{[g]}$ w.r.t. the
metric defined in \citep{Wu2019Sensitivity} is $O(\sum_{p\in\P\setminus\P'}\theta_p(T)).$
Here, to save space, we recommend readers to \citep{Wu2019Sensitivity} for a detailed definition of that metric.}

\WuThird{Let $\G_1^{[g]'}$ be the non-atomic congestion game that has the same components
as $\G_{1}^{[g]},$ but with the original scaled cost functions $\tau_a^{[g]}$ for
each arc $a\in A'.$ Similarly, let $\G_2^{[g]'}$ be the non-atomic congestion game
with all components of $\G_{2}^{[g]},$ but again with the original scaled cost functions 
$\tau_a^{[g]}$ for each arc $a\in A'.$ Then we obtain also that 
$\|\G_{1}^{[g]}-\G_{1}^{[g]'}\|\in O(\sum_{p\in\P\setminus\P'}\theta_p(T)),$
$\|\G_{2}^{[g]}-\G_{2}^{[g]'}\|\in O(\sum_{p\in\P\setminus\P'}\theta_p(T)),$
and $\|\G_{1}^{[g]'}-\G_2^{[g]'}\|\in O(\sum_{p\in\P\setminus\P'}\theta_p(T)).$}

\WuThird{Since $\G_{1}^{[g]}$ and $\G_{1}^{[g]'}$ differ only in their cost functions, Lemma~10d
of \citep{Wu2019Sensitivity} then yields that the total cost difference between 
the respective non-atomic NE flows of $\G_{1}^{[g]}$ and $\G_{1}^{[g]'}$
is in $O(\sqrt{\sum_{p\in\P\setminus\P'}\theta_p(T)}).$ Here, we observe that 
the cost functions of both $\G_{1}^{[g]}$ and $\G_{1}^{[g]'}$ are Lipschitz bounded by a 
constant independent of $T$ on $[0,1].$
Similarly, the cost difference between the respective non-atomic NE flows 
of $\G_2^{[g]}$ and $\G_{2}^{[g]'}$ is also in $O(\sqrt{\sum_{p\in\P\setminus\P'}\theta_p(T)}).$
Moreover, as $\G_{1}^{[g]'}$ and $\G_2^{[g]'}$ differ only at demands, Lemma~11a
of \citep{Wu2019Sensitivity} implies that the cost difference between their non-atomic NE
flows is again in $O(\sqrt{\sum_{p\in\P\setminus\P'}\theta_p(T)}).$
In summary, we have that 
\begin{displaymath}
	|C(\tilde{f}_{1,nat}^{[g]},\G^{[g]}_1)-
	C(\tilde{f}_{nat|\P'}^{[g]},\G_{2}^{[g]})|
	\in O(\sqrt{\sum_{p\in\P\setminus\P'}\theta_p(T)}),
\end{displaymath}
which, combined with inequality~\eqref{eq:Atomic-Nonatomic-NE-cost-dfference-Import2},
yields that 
\begin{displaymath}
	|C(\tilde{f}^{[g]}_{at},\G^{[g]})-C(\tilde{f}_{nat}^{[g]},\G)|
	\in O(T^{-\frac{1}{2\cdot \max_{a\in A}\beta_a}}).
\end{displaymath}
Here, we note that both $\theta_a(T)$ and $\theta_p(T)$ are 
$O(-\frac{1}{\max_{b\in A}\beta_b})$ for all non-tight arcs $a\in A$ and
all non-tight paths $p\in \P,$ that $\tilde{f}_{1,nat}^{[g]}$ is a non-atomic NE flow of 
$\G_1^{[g]},$ and that $\tilde{f}_{nat|\P'}^{[g]}$ is a non-atomic NE flow of $\G_{2}^{[g]}.$
Then Lemma~10 and Lemma~11 of \citep{Wu2019Sensitivity} apply here, since they
bound the non-atomic NE cost difference from above by the square root of the metric with  constant
multipliers in terms of the total demands, of the arc cost function values at the maximum feasible arc flows w.r.t. the total demands,
and of the Lipschitz constants of the cost functions, each of which
is bounded from above by a constant independent of $T$ in the four games $\G_{1}^{[g]},$ $\G_{1}^{[g]'},$
$\G_{2}^{[g]}$ and $\G_{2}^{[g]'}.$ Again, to save space, we recommend the readers to 
\citep{Wu2019Sensitivity} for details.}

\WuThird{This completes the proof of Theorem~\ref{thm:GeneralPoly}c.}

\textbf{Proof of Theorem~\ref{thm:GeneralPoly}a--b:}

\WuSecondRevision{The argument
	for the proof of Theorem~\ref{thm:GeneralPoly}c does \emph{not} carry over to
	Theorem~\ref{thm:GeneralPoly}a--b, since 
	the non-atomic SO flow of the resulting scaled game $\G^{[g]}$ could be of $o(1),$ and then
	the convergence rate of \citep{Colini2020OR} does not apply, when
	we still use the same scaling factor $g$ as above, and when the condition, that $\frac{d_k}{T}\ge \xi_k$ for all
	$k\in\K$ and some constants $\xi_k>0$ independent of $T,$ does not hold.
Interested readers may refer to \citep{Wu2019} for a detailed explanation.}
	
	\WuSecondRevision{To prove Theorem~\ref{thm:GeneralPoly}a--b, we now employ the technique
	of \emph{asymptotic decomposition} developed by \citep{Wu2019}, and show that 
Theorem~\ref{thm:GeneralPoly}a--b hold for an arbitrary infinite sequence of growing total demand,
which then directly implies the convergence in Theorem~\ref{thm:GeneralPoly}a--b.}

\WuSecondRevision{To that end, we now consider an arbitrary sequence
$(\S_n)_{n\in\N}$ s.t.}
each \WuSecondRevision{component}
$
\S_n$ is a \WuSecondRevision{tuple} $\big(\D^{(n)},d^{(n)},\tilde{f}^{(n)}_{ran},
\tilde{\Pi}^{(n)},
\tilde{f}^{(n)}_{nat},f^{*(n)}_{nat}\big)
$ \wu{satisfying}
properties (S1)--(S3) below:
\begin{itemize}
	\item (S1) $\D^{(n)}=\cup_{k\in \KK}\ \D_k^{(n)}$ is an agent set
	of the game $\G$, and $d^{(n)}=(d^{(n)}_{k,i})_{i\in\D_k^{(n)},\ k\in\KK}$
	 is a vector of demands for \WuSecondRevision{the}
	agents in $\D^{(n)}.$ Here, $\D_k^{(n)}$ is an agent set of  O/D pair $k\in\K,$ 
	$d_{k,i}^{(n)}\in (0,\ \upsilon]$ is the demand of agent $i\in \D_{k}$ of O/D pair $k\in\K,$
	and $\upsilon>0$ is a finite \emph{constant} upper bound
	of the maximum individual demand
	\WuSecondRevision{$d^{(n)}_{max}:=\max_{k\in\K,i\in\D^{(n)}_k}d_{k,i}^{(n)},$} which is independent of the sequence $(\S_n)_{n\in\N}$. \WuSecondRevision{To facilitate our discussion,
		we denote the resulting game $\G$ equipped with 
		$\D^{(n)}$ and $d^{(n)}$ by $\G_n:=(\tau,\D^{(n)},d^{(n)})$ for each $n\in\N.$}
	\item (S2)  
	$\tilde{f}^{(n)}_{ran}=(\tilde{f}^{(n)}_{ran,p})_{p\in\P},$
	$\tilde{f}^{(n)}_{nat}=(\tilde{f}^{(n)}_{nat,p})_{p\in\P}$
	and $f^{*(n)}_{nat}=(f^{*(n)}_{nat,p})_{p\in\P}$
	are \WuSecondRevision{an} \emph{arbitrary} mixed \WuSecondRevision{NE flow,} 
	\WuSecondRevision{an arbitrary} non-atomic NE flow,
	and an arbitrary non-atomic SO flow
	of \wu{the game} $\G_n,$ respectively. Moreover,
	$\tilde{\Pi}^{(n)}=(\tilde{\Pi}_{i,p}^{(n)})_{i\in\D_k^{(n)},
		p\in\P_k,k\in\KK}$
	is the mixed profile
	of $\tilde{f}^{(n)}_{ran}.$ 
	\item (S3) $\lim_{n\to \infty}T(\D^{(n)},d^{(n)})=\infty,$
	where $T(\D^{(n)},d^{(n)})=\sum_{k\in \KK}d_k^{(n)}$
	is the total demand of $\G_n,$
	and $d_k^{(n)}=\sum_{i\in \D_k^{(n)}}d_{k,i}^{(n)}$
	is the demand of O/D pair  $k\in\K.$
	\WuSecondRevision{To simplify notation, we write 
		$T_n:=T(\D^{(n)},d^{(n)})$ in this proof.}
\end{itemize}
Due to the arbitrary choice of $(\S_n)_{n\in\N}$,
Theorem~\ref{thm:GeneralPoly}a--b \WuSecondRevision{hold if and only if}
\begin{equation}\label{eq:GeneralPoly-Obj}
	\lim_{n\to\infty}
	\frac{C\left(\E_{\tilde{\Pi}^{(n)}}\big(\tilde{f}^{(n)}_{ran}\big),\G_n\right)}{C(f^{*(n)}_{nat},\G_n)}=1\
	\text{   and  }
	\lim_{n\to \infty}\frac{\E_{\tilde{\Pi}^{(n)}}\left[C(\tilde{f}^{(n)}_{ran},
		\G_n)\right]}{
		C(f^{*(n)}_{at},
		\G_n)}=1.
\end{equation}
Here, 
$f^{*(n)}_{at}$ \WuSecondRevision{is} an arbitrary
atomic SO flow 
of \WuSecondRevision{$\G_n.$}
%
Note that
\citep{Wu2019} 
\wu{have} proved \wu{that} 
$	\lim_{n\to \infty}\rho_{nat}(\G_n)=
\lim_{n\to \infty}\frac{C(\tilde{f}^{(n)}_{nat},\G_n)}{C(f^{*(n)}_{nat},\G_n)}=1$
as $n\to \infty$ (i.e., $T_n\to\infty$).
\wu{Hence,}  
we can obtain \eqref{eq:GeneralPoly-Obj}
\wu{with Lemma~\ref{lemma:SO-Cost-Comparision},}  if
\eqref{eq:AD-Obj-Mixed}
below holds.
\begin{equation}\label{eq:AD-Obj-Mixed}
	\lim_{n\to\infty}\!
	\frac{C\big(\E_{\tilde{\Pi}^{(n)}}(\tilde{f}^{(n)}_{ran}),\G_n\big)}{C(\tilde{f}^{(n)}_{nat},\G_n)}\\
	=\!
	\!\lim_{n\to\infty}\!
	\frac{\E_{\tilde{\Pi}^{(n)}}\!\big[C(\tilde{f}^{(n)}_{ran},\G_n)\big]}{C(\tilde{f}^{(n)}_{nat},\G_n)}\!=\!1
\end{equation}

\WuSecondRevision{\wu{Equation \eqref{eq:AD-Obj-Mixed} means that
	the expected flow
	$\E_{\tilde{\Pi}^{(n)}}(\tilde{f}_{ran}^{(n)})$ is
	\emph{asymptotically} as efficient as $\tilde{f}_{nat}^{(n)},$
	and thus almost as efficient as 
	$f_{nat}^{*(n)}$ when $n$ is large enough.
	Moreover, the mixed NE flow $\tilde{f}_{ran}^{(n)}$
	is also asymptotically as efficient as $f_{nat}^{*(n)}$ w.r.t. its expected total cost.
	Hence, all the flows, $\tilde{f}^{(n)}_{at}, f^{*(n)}_{at}, \E_{\tilde{\Pi}^{(n)}}(\tilde{f}_{ran}^{(n)}), \tilde{f}_{ran}^{(n)}, f_{ran}^{*(n)},\tilde{f}_{nat}^{(n)},$
	and $f_{nat}^{*(n)},$ are almost
	equally efficient, when $T_n$ gets large and \eqref{eq:AD-Obj-Mixed} holds.}}

\WuSecondRevision{To prove \eqref{eq:AD-Obj-Mixed}, we only need to consider
NE flows. This avoids the difficulties of discussing the SO flows
$\tilde{f}_{nat}^{*(n)}$ and $f^{*(n)}_{at}.$}
To facilitate \WuSecondRevision{our} discussion,
we assume, w.l.o.g., that
\begin{itemize}
	\item (S4) $\lim_{n\to \infty}d_k^{(n)}\in [0,\infty]$
	and $\lim_{n\to\infty}\frac{d_{k}^{(n)}}{d_{k'}^{(n)}}
	\in [0,\infty]$
	exist for all $k,k'\in\KK.$
\end{itemize}
Note that \eqref{eq:AD-Obj-Mixed}
holds for an arbitrary sequence
$(\S_n)_{n\in\N}$ satisfying (S1)--(S3)
if and only if \eqref{eq:AD-Obj-Mixed}
holds for an arbitrary sequence $(\S_n)_{n\in\N}$ satisfying (S1)--(S4).
This follows since every infinite subsequence
$(\S_{n_j})_{j\in\N}$ of 
a sequence $(\S_n)_{n\in\N}$ satisfying (S1)--(S3)
has \wu{an} infinite subsequence
$(\S_{n_{j_l}})_{l\in\N}$ fulfilling (S1)--(S4).
\WuSecondRevision{We will use similar subsequence arguments implicitly and repeatedly in this proof.}

\WuSecondRevision{We now show} \eqref{eq:AD-Obj-Mixed}
for an arbitrary sequence $(\S_{n})_{n\in\N}$ satisfying
(S1)--(S4) \WuSecondRevision{with} the  technique
of asymptotic decomposition
\WuSecondRevision{of} \citep{Wu2019}. 

\textbf{Step I: The asymptotic decomposition of $\G_{n}:$}

We put $\K_{reg}:=\{k\in\K:\ \lim_{n\to \infty}d_k^{(n)}=\infty\}$
and $\K\setminus\K_{reg}=:\K_{irreg}.$
We obtain by (S3)--(S4) that $\K_{reg}\ne \emptyset.$
We call $k$ in $\K_{reg}$ \emph{regular},
and $k'$ in $\K_{irreg}$ \emph{irregular}.
So $d_k^{(n)}$ is bounded for $k\in\KK_{irreg},$ and
unbounded for $k\in\KK_{reg}.$

We collect these $k\in\KK_{reg}$ with an \emph{equal} demand growth rate into one class, which, by property (S4), then results
in an \emph{ordered partition} $\K_1\prec \cdots\prec \K_m$ of $\K_{reg}$ satisfying conditions (AD1)--(AD2).
\begin{itemize}
	\item (AD1) $\lim_{n\to \infty}\frac{d_k^{(n)}}{d_{k'}^{(n)}}\in (0,\infty),$
	i.e., $d_k^{(n)}\in \Theta(d^{(n)}_{k'}),$
	for all $k,k'\in \K_u$ for each $u\in\mathcal{M}:=\{1,\ldots,m\},$
	\item (AD2) $\lim_{n\to \infty}\frac{d_k^{(n)}}{d_{k'}^{(n)}}=0,$
	i.e., $d_k^{(n)}\in o(d^{(n)}_{k'}),$
	for all $k\in \K_u,k'\in \K_l$ for all $u,l\in\mathcal{M}$
	with $l<u.$
\end{itemize}
Here, $m\ge 1$ is an integer, and 
$\K_{l}\prec\K_{u}$ means that these 
$k'\in \K_{l}$ have demands $d_{k'}^{(n)}$ converging to $\infty$ much \emph{faster} than
\wu{the demands $d_k^{(n)}$ of} those $k\in \K_u.$

\WuSecondRevision{W.r.t. this partition, $\G_n$ is decomposed into 
``subgames'' $\G_{n|\K_1},\ldots,\G_{n|\K_m},\G_{n|\K_{irreg}}.$
Here, we call $\G_{n|\K'}$ a \emph{subgame} of $\G_{n}$ if 
$\G_{n|\K'}$ is a restriction of $\G_{n}$ to the subset $\K'$ of O/D pairs, i.e.,
$\G_{n|\K'}$ is the game obtained by removing all O/D pairs $k\in\K\setminus \K',$ and all
agents $i\in \cup_{k\in\K\setminus \K'}\D_k^{(n)}$ together with their demands
$d_{k,i}^{(n)}$ from $\G_{n}.$ We thus ignore completely the influence
of all O/D pairs $k\in\K\setminus \K'$ when we consider the subgame 
$\G_{n|\K'}.$}

\WuSecondRevision{Clearly, each \emph{regular subgame} $\G_{n|\KK_{u}}$
has the agent set $\D^{(n)}_{|\K_u}:=\cup_{k\in \KK_u}\D_k^{(n)},$
the demand vector $d^{(n)}_{|\K_u}:=(d^{(n)}_{k,i})_{i\in\D_k^{(n)},
	k\in\KK_u}$
and  the total demand $T_{n|\K_u}$ $
:=$ $\sum_{k\in \KK_u}d_k^{(n)}$ that tends to $\infty$ as
$n\to\infty.$ The \emph{irregular subgame} 
$\G_{n|\K_{irreg}}$ has \wu{the} agent set 
$\D^{(n)}_{|\K_{irreg}}:=\cup_{k\in \KK_{irreg}}\D_k^{(n)},$
\wu{the} demand vector $d_{|\K_{irreg}}^{(n)}:=(d^{(n)}_{k,i})_{i\in\D_k^{(n)},
	k\in\KK_{irreg}}$
and \wu{the} total demand $T_{n|\KK_{irreg}}
:=\sum_{k\in \KK_{irreg}}d_k^{(n)}$ that tends to a \emph{bounded
	constant} as
$n\to\infty.$ Moreover, we obtain by condition (AD2) that 
\begin{equation}\label{eq:AD-Total-Demand-Comparison}
	\lim_{n\to \infty}\frac{T_{n|\K_u}}{T_{n|\K_l}}=
	\lim_{n\to \infty} \frac{T_{n|\K\setminus\cup_{l'=1}^l\K_{l'}}}{T_{n|\K_l}}
	=0\quad \forall u,l\in\mathcal{M} \text{ with }l<u.
\end{equation} 
Here, we observe that 
$T_n=T_{n|\KK_{irreg}}+\sum_{l=1}^m T_{n|\K_l}$
and $T_{n|\K\setminus\cup_{l'=1}^l\K_{l'}}=T_{n|\KK_{irreg}}+\sum_{l'=l+1}^m T_{n|\K_{l'}}.$}

\WuSecondRevision{Note that each flow $f^{(n)}$ of $\G_{n}$ 
induces a \emph{joint} total cost 
\begin{displaymath}
	C_{\K'}(f^{(n)},\G_{n})
	:=\sum_{k\in \KK'}\sum_{p\in\P_k}
	f^{(n)}_p\cdot \tau_p(f^{(n)})
\end{displaymath}
and an \emph{independent} total cost 
\begin{displaymath}
	C(f^{(n)}_{|\K'},\G_{n|\KK'})=\sum_{k\in \KK'}\sum_{p\in\P_k}
	f^{(n)}_p\cdot \tau_p(f_{|\K'}^{(n)})=
	\sum_{a\in A}f^{(n)}_{a|\KK'}\cdot \tau_a(f^{(n)}_{a|\K'})
\end{displaymath}
for an arbitrary subset $\K'$ of $\K,$ where
$f^{(n)}_{|\K'}=
(f^{(n)}_p)_{p\in \cup_{k\in\K'}\P_k}$ 
is the restriction of $f^{(n)}$ into the subgame $\G_{n|\K'},$
$f^{(n)}_{a|\K'}=\sum_{k\in\K'}\sum_{p\in\P_k:a\in p}f_p^{(n)}$ is the
arc flow induced \emph{independently} by the ``flow'' $f^{(n)}_{|\K'}$
of $\G_{n|\K'},$ and 
$\tau_p(f^{(n)}_{|\K'})=\sum_{a\in p} \tau_a(f^{(n)}_{a|\K'})$
is the \emph{independent} path cost under the flow $f^{(n)}_{|\K'}.$
Here, we use that $f^{(n)}_{|\K'}$ is indeed a flow of $\G_{n|\K'},$ and so
the independent total cost of $f^{(n)}$ is exactly the total cost of the flow
$f^{(n)}_{|\K'}$ in $\G_{n|\K'}.$}

\WuSecondRevision{W.r.t. the above asymptotic decomposition, we obtain
for an arbitrary flow $f^{(n)}$ of $\G_{n}$ that
\begin{displaymath}
	\begin{split}
		&C(f^{(n)},\G_{n})=C_{\K_{irreg}}(f^{(n)},\G_{n})+\sum_{u=1}^m C_{\K_u}(f^{(n)},\G_{n})\\
		&\hspace{1.5cm}\ge C(f_{|\K_{irreg}}^{(n)},\G_{n|\K_{irreg}})+\sum_{u=1}^m C(f_{|\K_u}^{(n)},\G_{n|\K_u}),\\
		&f^{(n)}_a=f^{(n)}_{a|\KK_{irreg}}+\sum_{u=1}^{m}f^{(n)}_{a|\K_u}\quad 
		\forall a\in A.
	\end{split}
\end{displaymath}
The above inequality follows since the \emph{joint} path cost $\tau_p(f^{(n)})$
considers all subgames and the independent path cost
$\tau_p(f^{(n)}_{|\K'})$ considers only flow induced by agents
from $k\in \K'$, and so  $\tau_p(f^{(n)})\ge \tau_p(f_{|\K'}^{(n)})$
for each subset $\K'$ of $\K.$}

\textbf{Step II: \WuSecondRevision{An equivalent transformation in the limit}}

\citep{Wu2019} \wu{have shown} for \wu{this decomposition of} non-atomic NE flows that 
\begin{align}
	&\lim_{n\!\to\! \infty} \frac{C(\!\tilde{f}_{nat}^{(n)},
		\G_n)}{C(\!\tilde{f}_{nat}^{(n,-)},\G_{n|\K_{irreg}})\!+\!\sum_{l\!=\!1}^{m} \!C(\!\tilde{f}_{nat}^{(n,l)}\!, \G_{n|\K_l})}\notag\\
	&=\!\lim_{n\!\to\! \infty}\! \frac{C_{\K_{irreg}}(\tilde{f}_{nat}^{(n)},
		\G_n)\!+\!\sum_{l=1}^m\!C_{\K_l}(\!\tilde{f}_{nat}^{(n)},
		\G_n)}{C(\!\tilde{f}_{nat}^{(n,-)},\G_{n|\K_{irreg}})\!+\!\sum_{l\!=\!1}^{m}\! C(\tilde{f}_{nat}^{(n,l)},
		\G_{n|\K_l})}\notag\\
	&=\!\lim_{n\to \infty} \frac{\sum_{l=1}^mC_{\K_l}(\tilde{f}_{nat}^{(n)},
		\G_n)}{\sum_{l=1}^{m} C(\tilde{f}_{nat}^{(n,l)},
		\G_{n|\K_l})}=1,\label{eq:Wu-AD}
\end{align}
where $\tilde{f}_{nat}^{(n,l)}$
and $\tilde{f}_{nat}^{(n,-)}$
are \wu{non-atomic NE flows}
of $\G_{n|\K_{l}}$ 
and $\G_{n|\K_{irreg}},$ respectively, 
and \WuSecondRevision{where} $C_{\K_l}(\tilde{f}_{nat}^{(n)},\G_{n})=
\sum_{k\in\K_l}\sum_{p\in\P_k} 
\tilde{f}_{nat,p}^{(n)}\cdot \tau_p(\tilde{f}_{nat}^{(n)})$ \wu{is} the joint
total cost of $\G_{n|\K_l}$
in the non-atomic NE flow $\tilde{f}_{nat}^{(n)}$
of $\G_n.$ Note that the restriction $\tilde{f}^{(n)}_{nat| \K_l}=
(\tilde{f}_{nat,p}^{(n)})_{p\in\P_k,k\in\K_l}$ of $\tilde{f}^{(n)}_{nat}$
is a non-atomic flow of $\G_{n|\K_l}$, but \WuSecondRevision{need} not be a non-atomic NE flow
of $\G_{n|\K_l},$ and so has \WuSecondRevision{a} total cost that may differ from 
$\tilde{f}_{nat}^{(n,l)}.$

The
irregular subgame vanishes in the limit
of \eqref{eq:Wu-AD}, since it has a bounded
total demand \WuSecondRevision{and thus a negligible influence on the limit}, see \citep{Wu2019} for details.

For each $n\in \N,$ let 
$\tilde{f}^{(n,l)}_{nat},$ $l\in \mathcal{M}=\{1,\ldots,m\},$ 
\WuSecondRevision{be arbitrary non-atomic NE flows of subgames $\G_{n|\K_{l}},$}
and \WuSecondRevision{let}
$\tilde{f}^{(n,-)}_{nat}$ be \WuSecondRevision{an arbitrary 
non-atomic NE flow} of $\G_{n|\K_{irreg}}.$ 

\WuSecondRevision{Then \eqref{eq:AD-Obj-Mixed} follows
from \eqref{eq:Wu-AD}} if and only if
	\begin{align}
	&
	\lim_{n\to \infty}\frac{C\left(\E_{\tilde{\Pi}^{(n)}}\big(\tilde{f}^{(n)}_{ran}\big),\G_n\right)}{C(\tilde{f}_{nat}^{(n,-)},\G_{n|\K_{irreg}})+\sum_{l=1}^{m} C(\tilde{f}_{nat}^{(n,l)},
		\G_{n|\K_l})}=1,\label{eq:Arb-Poly-Exp-Obj}\\
	&
	\lim_{n\to\infty}	\frac{\E_{\tilde{\Pi}^{(n)}}\left[C(\tilde{f}^{(n)}_{ran},\G_n)\right]}{C(\tilde{f}_{nat}^{(n,-)},\G_{n|\K_{irreg}})+\sum_{l=1}^{m} C(\tilde{f}_{nat}^{(n,l)},
		\G_{n|\K_l})}=1.\label{eq:Arb-Poly-Mixed-Obj}
\end{align}


\textbf{Step III: Further subsequence arguments}

We will prove \eqref{eq:Arb-Poly-Exp-Obj}--\eqref{eq:Arb-Poly-Mixed-Obj} by 
scaling each of the above regular subgames $\G_{n|\K_u}$ independently. \WuSecondRevision{We define 
 a scaling factor 
$g^{(u)}_n:=
T_{n|\K_u}^{\lambda_u}$ for each $u\in\mathcal{M}$,}
\wu{where} $\lambda_u:=\max_{k\in\KK_u}\min_{p\in \P_k}\max_{a\in p}
\beta_a$ $\ge 0.$ \WuSecondRevision{To facilitate the discussion,}
 we also call \WuSecondRevision{a path $p\in \cup_{k\in \KK_u}\P_k$ for $u\in\mathcal{M}$}
\emph{tight} if $\max_{a\in p}\beta_a\le \lambda_u,$ and \emph{non-tight} if $\max_{a\in p}\beta_a>\lambda_u.$
Clearly, every $k\in\K_{u}$ has \emph{at least} one tight
path.
Moreover, each tight path $p\in \cup_{k\in\KK_u}\P_k$
contains only arcs $a\in A$ with $\beta_a\le \lambda_u,$
while a non-tight path $p' \in\cup_{k\in\KK_u}\P_k$
contains \emph{at least} one arc $a\in A$ with 
$\beta_a>\lambda_u,$ for each 
$u\in\mathcal{M}.$  These simple facts will be very helpful in \WuSecondRevision{the further} discussion.


To simplify the proof, we 
assume further that the sequence $(\S_n)_{n\in\N}$ satisfies
properties (S5)--(S8) below.
\begin{description}
	\item[(S5)] $\lim_{n\to \infty}\frac{g_n^{(u)}}{g_n^{(l)}}\in [0,\infty]$
	exists for $u,l\in \mathcal{M}.$
	\wu{We call $g_n^{(u)}$
		and $g_n^{(l)}$} \emph{mutually comparable}.
	\item[(S6)] 
	$
	\lim_{n\to \infty}\frac{\tilde{f}_{nat}^{(n,u)}}{T_{n|\K_u}}
	=\lim_{n\to \infty}\frac{\big(\tilde{f}_{nat,p}^{(n,u)}\big)_{p\in\P_k,k\in\KK_u}}{T_{n|\K_u}}
	=: \tilde{f}^{(\infty,u)}_{nat}=
	(\tilde{f}^{(\infty,u)}_{nat,p})_{p\in\P_k,k\in\KK_u}
	$
	for $u\in\mathcal{M}.$
	\item[(S7)] For $u\in\mathcal{M},$
	$
	\lim_{n\to \infty}\frac{\E_{\tilde{\Pi}^{(n)}}(\tilde{f}^{(n)}_{ran})_{|\K_u}}{T_{n|\K_u}}
	=\lim_{n\to \infty}\frac{\big(\E_{\tilde{\Pi}^{(n)}}(\tilde{f}^{(n)}_{ran,p})\big)_{p\in\P_k,k\in\KK_u}}{T_{n|\K_u}}
	=: \tilde{f}^{(\infty,u)}_{exp}$ $=$ $
	(\tilde{f}^{(\infty,u)}_{exp,p})_{p\in\P_k,k\in\KK_u}
	$.
	Here,  $\E_{\tilde{\Pi}^{(n)}}(\tilde{f}^{(n)}_{ran})_{|\K_u}
	=\E_{\tilde{\Pi}^{(n)}}(\tilde{f}^{(n)}_{ran|\K_u})$
	is the restriction of the expected flow $\E_{\tilde{\Pi}^{(n)}}(\tilde{f}_{ran}^{(n)})=(\E_{\tilde{\Pi}^{(n)}}(\tilde{f}_{ran,p}^{(n)}))_{p\in\P}$ of the mixed NE flow $\tilde{f}^{(n)}_{ran}$ of $\G_n$
	to the subgame $\G_{n|\K_u},$ which is a non-atomic flow 
	of $\G_{n|\K_u}$.
	\item[(S8)]
	$\lim_{n\to \infty}\frac{d_k^{(n)}}{T_{n|\K_u}}
	=:d_k^{(\infty,u)}\in (0,1]$
	for each $k\in\KK_u$ \wu{and} each 
	$u\in\mathcal{M}.$
	This actually follows directly
	from \WuSecondRevision{property (S4) and decomposition}
	condition (AD1).
\end{description}

Note that \eqref{eq:Arb-Poly-Exp-Obj}--\eqref{eq:Arb-Poly-Mixed-Obj} hold
for an arbitrary sequence $(\S_n)_{n\in\N}$
fulfilling (S1)--(S4) if and only if
they hold for an arbitrary sequence $(\S_n)_{n\in\N}$
satisfying (S1)--(S8). This follows again
since every infinite subsequence $(\S_{n_j})_{j\in\N}$
of an sequence $(\S_n)_{n\in\N}$ fulfilling (S1)--(S4)
contains an infinite subsequence $(\S_{n_{j_l}})_{l\in\N}$ fulfilling (S1)--(S8).

\textbf{Step IV: \WuSecondRevision{The inductive assumptions}}

We will  prove \eqref{eq:Arb-Poly-Exp-Obj}--\eqref{eq:Arb-Poly-Mixed-Obj} by
showing that the statements IA1--IA7 below hold for each 
$u\in\mathcal{M},$ using an induction
	on $u$ over \wu{the} set $\{0,\ldots,m\}=\{0\}\bigcup \mathcal{M}.$ Here, we put 
$\K_0:=\emptyset,$ $g_n^{(0)}:=0$ and identify $\G_{n|\K_0}$ as \wu{the empty} subgame and employ \wu{the} convention that 
IA1--IA7 hold for $u=0.$
\begin{description}
	\item[\textbf{IA1}]  $\max_{p\in \P_k:\ \E_{\tilde{\Pi}^{(n)}}(\tilde{f}^{(n)}_{ran,p})>0}
	\tau_p(\E_{\tilde{\Pi}^{(n)}}(\tilde{f}^{(n)}_{ran}))\in O\big(\max_{l=0}^{u}\ g_n^{(l)}\big)$ for $k\in \cup_{l=0}^{u}\KK_l,$
	i.e., the \emph{most costly} path \wu{used} by 
	agents of the subgame $\G_{n|\cup_{l=0}^{u}\K_l}$
	has a cost of \emph{at most}  $O\big(\max_{l=0}^{u}\ g_n^{(l)}\big)$
	in the expected flow $\E_{\tilde{\Pi}^{(n)}}(\tilde{f}_{ran}^{(n)})$.
	\item[\textbf{IA2}] The joint total cost of 
	$\G_{n|\cup_{l=0}^{u}\K_l}$
	in $\E_{\tilde{\Pi}^{(n)}}(\tilde{f}_{ran}^{(n)})$
	is $\Theta\big(\max_{l=0}^{u}\ g_n^{(l)}\cdot 
	T_{n|\K_l}\big),$ i.e., 
	\begin{displaymath}
		\begin{split}
			\sum_{l=0}^{u}C_{\K_l} (\E_{\tilde{\Pi}^{(n)}}(\tilde{f}_{ran}^{(n)}),\G_{n})&=
			\sum_{l=0}^{u}\sum_{k\in\K_l}\sum_{p\in \P_k}
			\E_{\tilde{\Pi}^{(n)}}(\tilde{f}_{ran,p}^{(n)})\cdot \tau_p(\E_{\tilde{\Pi}^{(n)}}(\tilde{f}_{ran}^{(n)}))
			\\
			&\in \Theta\big(\max_{l=0}^{u}\ g_n^{(l)}\cdot 
			T_{n|\K_l}\big).
		\end{split}
	\end{displaymath}
	\item[\textbf{IA3}] $\lim_{n\!\to\! \infty}\!\frac{\sum_{l=0}^{u}C_{\K_l} (\E_{\tilde{\Pi}^{(n)}}(\tilde{f}_{ran}^{(n)}),\G_{n})}{\sum_{l=0}^{u}\!C(\tilde{f}^{(n,l)}_{nat},\G_{n|\K_l})}\!=\!1.$
	\item[\textbf{IA4}]  $\max_{p\in \P_k:\ \E_{\tilde{\Pi}^{(n)}}(\tilde{f}^{(n)}_{ran,p})>0}
	\E_{\tilde{\Pi}^{(n)}}(\tau_p(\tilde{f}^{(n)}_{ran}))\in O\big(\max_{l=0}^{u}\ g_n^{(l)}\big)$ for $k\in \cup_{l=0}^{u}\KK_l.$
	\item[\textbf{IA5}]  The expected joint total cost of 
	$\G_{n|\cup_{l=0}^{u}\K_l}$
	in $\tilde{f}_{ran}^{(n)}$
	is also $\Theta\big(\max_{l=0}^{u}\ g_n^{(l)}\cdot 
	T_{n|\K_l}\big),$ i.e.,
	\begin{displaymath}
		\begin{split}
			\E_{\tilde{\Pi}^{(n)}}[\sum_{l=0}^u C_{\K_l} (\tilde{f}^{(n)},\G_{n})]&=
			\sum_{l\!=\!0}^{u}\sum_{k\in\K_l}\!\sum_{p\in \P_k}\!
			\E_{\tilde{\Pi}^{(n)}}(\tilde{f}_{ran,p}^{(n)}\cdot \tau_p(\tilde{f}_{ran}^{(n)}))\\
			&\in\! \Theta\big(\max_{l\!=\!0}^{u}\ g_n^{(l)}\cdot 
			T_{n|\K_l}\big).
		\end{split}
	\end{displaymath}
	\item[\textbf{IA6}] $\lim_{n\!\to\! \infty}\!\frac{\E_{\tilde{\Pi}^{(n)}}[\sum_{l=0}^u C_{\K_l} (\tilde{f}^{(n)},\G_{n})]}{\sum_{l=0}^{u}C(\tilde{f}^{(n,l)}_{nat},\G_{n|\K_l})}\!=\!1.$
	\item[\textbf{IA7}]  For each $k\in \KK_l$ and each
	$l=0,\ldots,u,$
	\[
	\max_{p\in \P_k:\ \E_{\tilde{\Pi}^{(n)}}(\tilde{f}^{(n)}_{ran,p})>0}
	\E_{\tilde{\Pi}^{(n)}}[\tilde{f}_{ran,p}^{(n)}\cdot \tau_p(\tilde{f}^{(n)}_{ran})]\in O\big(T_{n|\K_l}\cdot \max_{l'=0}^{l}\ g_n^{(l')}\big).
	\]
\end{description}

\WuSecondRevision{Among these inductive assumptions,}
IA3 and IA6
\wu{are the most crucial}. We obtain trivially that 
\wu{\begin{align}
&\lim_{n\to \infty}\!\frac{\sum_{l=0}^{m}
	\sum_{k\in\K_l} \sum_{p\in \P_k} 
	\E_{\tilde{\Pi}^{(n)}}\big(\tilde{f}_{ran,p}^{(n)}\big)\!\cdot\! \tau_p\big(\E_{\tilde{\Pi}^{(n)}}(\tilde{f}^{(n)}_{ran})\big)}{
	\sum_{l=0}^{m}\!C(\tilde{f}^{(n,l)}_{nat},\G_{n|\K_{l}})}
\!=\!1,\label{eq:Regular-OD-Expt}\\
&\lim_{n\to\infty} \frac{\sum_{l=0}^{m}\sum_{k\in\K_l}\sum_{p\in \P_k}
	\E_{\tilde{\Pi}^{(n)}}\big(\tilde{f}_{ran,p}^{(n)}\cdot \tau_p(\tilde{f}_{ran}^{(n)})\big)}{\sum_{l=0}^{m}C(\tilde{f}^{(n,l)}_{nat},\G_{n|\K_{l}})}
\!=\!1,\label{eq:Regular-OD-Sto}
\end{align}}
when IA3 and IA6
hold for all $u\in\mathcal{M}=\{1,\ldots,m\}.$
Then \eqref{eq:Arb-Poly-Exp-Obj}--\eqref{eq:Arb-Poly-Mixed-Obj} follow immediately
from \eqref{eq:Regular-OD-Expt}--\eqref{eq:Regular-OD-Sto},
since
\wu{the} subgame $\G_{n|\K_{irreg}}$ \wu{has} a bounded total demand and thus can be neglected
in the limits by \wu{an} argument similar to that in the proof
of Fact~\ref{claim:CaseOfNegligibility} below.

Moreover, \WuSecondRevision{IA4 implies IA7.}
This follows since the random event 
``$\tilde{f}_{ran,p}^{(n)}\le T_{n|\K_l}$'' occurs almost surely
for each $p\in\cup_{k\in \KK_l}\P_k$
and each $l\in \mathcal{M}.$
\WuSecondRevision{In fact, IA4 also implies IA1, which we will claim later
in Fact~\ref{claim:User-Cost-Bound}.}

Now, we consider an arbitrary $u\in \{0,\ldots,m-1\}$ such that 
IA1--IA7 hold for each 
non-negative integer $l\le u.$
We will prove IA1--IA7 for $u+1,$
which then implies \eqref{eq:Regular-OD-Expt}--\eqref{eq:Regular-OD-Sto}
by induction, and so completes the proof of Theorem~\ref{thm:GeneralPoly}a--b.

\textbf{Step V: Validating IA1--IA7 for $u+1$}



\wu{For each $k\in\KK_{u+1}$ and each $p\in\P_k,$ $\E_{\tilde{\Pi}^{(n)}}(\tilde{f}_{ran,p}^{(n)})=0$}  implies that 
$\tilde{\Pi}_{i,p}^{(n)}=0$
for every $i\in\D_k^{(n)}$
\wu{because of} \eqref{eq:ExpectedPathFlow}. So, \wu{for each $p\in\P_k,$} $\E_{\tilde{\Pi}^{(n)}}(\tilde{f}^{(n)}_{ran,p})
=0$ is equivalent to \wu{the fact} that the
random event ``$\tilde{f}_{ran,p}^{(n)}=0$''
occurs \emph{almost surely},
i.e., $\PP_{\tilde{\Pi}^{(n)}}(\tilde{f}_{ran,p}^{(n)}=0)=1.$
Similarly, for each $a\in A,$
$\E_{\tilde{\Pi}^{(n)}}(\tilde{f}^{(n)}_{ran,a|\cup_{l=0}^{u}\K_l})
=0$ is equivalent to \wu{the fact} that
the random event ``$\tilde{f}_{ran,a|\cup_{l=0}^{u}\K_l}^{(n)}=0$''
\wu{occurs almost surely.}
Therefore, we can directly remove $\tilde{f}_{ran,a|\cup_{l=0}^{u}\K_l}^{(n)}$
from the respective expectations of \wu{the random variables} $\tau_a(\tilde{f}_{ran,a}^{(n)})$
\wu{and} $\tilde{f}_{ran,a}^{(n)}\cdot \tau_a(\tilde{f}_{ran,a}^{(n)})$
when $\E_{\tilde{\Pi}^{(n)}}(\tilde{f}^{(n)}_{ran,a|\cup_{l=0}^{u}\K_l})
=0.$

With the above observations and the inductive assumptions IA1
and IA4 of step $u,$ we obtain 
\eqref{eq:Cost-Split-Expected}--\eqref{eq:Cost-Split-Sto} for every arc $a\in A$ and every path $p\in\cup_{k\in \KK_{u+1}}\P_k.$

\begin{align}
&\tau_a\big[\E_{\tilde{\Pi}^{(n)}}(\tilde{f}_{ran,a}^{(n)})\big]
=\tau_a\big[\E_{\tilde{\Pi}^{(n)}}(\tilde{f}^{(n)}_{ran,a|\cup_{l=0}^{u}\K_l})
+\E_{\tilde{\Pi}^{(n)}}(\tilde{f}^{(n)}_{ran,a|\K\setminus \cup_{l=0}^{u}\K_l})\big]\notag\\
&\hspace{2cm}=\begin{cases}
O\big(\max_{l=0}^{u}\ g_n^{(l)}\big)&\text{if }\E_{\tilde{\Pi}^{(n)}}[\tilde{f}^{(n)}_{ran,a|\cup_{l=0}^{u}\K_l}]>0,\\
\tau_a\big[\E_{\tilde{\Pi}^{(n)}}(\tilde{f}^{(n)}_{ran,a|\K\setminus \cup_{l=0}^{u}\K_l})\big]
&\text{if }\E_{\tilde{\Pi}^{(n)}}[\tilde{f}^{(n)}_{ran,a|\cup_{l=0}^{u}\K_l}]=0,
\end{cases}\label{eq:Cost-Split-Expected}\\
&\E_{\tilde{\Pi}^{(n)}}\big[\tau_a(\tilde{f}^{(n)}_{ran,a})
\big]
=\E_{\tilde{\Pi}^{(n)}}\big[\tau_a(\tilde{f}^{(n)}_{ran,a|\cup_{l=0}^u\K_l}+\tilde{f}^{(n)}_{ran,a|\K\setminus\bigcup_{l=0}^u\K_l})
\big]\notag\\
&\hspace{2cm}=\begin{cases}
O(\max_{l=0}^{u}\ g_n^{(l)})&\text{if }
\E_{\tilde{\Pi}^{(n)}}\big[\tilde{f}^{(n)}_{ran,a|\cup_{l=0}^{u}\K_l}\big]>0,\\
\E_{\tilde{\Pi}^{(n)}}\big[\tau_a(\tilde{f}^{(n)}_{ran,a|\K\setminus\cup_{l=1}^{u}\K_l})\big]&\text{if }\E_{\tilde{\Pi}^{(n)}}\big[\tilde{f}^{(n)}_{ran,a|\cup_{l=0}^{u}\K_l}\big]=0,
\end{cases}\label{eq:Cost-Split-random-Epected}\\
&\E_{\tilde{\Pi}^{(n)}}\big[
\tilde{f}_{ran,p}^{(n)}
\cdot \tau_a(\tilde{f}_{ran,a}^{(n)})
\big]=\E_{\tilde{\Pi}^{(n)}}\big[
\tilde{f}_{ran,p}^{(n)}
\cdot \tau_a(\tilde{f}_{ran,a|\cup_{l=0}^{u}\K_l}^{(n)}+
\tilde{f}_{ran,a|\K\setminus\cup_{l=0}^{u}\K_l}^{(n)})
\big]\notag\\
&\hspace{0.2cm}=\begin{cases}
O\big(T_{n|\KK_{u\!+\!1}}\cdot \max_{l=0}^{u}\ g_n^{(l)}\big)&\text{if }\E_{\tilde{\Pi}^{(n)}}[\tilde{f}^{(n)}_{ran,a|\cup_{l=0}^{u}\K_l}]\!>\!0,\\
\E_{\tilde{\Pi}^{(n)}}\big[\tilde{f}_{ran,p}^{(n)}
\cdot \tau_a(\tilde{f}_{ran,a|\K\setminus\cup_{l\!=\!0}^{u}\K_l}^{(n)})\big]&\text{if }\E_{\tilde{\Pi}^{(n)}}[\tilde{f}^{(n)}_{ran,a|\cup_{l\!=\!0}^{u}\K_l}]\!=\!0.
\end{cases}\label{eq:Cost-Split-Sto}
\end{align}

 \WuSecondRevision{\eqref{eq:Cost-Split-Expected} and \eqref{eq:Cost-Split-random-Epected}
 follow} since IA1 and IA4
hold in steps $l\le u,$ and \WuSecondRevision{since the expected arc flow} $\E_{\tilde{\Pi}^{(n)}}(\tilde{f}^{(n)}_{ran,a|\cup_{l=0}^{u}\K_l})$ $>$ $0$
implies that arc $a$ belongs to some path $p\in \cup_{l=0}^u\cup_{k\in\K_l}\P_k$
with $\E_{\tilde{\Pi}^{(n)}}(\tilde{f}_{ran,p}^{(n)})>0.$
\eqref{eq:Cost-Split-Sto} follows immediately
from \eqref{eq:Cost-Split-random-Epected} and
\wu{the fact} that $\tilde{f}_{ran,p}^{(n)}\le T_{n|\KK_{u\!+\!1}}$
for every path $p\in\cup_{k\in \KK_{u+1}}\P_k.$
Here, we observe that \eqref{eq:Cost-Split-Expected}--\eqref{eq:Cost-Split-Sto} hold trivially
when $u=0,$ i.e., \WuSecondRevision{when} $\cup_{l=0}^{u}\K_l=\emptyset.$

With  \eqref{eq:Cost-Split-Expected}--\eqref{eq:Cost-Split-random-Epected},
we \wu{now} show  IA1, IA4 and IA7
for step $u+1$.
\begin{fact}\label{claim:User-Cost-Bound}
	IA1, IA4 and IA7 hold
	for step $u+1.$
\end{fact}

\textbf{Proof of Fact~\ref{claim:User-Cost-Bound}:}
We only need to show IA4 \WuSecondRevision{and IA1}, as 
IA4 implies IA7.

\WuSecondRevision{\textbf{Proof of IA4:}} We obtain by \eqref{eq:Cost-Split-random-Epected} that $\E_{\tilde{\Pi}^{(n)}}(\tau_p(\tilde{f}^{(n)}_{ran}))\in O(\max_{l=0}^{u+1}\ g_n^{(l)})$ for every \emph{tight} path $p\in\cup_{k\in \KK_{u+1}}\P_k.$
This follows since \WuSecondRevision{a} tight path consists of arcs $a$ with degrees $\beta_a\le \lambda_{u+1},$
$g_n^{(u+1)}=T_{n|\KK_{u\!+\!1}}^{\lambda_{u+1}},$
and $T_{n|\KK_{u\!+\!1}}\in \Theta(T_{n|\K_l\setminus\cup_{l=0}^u\K_l})$, see \eqref{eq:AD-Total-Demand-Comparison}. 
Then IA4 of step $u+1$ follows immediately from \wuu{the facts} that every $k\in\K_{u+1}$ has \emph{at least} one tight path, that $\tilde{f}_{ran}^{(n)}$ is a mixed NE flow,
and that $d_{k,i}^{(n)}\le \upsilon$ for all $k$ and $i.$
Here, we \WuSecondRevision{use} that the \WuSecondRevision{choice of a \WuSecondRevision{single} agent}
has a \emph{negligible} influence on the expected cost of a path 
(compared to $\max_{l=0}^{u+1}\ g_n^{(l)}$) when $n$
is large enough, since his 
demand is bounded from above by the constant $\upsilon$ and 
$T_{n|\KK_{u\!+\!1}}\to \infty$ as $n\to\infty.$
In fact, we can even think of $\tilde{f}_{ran}^{(n)}$ as 
a mixed WE flow (see Remark~\ref{remark:Mixed-WE}) in this proof.

\WuSecondRevision{\textbf{Proof of IA1:}} 
	\WuSecondRevision{We show for each $a\in A$ that 
\begin{equation}\label{eq:A1_proofs}
	\tau_a[\E_{\tilde{\Pi}^{(n)}}(\tilde{f}^{(n)}_{ran,a})]
	\le \E_{\tilde{\Pi}^{(n)}}
	[\tau_a(\tilde{f}^{(n)}_{ran,a})]+O(1),
\end{equation}
which, combined with IA4 of step $u+1,$ implies 
IA1 in step $u+1,$ since
$\max_{l=0}^{u+1}\ g_n^{(l)}\in \Omega(1)$ for every $u=0,\ldots,m-1.$
	Note that $\tau_a(\cdot)$ is convex on $[M_a,\infty)$
	for some constant $M_a>0,$ since $\tau_a(\cdot)$ is a non-decreasing
	polynomial with 
	an integer degree $\beta_a\ge 0.$ Jensen's inequality from Lemma~\ref{lemma:Markov}c
	then yields that
\begin{displaymath}
	\begin{split}
		\tau_a[\E_{\tilde{\Pi}^{(n)}}&(\tilde{f}^{(n)}_{ran,a})]
		=
		\tau_a[\E_{\tilde{\Pi}^{(n)}}(\tilde{f}^{(n)}_{ran,a}|
		\tilde{f}^{(n)}_{ran,a}\ge M_a)]\cdot \PP_{\tilde{\Pi}^{(n)}}[\tilde{f}^{(n)}_{ran,a}\ge M_a]\\
		&\hspace{1cm}+\tau_a[\E_{\tilde{\Pi}^{(n)}}(\tilde{f}^{(n)}_{ran,a}|
		\tilde{f}^{(n)}_{ran,a}< M_a)]\cdot \PP_{\tilde{\Pi}^{(n)}}[\tilde{f}^{(n)}_{ran,a}< M_a]\\
		&\le \E_{\tilde{\Pi}^{(n)}}[\tau_a(\tilde{f}^{(n)}_{ran,a})|
		\tilde{f}^{(n)}_{ran,a}\ge M_a]\cdot \PP_{\tilde{\Pi}^{(n)}}[\tilde{f}^{(n)}_{ran,a}\ge M_a]\\
		&\hspace{1cm}+\tau_a[\E_{\tilde{\Pi}^{(n)}}(\tilde{f}^{(n)}_{ran,a}|
		\tilde{f}^{(n)}_{ran,a}< M_a)]\cdot \PP_{\tilde{\Pi}^{(n)}}[\tilde{f}^{(n)}_{ran,a}< M_a]\\
		&\le \E_{\tilde{\Pi}^{(n)}}[\tau_a(\tilde{f}^{(n)}_{ran,a})|
		\tilde{f}^{(n)}_{ran,a}\ge M_a]\cdot \PP_{\tilde{\Pi}^{(n)}}[\tilde{f}^{(n)}_{ran,a}\ge M_a]\\
		&\hspace{1cm}+\E_{\tilde{\Pi}^{(n)}}[\tau_a(\tilde{f}^{(n)}_{ran,a})|
		\tilde{f}^{(n)}_{ran,a}< M_a]\cdot \PP_{\tilde{\Pi}^{(n)}}[\tilde{f}^{(n)}_{ran,a}< M_a]
		+\tau_a(M_a)\\
		&=\E_{\tilde{\Pi}^{(n)}}
		[\tau_a(\tilde{f}^{(n)}_{ran,a})]+O(1).
	\end{split}
\end{displaymath}
This proves IA1 for step $u+1,$ and completes the proof of Fact~\ref{claim:User-Cost-Bound}\footnote{\WuSecondRevision{There is an alternative proof
that does not need the convexity of the polynomial cost functions. The random variable 
$X_n:=\tilde{f}^{(n)}_{ran,a|\K\setminus\cup_{l=0}^{u}\K_l}$ has a variance
of at most $\upsilon\cdot \E_{\tilde{\Pi}^{(n)}}[X_n],$
and so the random event ``$X_n\le 
\E_{\tilde{\Pi}^{(n)}}[X_n]-\sqrt{2\cdot \upsilon\cdot \E_{\tilde{\Pi}^{(n)}}[X_n]}$''
occurs with a probability of at most $\frac{1}{2}$ by Chebyshev's inequality
from Lemma~\ref{lemma:Markov}b. This then implies
$\tau_a(\E_{\Pi^{(n)}}[X_n])\in O(\E_{\tilde{\Pi}^{(n)}}[\tau_a(X_n)]),$ and so 
IA1 in step $u+1$ holds by IA4 of step $u+1$ and \eqref{eq:Cost-Split-Expected}.
Hence, Theorem~\ref{thm:GeneralPoly} carries also over to non-decreasing polynomial cost
functions with arbitrary non-negative
real-valued degrees, since only the above proof of IA1 for step $u+1$ involves 
the convexity of the cost functions.}}.\hfill$\square$}

Note that either $g_n^{(u+1)}\in O(\max_{l=0}^{u}\ g_n^{(l)})$
or $g_n^{(u+1)}\in \omega(\max_{l=0}^{u}\ g_n^{(l)}),$ since \wu{the scaling factors are mutually comparable, i.e.,}
the sequence $(\S_n)_{n\in\N}$ satisfies property (S5).
To validate IA2--IA3 and IA5--IA6, we thus
distinguish \emph{two} subcases.

\textbf{Subcases I: $g_n^{(u+1)}\in O(\max_{l=0}^{u}\ g_n^{(l)})$}

Fact~\ref{claim:CaseOfNegligibility} shows IA2--IA3,
and IA5--IA6 for step $u+1$
when $g_n^{(u+1)}\in O(\max_{l=0}^{u}\ g_n^{(l)}).$
Then  Fact~\ref{claim:User-Cost-Bound}--Fact~\ref{claim:CaseOfNegligibility}
together imply IA1--IA7 for step $u+1$ when $g_n^{(u+1)}\in O(\max_{l=0}^{u}\ g_n^{(l)}).$ 
Here, we observe that $g_n^{(u+1)}\in O(\max_{l=0}^{u}\ g_n^{(l)})$
happens only \wu{when} $u>0,$ since  $g_n^{(0)}=0$
and $g_n^{(u+1)}\in \Omega(1)$
for each $u\in \{0,\ldots,m-1\}.$
\begin{fact}\label{claim:CaseOfNegligibility}
	If $g_n^{(u+1)}\in O(\max_{l=0}^{u}\ g_n^{(l)}),$
	then IA2--IA3,
	and IA5--IA6 hold at step $u+1$.
\end{fact}

\textbf{Proof of Fact~\ref{claim:CaseOfNegligibility}:}
IA1 of step $u+1$ \wu{yields}
\begin{align*}
\E_{\tilde{\Pi}^{(n)}}[\tilde{f}_{ran,p}^{(n)}]\cdot \tau_p\big[\E_{\tilde{\Pi}^{(n)}}(\tilde{f}_{ran}^{(n)})\big]
\in O\big(T_{n|\KK_{u+1}}\cdot 
\max_{l=0}^{u+1}\ g_n^{(l)}\big)
=O\big(T_{n|\KK_{u\!+\!1}}\cdot 
\max_{l=0}^{u}\ g_n^{(l)}\big)
\end{align*} 
for every $p\in\cup_{k\in \KK_{u+1}}\P_k$ with 
$\E_{\tilde{\Pi}^{(n)}}[\tilde{f}_{ran,p}^{(n)}]>0$ when $g_n^{(u+1)}\in O(\max_{l=0}^{u}\ g_n^{(l)}).$
This in turn implies with \eqref{eq:AD-Total-Demand-Comparison} that
\begin{equation}\label{eq:Ato-IA2}
\begin{split}
C_{\KK_{u+1}}\big[\E_{\tilde{\Pi}^{(n)}}(\tilde{f}_{ran}^{(n)}),\G_{n}\big]&=\sum_{k\in\KK_{u+1}}\sum_{p\in\!\P_k}\!\E_{\tilde{\Pi}^{(n)}}[\tilde{f}_{ran,p}^{(n)}]\cdot \tau_p\big[\E_{\tilde{\Pi}^{(n)}}(\tilde{f}_{ran}^{(n)})\big]\\
&\in\! O\big(T_{n|\KK_{u+1}}\!\cdot\! 
\max_{l=0}^{u}\ g_n^{(l)}\big)
\subseteq\! o\big( 
\max_{l=0}^{u}\ T_{n|\KK_{l}}\!\cdot\! g_n^{(l)}\big).
\end{split}
\end{equation}
Then IA2 of step $u+1$ follows from \eqref{eq:AD-Total-Demand-Comparison},
$g_n^{(u+1)}\in O(\max_{l=0}^u\ g_n^{(l)}),$ and
\wu{IA2} of step $u.$

IA3 of step $u+1$ \wu{then} follows \wu{from} \eqref{eq:Ato-IA2}, 
IA3 of step $u,$ and \eqref{eq:Neg1},
\begin{equation}\label{eq:Neg1}
\begin{split}
C(\tilde{f}^{(n\!,u\!+\!1)}_{nat},\G_{n|\K_{u+1}})
\in\! \Theta(T_{n|\KK_{u+1}}\!\cdot\! g_n^{(u\!+\!1)})
\subseteq\! o\big( 
\max_{l=0}^{u}\ T_{n|\KK_{l}}\!\cdot\! g_n^{(l)}\big),
\end{split}
\end{equation}
see \eqref{eq:NonNE-Limit-Stepu} of Fact~\ref{fact:Wu2019} below.

\eqref{eq:Ato-IA2}--\eqref{eq:Neg1}
show that $\G_{n|\K_{u+1}}$ is negligible when we compute its 
respective total cost in  the
expected flow of $\tilde{f}_{ran}^{(n)}$ and $\tilde{f}_{nat}^{(n,u+1)},$
and
when $g_n^{(u+1)}\in O(\max_{l= 0}^u\ g_n^{(l)}).$ 
Similarly, we can obtain 
IA5--IA6 of step $u+1$ by
showing that $\G_{n|\K_{u+1}}$
is again negligible when we compute its joint expected total cost in $\tilde{f}^{(n)}_{ran}$
and when $g_n^{(u+1)}\in O(\max_{l=0}^{u}\ g_n^{(l)}),$
\wu{where} we use {IA4} 
and IA7 of step $u+1$. 

This completes \wu{the} proof of Fact~\ref{claim:CaseOfNegligibility}.
\hfill$\square$

\textbf{Subcase II: $g_n^{(u+1)}\in \omega(\max_{l=1}^{u}\ g_n^{(l)})$}

We \wu{now show} 
IA2--IA3
and IA5--IA6 for step $u+1$
\wu{when} $g_n^{(u+1)}\in \omega(\max_{l=1}^{u}\ g_n^{(l)}).$
\wu{This,} together with Fact~\ref{claim:User-Cost-Bound} \wu{and} Fact~\ref{claim:CaseOfNegligibility},
\WuSecondRevision{completes} the proof of IA1--IA7 for step $u+1$.

Fact~\ref{fact:Wu2019} below \wu{states} a helpful result 
from \citep{Wu2019},
which \wu{shows} that the limit $\tilde{f}_{nat}^{(\infty,u+1)}
=\lim_{n\to \infty}$ $\frac{\tilde{f}_{nat}^{(n,u+1)}}{T_{n|\KK_{u+1}}}$
in (S6)
is a non-atomic NE flow of a limit game $\G_{|\K_{u+1}}^{(\infty)},$
and the scaled non-atomic NE cost $\frac{C(\tilde{f}_{nat}^{(n,u\!+\!1)},\G_{n|\K_{u+1}})}{
	T_{n|\KK_{u+1}}\cdot g_n^{(u+1)}}$
of
subgame $\G_{n|\K_{u+1}}$ converges to the total cost of the non-atomic NE flow $\tilde{f}_{nat}^{(\infty,u+1)}$ of $\G_{|\K_{u+1}}^{(\infty)}.$
Here, $\G_{|\K_{u+1}}^{(\infty)}$ is a \wu{(non-atomic)} congestion game  with (O/D pair) demand vector 
$
d^{(\infty,u\!+\!1)}\!=\!(d_k^{(\infty,u\!+\!1)})_{k\in\KK_{u\!+\!1}}\!=
\lim_{n\!\to\! \infty}\frac{d_{|\KK_{u+1}}^{(n)}}{T_{n|\KK_{u+1}}}\!\lim_{n\to \infty}
\frac{(d_k^{(n,u\!+\!1)})_{k\in\KK_{u\!+\!1}}}{T_{n|\KK_{u+1}}}
$
and cost function
\begin{equation}\label{eq:Limit-Cost-Function-SubgameKu}
\tau_a^{(\infty\!,u\!+\!1)}(x)\!=\!\lim_{y\to x^+}\!
\lim_{n\!\to\!\infty}\!\frac{\tau_a\big(T_{n|\KK_{u+1}}\!\cdot\! y\big)}{g_n^{(u\!+\!1)}}
\!=\!
\begin{cases}
\infty&\text{if }\beta_a\!>\!\lambda_{u\!+\!1},\\
\eta_a\cdot x^{\beta_a}&\text{if }\beta_a\!=\!\lambda_{u\!+\!1},\\
0&\text{if }\beta_a\!<\!\lambda_{u\!+\!1},
\end{cases}
\end{equation}
for every $x\in [0,1]$ and every arc $a\in A.$ 
\begin{fact}[See \citep{Wu2019}]\label{fact:Wu2019}
	For each $u=\{0,\ldots,m-1\}=\{0\}\cup (\mathcal{M}\setminus\{m\}),$
	\begin{align}\label{eq:NonNE-Limit-Stepu}
	\lim_{n\to\infty}	 &\frac{C(\tilde{f}^{(n,u+1)}_{nat},\G_{n|\K_{u+1}})}{T_{n|\KK_{u+1}}\cdot g_n^{(u+1)}}
	=\lim_{n\to\infty}	\sum_{k\in \KK_{u+1}}\sum_{p\in \P_k}
	\frac{\tilde{f}_{nat,p}^{(n,u+1)}}{T_{n|\KK_{u+1}}}\cdot \frac{\tau_p(\tilde{f}_{nat}^{(n,u+1)})}{g_n^{(u+1)}}\notag\\
	&=\sum_{a\in A}f^{(\infty,u+1)}_{nat,a}\cdot \tau_a^{(\infty,u+1)}(f^{(\infty,u+1)}_{nat,a})
	\in (0,\infty)
	\end{align}
	and $f^{(\infty,u+1)}_{nat}$
	is a non-atomic NE flow of $\G_{|\K_{u+1}}^{(\infty)}$
	s.t. $f^{(\infty,u+1)}_{nat.p}=0$
	for each non-tight $p\in \cup_{k\in \K_{u+1}}\P_k.$
	Here, we employ the convention
	\wu{that} $0\cdot \infty=0.$
\end{fact}

\wu{Properties similar} to Fact~\ref{fact:Wu2019} actually
carry over to the expected flow $\E_{\tilde{\Pi}^{(n)}}(\tilde{f}_{ran|\KK_{u+1}}^{(n)})$
when $g_n^{(u+1)}\in \omega(\max_{l=0}^{u}\ g_n^{(l)}).$
Here, we recall that 
$\lim_{n\to \infty}\frac{\E_{\tilde{\Pi}^{(n)}}(\tilde{f}_{ran|\KK_{u+1}}^{(n)})}{T_{n|\KK_{u+1}}}
=\tilde{f}^{(\infty,u+1)}_{exp},$ see (S7).

\WuSecondRevision{Consider an arbitrary arc $a\in A$ with
$\beta_a\le \lambda_{u+1}.$} \WuSecondRevision{Then} \eqref{eq:Cost-Split-Expected} 
\WuSecondRevision{yields}
\begin{align}
&\lim_{n\to\infty}	\frac{\tau_a\big(\E_{\tilde{\Pi}^{(n)}}(\tilde{f}_{ran,a}^{(n)})\big)}{g_n^{(u+1)}}
=\lim_{n\to \infty} \frac{\tau_a\big(\E_{\tilde{\Pi}^{(n)}}(\tilde{f}_{ran,a}^{(n)})\big)}{g_n^{(u+1)}}
\cdot \1_{(0,\infty)}\left(\E_{\tilde{\Pi}^{(n)}}[\tilde{f}_{ran,a|\cup_{l=0}^u\K_l}^{(n)}]\right)\notag\\
&\hspace{0.5cm}+\lim_{n\to\infty}\frac{\tau_a\left(
	\E_{\tilde{\Pi}^{(n)}}\left(\tilde{f}_{ran,a|\K_{u+1}}^{(n)}
	+\tilde{f}_{ran,a|\K\setminus \cup_{l=0}^{u+1}\K_l}^{(n)}\right)
	\right)}{g_n^{(u+1)}}\cdot\1_{\{0\}}\left(\E_{\tilde{\Pi}^{(n)}}[\tilde{f}_{ran,a|\cup_{l=0}^u\K_l}^{(n)}]\right)\notag\\
&=\lim_{n\to\infty}\frac{\tau_a\left(T_{n|\KK_{u+1}}\cdot 
	\E_{\tilde{\Pi}^{(n)}}\left(\frac{\tilde{f}_{ran,a|\K_{u+1}}^{(n)}}{T_{n|\KK_{u+1}}}
	+\frac{\tilde{f}_{ran,a|\K\setminus \cup_{l=0}^{u+1}\K_l}^{(n)}}{T_{n|\KK_{u+1}}}\right)
	\right)}{T_{n|\KK_{u+1}}^{\lambda_{u+1}}}\notag\\
&=\tau_a^{(\infty,u+1)}\Big(\sum_{k\in \KK_{u+1}}\sum_{p\in\P_k:a\in p}\tilde{f}_{exp,p}^{(\infty,u+1)}\Big)=\tau_a^{(\infty,u+1)}(\tilde{f}_{exp,a}^{(\infty,u+1)}).\label{eq:Arc-Cost-Limit-expected}
\end{align}
\WuSecondRevision{Here, we use \eqref{eq:AD-Total-Demand-Comparison}
to remove the influence of subgame $\G_{n|\K\setminus\cup_{l=0}^{u+1}\K_l},$
and use \eqref{eq:Limit-Cost-Function-SubgameKu} to obtain the limit.
The subgame $\G_{n|\cup_{l=0}^u\K_{l}}$ vanishes in the limit
since $g^{(u+1)}_n\in \omega(\max_{l=0}^u\ g_n^{(l)})$ and \eqref{eq:Cost-Split-Expected}.}

Hence, we obtain for each \emph{tight} path $p\in\cup_{k\in \KK_{u+1}}\P_k$ that
\begin{equation}\label{eq:Path-Cost-Limit-atomic-tight}
\begin{split}
\lim_{n\!\to\!\infty}\!	\frac{\tau_p(\E_{\tilde{\Pi}^{(n)}}(\tilde{f}_{ran}^{(n)}))}{g_n^{(u+1)}}
\!=\!\lim_{n\!\to\!\infty}\!	\frac{\tau_p(\E_{\tilde{\Pi}^{(n)}}(\tilde{f}^{(n)}_{ran|\K_{u+1}}))}{g_n^{(u+1)}}
\!=\!\tau_p^{(\infty,\!u+1)}(\tilde{f}_{exp}^{(\infty,\!u+1)})\!\in\! [0,\infty),
\end{split}
\end{equation}
\wu{since} a tight path $p\in\cup_{k\in \KK_{u+1}}\P_k$ contains \emph{only}
arcs $a\in A$ with $\beta_a\le \lambda_{u+1}.$

 Lemma~\ref{lemma:Subgame-U-Sto-Exp} shows another helpful
 result when we justify
 IA2--IA3 and IA5--(Sto--IA6)
 for the case that $g_n^{(u+1)}\in \omega(\max_{l=0}^{u} g_n^{(l)})$.    We move \wu{the} long proof of 
Lemma~\ref{lemma:Subgame-U-Sto-Exp}
to Appendix~\ref{proof:Subgame-U-Sto-Exp}.
\begin{lemma}\label{lemma:Subgame-U-Sto-Exp}
	Consider an arbitrary $a\in A,$ an arbitrary
	\WuSecondRevision{$u\in \{1,\ldots,m\},$} an
	arbitrary polynomial function 
	$h(\cdot)$ with degree $\beta\ge 0$ 
	and a constant $g_n:=T_{n|\K_{u}}^{\lambda}$
	with an arbitrary constant exponent $\lambda>0.$
	Assume that $h(x)$ is non-decreasing
	on $[0,\infty).$
	Then
	\begin{displaymath}
		\lim_{n\!\to\! \infty}\frac{\E_{\tilde{\Pi}^{(n)}}(h(\tilde{f}_{ran,a|\K\setminus\cup_{l=0}^{u-1}\K_l}^{(n)}))}{g_n}=\lim_{n\!\to\! \infty}\frac{h(\E_{\tilde{\Pi}^{(n)}}(\tilde{f}_{ran,a|\K\setminus\cup_{l=0}^{u-1}\K_l}^{(n)}))}{g_n}\in [0,\infty]
	\end{displaymath}
	 if 
	either of the two limits exist.
\end{lemma}

With Lemma~\ref{lemma:Subgame-U-Sto-Exp}, Fact~\ref{claim:SubcaseII-Sto-Lambda>0}
confirms IA2--IA3 and IA5--IA6
for the case that $g_n^{(u+1)}\in \omega(\max_{l= 0}^{u}\ g_n^{(l)}).$
\begin{fact}\label{claim:SubcaseII-Sto-Lambda>0}
	IA2--IA3, 
	IA5--IA6 hold at step $u+1$ when $g_n^{(u+1)}\in \omega(\max_{l= 0}^{u}\ g_n^{(l)}).$
\end{fact}

\textbf{Proof of Fact~\ref{claim:SubcaseII-Sto-Lambda>0}:}

We obtain by \wu{IA1 of step $u+1$} that
$
	\E_{\tilde{\Pi}^{(n)}}\big(\tilde{f}^{(n)}_{ran,p}\big)
	\in o(T_{n|\K_{u+1}})
$
for an arbitrary \emph{non-tight} path $p\in\cup_{k\in \KK_{u+1}}\P_k$.
Otherwise, \wu{there is} a non-tight path $p\in\cup_{k\in \KK_{u+1}}\P_k$
\wu{with} 
\wu{\begin{displaymath}
\tau_{p}(\E_{\tilde{\Pi}^{(n)}}(\tilde{f}_{ran}^{(n)}))=\sum_{a\in p}\tau_{a}(\E_{\tilde{\Pi}^{(n)}}(\tilde{f}_{ran,a}^{(n)}))
\ge \sum_{a\in p}\tau_{a}(\E_{\tilde{\Pi}^{(n)}}(\tilde{f}_{ran,p}^{(n)}))
\in \omega(g_n^{(u+1)})
\end{displaymath}}
and $\E_{\tilde{\Pi}^{(n)}}\big(\tilde{f}^{(n)}_{ran,p}\big)$ $
\in$ $\Omega(T_{n|\KK_{u+1}}).$
This \WuSecondRevision{contradicts}
IA1 of step $u+1,$ i.e.,
$\tau_{p'}\big(\E_{\tilde{\Pi}^{(n)}}(\tilde{f}_{ran}^{(n)})\big)\in O(g_n^{(u+1)})$
for every $p'\in \cup_{k\in\KK_{u+1}}\P_k$ with 
$\E_{\tilde{\Pi}^{(n)}}(\tilde{f}_{ran,p'}^{(n)})>0$ when 
$g_n^{(u+1)}\in \omega(\max_{l=0}^{u}\ g_n^{(l)}).$
Here, we recall again that every non-tight path
$p\in\cup_{k\in \KK_{u+1}}\P_k$ contains at least one arc
$a\in A$ \wu{whose cost function has a} degree $\beta_a>\lambda_{u+1}.$

Consequently, \wu{we obtain for each \emph{non-tight} path
$p\in\cup_{k\in \KK_{u+1}}\P_k$ that
\begin{equation}\label{eq:Non-tight-Mixed-Expected}
\begin{split}
&\tilde{f}_{exp,p}^{(\infty,u+1)}=\lim_{n\to \infty}\frac{\E_{\tilde{\Pi}^{(n)}}(\tilde{f}_{ran,p}^{(n)})}{
T_{n|\KK_{u+1}}}=0,\\
&\lim_{n\to \infty}\frac{\E_{\tilde{\Pi}^{(n)}}(\tilde{f}_{ran,p}^{(n)})}{
	T_{n|\KK_{u+1}}}
\cdot \frac{\tau_p\big(\E_{\tilde{\Pi}^{(n)}}\big(\tilde{f}^{(n)}_{ran}\big)\big)}{g_n^{(u+1)}}\\
&\hspace{2cm}=\lim_{n\to \infty}\frac{\E_{\tilde{\Pi}^{(n)}}(\tilde{f}_{ran,p}^{(n)})
\cdot \1_{(0,\infty)}(\E_{\tilde{\Pi}^{(n)}}(\tilde{f}_{ran,p}^{(n)}))}{
	T_{n|\KK_{u+1}}}
\cdot \frac{\tau_p\big(\E_{\tilde{\Pi}^{(n)}}\big(\tilde{f}^{(n)}_{ran}\big)\big)}{g_n^{(u+1)}}\\
&\hspace{2cm}=0.
\end{split}
\end{equation} 
Here, we used again IA1 of step $u+1.$ }

\eqref{eq:Path-Cost-Limit-atomic-tight} and \eqref{eq:Non-tight-Mixed-Expected}
together imply that 
\begin{align}\label{eq:Expected-Limit-U}
\lim_{n\to\infty}	 &\frac{C_{\KK_{u+1}}\big[\E_{\tilde{\Pi}^{(n)}}(\tilde{f}_{ran}^{(n)}),\G_{n}\big]}{T_{n|\KK_{u+1}}\cdot g_n^{(u+1)}}=\lim_{n\to\infty}\frac{\sum_{k\in\KK_{u+1}}\sum_{p\in\P_k
	}
	\E_{\tilde{\Pi}^{(n)}}(\tilde{f}_{ran,p}^{(n)})
	\cdot \tau_p(\E_{\tilde{\Pi}^{(n)}}(\tilde{f}_{ran}^{(n)}))}{T_{n|\KK_{u+1}}\cdot g_n^{(u+1)}}\notag\\
&=\lim_{n\to\infty}	
\frac{\sum_{k\in\KK_{u+1}}\sum_{p\in\P_k:
		p \text{ is tight}}
	\E_{\tilde{\Pi}^{(n)}}(\tilde{f}_{ran,p}^{(n)})
	\cdot \tau_p(\E_{\tilde{\Pi}^{(n)}}(\tilde{f}_{ran|\K_{u+1}}^{(n)}))}{T_{n|\KK_{u+1}}\cdot g_n^{(u+1)}}\notag\\
&=\sum_{a\in A}f^{(\infty,u+1)}_{exp,a}\cdot \tau_a^{(\infty,u+1)}(f^{(\infty,u+1)}_{exp,a}),
\end{align}
\wu{where} we again \wu{use} the convention 
\wu{that} $0\cdot \infty=0.$
So IA2 of step $u+1$ \wu{holds.}

When $\lambda_{u+1}>0,$ then 
we obtain by Lemma~\ref{lemma:Subgame-U-Sto-Exp}
that 
$\tilde{f}_{exp}^{(\infty,u+1)}$
is a non-atomic NE flow of $\G^{(\infty)}_{|\K_{u+1}}.$
This follows since $g_n^{(u+1)}\in \omega(\max_{l= 0}^{u}\ g_n^{(l)})$ and
\begin{equation}\label{eq:Expected-NE}
\begin{split}
\tau^{(\infty,u+1)}_p(\tilde{f}_{exp}^{(\infty,u+1)})
&=\lim_{n\to \infty}\frac{\tau_p(
	\E_{\tilde{\Pi}^{(n)}}(\tilde{f}_{ran}))}{g_n^{(u+1)}}\\
&=\lim_{n\to \infty}\frac{\E_{\tilde{\Pi}^{(n)}}(\tau_{p}
	(\tilde{f}_{ran}^{(n)}))}{g_n^{(u+1)}}\le 
\lim_{n\to \infty}\frac{\E_{\tilde{\Pi}^{(n)}}(\tau_{p'}
	(\tilde{f}_{ran}^{(n)}))}{g_n^{(u+1)}}\\
&=\lim_{n\to \infty}\frac{\tau_{p'}(\E_{\tilde{\Pi}^{(n)}}
	(\tilde{f}_{ran}))}{g_n^{(u+1)}}
=\tau_{p'}^{(\infty,u+1)}(\tilde{f}_{exp}^{(\infty,u+1)})
\end{split}
\end{equation}
for an arbitrary $k\in\KK_{u+1}$ and two arbitrary
\emph{tight} paths $p,p'\in\P_k$ with
$
\tilde{f}_{exp,p}^{(\infty,u+1)}>0.
$
We used Lemma~\ref{lemma:Subgame-U-Sto-Exp}
to exchange the \WuSecondRevision{expectation and the function $\tau_p(\cdot)$} in \eqref{eq:Expected-NE},
used \eqref{eq:Path-Cost-Limit-atomic-tight} to obtain the limits on both sides,
and used \eqref{eq:Cost-Split-Expected}--\eqref{eq:Cost-Split-random-Epected} to remove \wu{the} influence of \wu{subgame
$\G_{|\cup_{l=0}^{u}\K_{l}}$} in the limits
when $g_n^{(u+1)}\in \omega(\max_{l=0}^{u}\ g_n^{(l)})$
and the paths $p$ and $p'$ are tight. Moreover,
the inequality in \eqref{eq:Expected-NE} follows since $\tilde{f}^{(n)}_{ran}$
is a mixed NE flow, which has a similar behavior with a mixed WE flow when 
we scale the path cost with $g^{(u+1)}$ and the maximum individual demand
is bounded from above by $\upsilon.$

When $\lambda_{u+1}=0,$ then every tight path 
has constant cost.
So \eqref{eq:Expected-NE}
holds trivially and $\tilde{f}_{exp}^{(\infty,u+1)}$
is also a non-atomic NE flow of $\G_{|\K_{u+1}}^{(\infty)}.$ Here, we recall
\eqref{eq:Non-tight-Mixed-Expected}, i.e.,  $\tilde{f}_{exp,p}^{(\infty,u+1)}>0$
only if $p\in\cup_{k\in\KK_{u+1}}\P_k$ is tight.

The above \wu{arguments} together with 
Fact~\ref{fact:Wu2019} and IA3 of step $u$ imply IA3 for step $u+1.$
%

\wu{Below we show} IA5--\wu{IA6}
for step $u+1$ when $g_n^{(u+1)}\in \omega(\max_{l=0}^{u}\ g_n^{(l)}).$


Lemma~\ref{lemma:Subgame-U-Sto-Exp}
implies for each
$\theta_n\in o(T_{n|\KK_{u+1}})$
and each $a\in A$ with $\beta_a\le \lambda_{u+1}$
that 
\begin{equation}\label{eq:Sto-LIMIT-Tight}
\begin{split}
&\lim_{n\to \infty}\frac{\E_{\tilde{\Pi}^{(n)}}\big[(\tilde{f}_{ran,a|\K\setminus\cup_{l=0}^{u}\K_l}^{(n)}\pm \theta_n)\cdot \tau_a(\tilde{f}^{(n)}_{ran,a|\K\setminus\cup_{l=0}^{u}\K_l})\big]}{
	T_{n|\KK_{u+1}}\cdot 
	g_n^{(u+1)}}\\
&=\lim_{n\to \infty}\frac{\E_{\tilde{\Pi}^{(n)}}\big[\tilde{f}_{ran,a|\K\setminus\cup_{l=0}^{u}\K_l}^{(n)}\cdot \tau_a(\tilde{f}^{(n)}_{ran,a|\K\setminus\cup_{l=0}^{u}\K_l})\big]}{
	T_{n|\K_{u+1}}\cdot 
	g_n^{(u+1)}}\\
&=\lim_{n\to \infty}\frac{\E_{\tilde{\Pi}^{(n)}}\big[\tilde{f}_{ran,a|\K\setminus\cup_{l=0}^{u}\K_l}^{(n)}\big]\cdot  \tau_a\big[\E_{\tilde{\Pi}^{(n)}}(\tilde{f}^{(n)}_{ran,a|\K\setminus\cup_{l=0}^{u}\K_l})\big]}{
	T_{n|\K_{u+1}}\cdot 
	g_n^{(u+1)}}\\
&=\lim_{n\to \infty}\frac{\E_{\tilde{\Pi}^{(n)}}\big[\tilde{f}_{ran,a|\K_{u+1}}^{(n)}\big]\cdot  \tau_a\big[\E_{\tilde{\Pi}^{(n)}}(\tilde{f}^{(n)}_{ran,a|\K_{u+1}})\big]}{
	T_{n|\K_{u+1}}\cdot 
	g_n^{(u+1)}}\\
&=\tilde{f}_{exp,a}^{(\infty,u+1)}\cdot \tau_a^{(\infty,u+1)}(\tilde{f}_{exp,a}^{(\infty,u+1)}).
\end{split}
\end{equation}
Here, 
$\E_{\tilde{\Pi}^{(n)}}[\theta_n\cdot \tau_a(\tilde{f}^{(n)}_{ran,a|\K\setminus\cup_{l=1}^{u}\K_l})]\in o(T_{n|\K_{u+1}}\cdot
	g_n^{(u+1)}),$
 as $\tau_a(\tilde{f}^{(n)}_{ran,a|\K\setminus\cup_{l=1}^{u}\K_l})$ $\in O\big(g_n^{(u+1)}\big)$
holds \wu{almost surely} when $\beta_a\le \lambda_{u+1}$.

Lemma~\ref{lemma:Subgame-U-Sto-Exp}, \eqref{eq:Cost-Split-random-Epected}--\eqref{eq:Cost-Split-Sto}, $g_n^{(u+1)}\in\omega( \max_{l=0}^{u}g_n^{(l)})$ and  \eqref{eq:Non-tight-Mixed-Expected} together \wu{imply} for each  \emph{non-tight} path $p\in \cup_{k\in \KK_{u+1}}\P_k$   that
\begin{align}
&\lim_{n\!\to\! \infty}\!\frac{\E_{\tilde{\Pi}^{(n)}}
	\!\big[\tilde{f}_{ran,p}^{(n)}
	\!\cdot\! \tau_p(\tilde{f}_{ran}^{(n)})\big]}{T_{n|\KK_{u+1}}\!\cdot\! 
	g_n^{(u\!+\!1)}}\!=\!\lim_{n\!\to\! \infty}\!\frac{\1_{(0,\infty)}\![\E_{\tilde{\Pi}^{(n)}}(\tilde{f}_{ran,p}^{(n)})]\!\cdot\! \E_{\tilde{\Pi}^{(n)}}\!\big[\tilde{f}_{ran,p}^{(n)}
	\!\cdot\! \tau_p(\tilde{f}_{ran}^{(n)})\big]}{T_{n|\KK_{u+1}}\!\cdot\! 
	g_n^{(u\!+\!1)}}\notag\\
&=\lim_{n\to \infty}\frac{\sum_{a\in p:\beta_{a}> \lambda_{u+1}}\1_{(0,\infty)}[\E_{\tilde{\Pi}^{(n)}}(\tilde{f}_{ran,p}^{(n)})]\cdot \E_{\tilde{\Pi}^{(n)}}\big[\tilde{f}_{ran,p}^{(n)}
	\cdot \tau_a(\tilde{f}_{ran,a}^{(n)})\big]}{T_{n|\K_{u+1}}\cdot
	g_n^{(u+1)}}\notag\\
&\!=\!\lim_{n\!\to\! \infty}\!\frac{\sum_{a\!\in\! p:\beta_{a}\!>\! \lambda_{u\!+\!1}}\!\1_{(\!0\!,\infty\!)}\![\E_{\tilde{\Pi}^{(n)}}\!(\tilde{f}_{ran,p}^{(n)})]\!\cdot \!\E_{\tilde{\Pi}^{(n)}}\!\big[\tilde{f}_{ran,p}^{(n)}
	\!\cdot\! \tau_a(\tilde{f}_{ran,a|\K\setminus\cup_{l\!=\!0}^{u}\K_l}^{(n)})\big]}{T_{n|\KK_{u+1}}\!\cdot\! 
	g_n^{(u\!+\!1)}}\label{eq:Sto-LIMIT-NonTight}\\
&\le\! \lim_{n\!\to\! \infty}\!\frac{\sum_{a\!\in\! p:\beta_{a}\!>\! \lambda_{u\!+\!1}}\!\1_{(\!0,\!\infty\!)}\![\E_{\tilde{\Pi}^{(n)}}\!(\tilde{f}_{ran,p}^{(n)})]\!\cdot\! \E_{\tilde{\Pi}^{(n)}}\!\big[\tilde{f}_{ran,a\!|\!\K\!\setminus\!\cup_{l\!=\!0}^{u}\K_l}^{(n)}
	\!\cdot\! \tau_a(\tilde{f}_{ran,a\!|\!\K\!\setminus\!\cup_{l\!=\!0}^{u}\K_l}^{(n)})\big]}{T_{n|\KK_{u+1}}\!\cdot\! 
	g_n^{(u\!+\!1)}}\notag\\
&=\!\lim_{n\!\to\! \infty}\!\frac{\sum_{a\!\in\! p:\beta_{a}\!> \!\lambda_{u\!+\!1}}\!\1_{(\!0\!,\infty\!)}\![\E_{\tilde{\Pi}^{(\!n\!)}}\!(\!\tilde{f}_{ran,p}^{(n)}\!)\!]\!\cdot\! \E_{\tilde{\Pi}^{(\!n\!)}}\!\big[\!\tilde{f}_{ran,a\!|\!\K\!\setminus\!\cup_{l\!=\!0}^{u}\K_l}^{(\!n\!)}\!\big]
	\!\cdot \!\tau_a\!\big[\E_{\tilde{\Pi}^{(\!n\!)}}\!(\!\tilde{f}_{ran\!,a\!|\!\K\!\setminus\cup_{l\!=\!0}^{u}\K_l}^{(n)}\!)\big]}{T_{n|\KK_{u+1}}\!\cdot\! 
	g_n^{(u\!+\!1)}}\notag\\
&\le\! \lim_{n\!\to\! \infty}\!\frac{\sum_{a\!\in\! p:\beta_{a}\!>\! \lambda_{u\!+\!1}}\!\1_{(\!0\!,\infty\!)}\![\E_{\tilde{\Pi}^{(n)}}\!(\tilde{f}_{ran,p}^{(n)})]\!\cdot\! \E_{\tilde{\Pi}^{(n)}}\!\big[\tilde{f}_{ran,a|\K\!\setminus\!\cup_{l\!=\!0}^{u}\K_l}^{(n)}\big]
	\!\cdot\! \tau_p\!\big[\E_{\tilde{\Pi}^{(n)}}(\tilde{f}_{ran}^{(n)})\big]}{T_{n|\KK_{u+1}}\!\cdot\! 
	g_n^{(u\!+\!1)}}\notag\\
&=\lim_{n\to \infty}\frac{\sum_{a\in p:\beta_{a}> \lambda_{u+1}}\1_{(0,\infty)}[\E_{\tilde{\Pi}^{(n)}}(\tilde{f}_{ran,p}^{(n)})]\cdot \E_{\tilde{\Pi}^{(n)}}\big[\tilde{f}_{ran,a|\K_{u+1}}^{(n)}\big]
}{T_{n|\K_{u+1}}}\cdot O(1)\notag\\
&\!=\!\lim_{n\!\to\! \infty}\!\frac{\sum_{a\!\in\! p:\beta_{a}\!>\! \lambda_{u+1}}\!\sum_{p'\in\cup_{k\in \KK_{u+1}\P_k}: a\!\in\! p'}\!\1_{(\!0\!,\infty\!)}\![\E_{\tilde{\Pi}^{(n)}}\!(\!\tilde{f}_{ran,p}^{(n)}\!)]\!\cdot\! \E_{\tilde{\Pi}^{(n)}}\!\big[\tilde{f}_{ran,p'}^{(n)}\big]
}{T_{n|\KK_{u+1}}}\cdot O(1)\notag\\
&=0.\notag
\end{align}
Here, we \wu{used} that 
$T_{n|\K\setminus\cup_{l=1}^{u+1}\K_l}\in o\big(T_{n|\KK_{u+1}}\big),$
that $\E_{\tilde{\Pi}^{(n)}}(\tilde{f}_{ran,p}^{(n)})=0$
implies $\PP_{\tilde{\Pi}^{(n)}}(\tilde{f}_{ran,p}^{(n)}\cdot \tau_p(\tilde{f}_{ran,a}^{(n)})=0)=1$
for every $a\in A,$ that $\E_{\tilde{\Pi}^{(n)}}(\tilde{f}_{ran,p}^{(n)})>0$
implies $\tau_p(\E_{\tilde{\Pi}^{(n)}}(\tilde{f}_{ran}^{(n)}))
\in O(g_n^{(u+1)}),$ that $p'$ is non-tight
if $p'$ contains an arc $a$ with $\beta_a>\lambda_{u+1},$
and that
\[
\begin{split}
&\lim_{n\to \infty}\frac{\sum_{a\in p:\beta_{a}\le  \lambda_{u+1}}\1_{(0,\infty)}(\E_{\tilde{\Pi}^{(n)}}(\tilde{f}_{ran,p}^{(n)}))\cdot \E_{\tilde{\Pi}^{(n)}}\big(\tilde{f}_{ran,p}^{(n)}
	\cdot \tau_a(\tilde{f}_{ran,a}^{(n)})\big)}{T_{n|\KK_{u+1}}\cdot
	g_n^{(u+1)}}\\
&=\lim_{n\to \infty}\frac{\sum_{a\in p:\beta_{a}\le  \lambda_{u+1}}\1_{(0,\infty)}(\E_{\tilde{\Pi}^{(n)}}(\tilde{f}_{ran,p}^{(n)}))\cdot \E_{\tilde{\Pi}^{(n)}}\big(\tilde{f}_{ran,p}^{(n)}
	\cdot \tau_a(\tilde{f}_{ran,a|\K\setminus\cup_{l=0}^{u}\K_l}^{(n)})\big)}{T_{n|\KK_{u+1}}
	\cdot
	g_n^{(u+1)}}\\
&\le\! \lim_{n\!\to\! \infty}\!\frac{\sum_{a
		\!\in\! p: \beta_{a}\!\le\!  \lambda_{u\!+\!1}}\!\1_{(\!0,\infty\!)}\!(\!\E_{\tilde{\Pi}^{(n)}}\!(\!\tilde{f}_{ran,p}^{(n)}\!))\!\cdot\! \E_{\tilde{\Pi}^{(n)}}\!\big(\!\tilde{f}_{ran,p}^{(n)}\!\big)
	\!\cdot\! \tau_a(T_{n|\K\!\setminus\!\cup_{l\!=\!0}^{u}\K_l}
	)}{T_{n|\KK_{u+1}}\cdot
	g_n^{(u\!+\!1)}}\\
&=\lim_{n\to \infty}\frac{\sum_{a\in p:\beta_{a}\le  \lambda_{u+1}}\1_{(0,\infty)}(\E_{\tilde{\Pi}^{(n)}}(\tilde{f}_{ran,p}^{(n)}))\cdot \E_{\tilde{\Pi}^{(n)}}\big(\tilde{f}_{ran,p}^{(n)}\big)
}{T_{n|\KK_{u+1}}}\cdot O(1)=0
\end{split}
\]
 when $p\in\cup_{k\in \KK_{u+1}}\P_k$ is non-tight. 
 
 \wu{\eqref{eq:Sto-LIMIT-NonTight} means
 that non-tight paths are also negligible in the limit when we scale the
 joint (expected) total cost of the subgame $\G_{n|\KK_{u+1}}$ in the mixed NE flow 
$\tilde{f}_{ran}^{(n)}$ with the factor 
$T_{n|\KK_{u+1}}\cdot g_n^{(u+1)}.$}

\eqref{eq:Sto-LIMIT-Tight}--\eqref{eq:Sto-LIMIT-NonTight},
\eqref{eq:Cost-Split-Sto} 
and $g_n^{(u+1)}\in \omega(\max_{l=0}^{u}\ g_n^{(l)})$ together
imply \wu{that}
\begin{align}\label{eq:Sto-LIMIT-TOTAL}
&\lim_{n\to\infty}\frac{\sum_{k\in\KK_{u+1}}\sum_{p\in\P_k}
	\E_{\tilde{\Pi}^{(n)}}(\tilde{f}_{ran,p}^{(n)}
	\cdot \tau_p(\tilde{f}_{ran}^{(n)}))}{T_{n|\KK_{u+1}}\cdot g_n^{(u+1)}}\notag\\
&=\lim_{n\to\infty}\frac{\sum_{k\in \KK_{u+1}}\sum_{p\in\P_k:
		p\text{ is tight}}
	\E_{\tilde{\Pi}^{(n)}}(\tilde{f}_{ran,p}^{(n)}
	\cdot \tau_p(\tilde{f}_{ran|\K\setminus \cup_{l=0}^{u}\K_l}^{(n)}))}{T_{n|\KK_{u+1}}\cdot g_n^{(u+1)}}\notag\\
&=\lim_{n\to\infty}\frac{\sum_{a\in A:\beta_a\le \lambda_{u+1}}
	\E_{\tilde{\Pi}^{(n)}}(\tilde{f}_{ran,a|\text{ tight } p}^{(n)}
	\cdot \tau_a(\tilde{f}_{ran,a|\K\setminus\cup_{l=0}^{u}\K_l}^{(n)}))}{T_{n|\KK_{u+1}}\cdot g_n^{(u+1)}}\notag\\
&=\lim_{n\to\infty}\frac{\sum_{a\in A:\beta_a\le \lambda_{u+1}}
	\E_{\tilde{\Pi}^{(n)}}(\tilde{f}_{ran,a|
		\K\setminus \cup_{l=0}^{u}\K_l}^{(n)}
	\cdot \tau_a(\tilde{f}_{ran,a|\K\setminus\cup_{l=0}^{u}\K_l}^{(n)}))}{T_{n|\KK_{u+1}}\cdot g_n^{(u+1)}}\\
&=\sum_{a\in A:\beta_a\le \lambda_{u+1}}\tilde{f}_{exp,a}^{(\infty,u+1)}
\cdot \tau_a^{(\infty,u+1)}(\tilde{f}_{exp,a}^{(\infty,u+1)})\notag\\
&=\sum_{a\in A}\tilde{f}_{exp,a}^{(\infty,u+1)}
\cdot \tau_a^{(\infty,u+1)}(\tilde{f}_{exp,a}^{(\infty,u+1)}),\notag
\end{align}
where we  put
$
\tilde{f}_{ran,a|\text{tight } p}^{(n)}\!:=\!
\sum_{p'\in \cup_{k\in\KK_{u\!+\!1}}\!\P_k: a\in p', p'\text{ is tight} }
\tilde{f}_{ran,p'}^{(n)}
$ for
each $a\in A$ with $\beta_a\le \lambda_{u+1}.$
\wu{We also used that
\[
\begin{split}
\tilde{f}^{(n)}_{ran,a|\K\setminus\cup_{l=1}^{u}\K_l}-&\tilde{f}_{ran,a|\text{ tight } p}^{(n)}
\le \sum_{k\in\KK_{u+1}}\ \sum_{p'\in \P_k:\ p'\text{ is non-tight} }
\tilde{f}_{ran,p'}^{(n)}+T_{n|\K\setminus\cup_{l=1}^{u+1}\K_l}\\
&\hspace{1.7cm}\in 
o(T_{n|\KK_{u+1}}),
\end{split}
\]}
\wu{and that
\begin{displaymath}
	\begin{split}
	\lim_{n\to \infty}\frac{\E_{\tilde{\Pi}^{(n)}}(\tilde{f}_{ran,p'}^{(n)}\cdot \tau_a(\tilde{f}_{ran,a|\K\setminus\cup_{l=0}^{u}\K_l}^{(n)}))}{T_{n|\KK_{u+1}}\cdot g_n^{(u+1)}}
	\le \lim_{n\to \infty}\frac{\E_{\tilde{\Pi}^{(n)}}(\tilde{f}_{ran,p'}^{(n)})}{T_{n|\KK_{u+1}}}\cdot O(1)=0
	\end{split}
\end{displaymath}}
\wu{when $\beta_a\le \lambda_{u+1}$ and 
$p'\in \cup_{k\in \KK_{u+1}}\P_k$ is non-tight.
Here, we observe that the random event ``$\tau_a(\tilde{f}_{ran,a|\K\setminus\cup_{l=0}^{u}\K_l}^{(n)})
\in O(g_n^{(u+1)})$'' occurs almost surely when 
$\beta_a\le \lambda_{u+1}.$}

\eqref{eq:Sto-LIMIT-TOTAL} together with 
Fact~\ref{fact:Wu2019} proves 
IA5--IA6 for step $u+1.$
\wu{Note that we have already shown that} $\tilde{f}_{exp}^{(\infty,u+1)}$
is a non-atomic NE flow of $\G_{|\K_{u+1}}^{(\infty)}.$
This completes the proof of Fact~\ref{claim:SubcaseII-Sto-Lambda>0}.\hfill$\square$

Therefore, IA1--IA7 hold
for all $u\in\mathcal{M}.$ 
This completes the whole proof by induction.
\hfill$\square$

\subsection{Proof of Lemma~\ref{lemma:Subgame-U-Sto-Exp}}
\label{proof:Subgame-U-Sto-Exp}

Consider an arbitrary 
arc $a\in A,$ and
an arbitrary $u\in\mathcal{M}=\{1,\ldots,m\}.$
\wu{Let} $g_n=T_{n|\K_u}^{\lambda}$
\wu{be a factor} with an arbitrary exponent 
$\lambda>0,$
and \wu{let $h: [0,\infty)\to [0,\infty)$ be} an arbitrary
\wu{non-decreasing} polynomial function
 with degree $\beta\ge 0$.
To \wu{simplify} notation, we assume that
$\K_u=\K\setminus \cup_{l=0}^{u-1}\K_l.$
\wu{The proof still holds
	when $\K_u$ is replaced by $\K\setminus \cup_{l=0}^{u-1}\K_l,$}
since \WuSecondRevision{\eqref{eq:AD-Total-Demand-Comparison} holds and}
$
	\lim_{n\to \infty}\frac{g_n}{T_{n|\K\setminus\cup_{l=0}^{u-1}\K_l}^{\lambda}}
	=\lim_{n\to \infty}\frac{g_n}{T_{n|\K_u}^{\lambda}}= 1.
$

\wu{We assume, w.o.l.g., that \wu{the} limit 
$
\lim_{n\!\to\! \infty}\frac{h(\E_{\tilde{\Pi}^{(n)}}(\tilde{f}_{ran,a|\K_u}^{(n)}))}{g_n}\in [0,\infty]
$
exists.} 

\wu{To prove Lemma~\ref{lemma:Subgame-U-Sto-Exp}, we} need  
\wu{tight} probability lower and upper bounds
\wu{for} the random event 
$
	|\tilde{f}^{(n)}_{ran,a|\K_u}-\E_{\tilde{\Pi}^{(n)}}(\tilde{f}^{(n)}_{ran,a|\K_u})| \in O(\E_{\tilde{\Pi}^{(n)}}(\tilde{f}^{(n)}_{ran,a|\K_u})),
$
for which \WuSecondRevision{we will need} \emph{Markov's inequality} \WuSecondRevision{from} Lemma~\ref{lemma:Markov}a.

Note that 
$
	\tilde{f}^{(n)}_{ran,a|\K_u}
	=\sum_{k\in \KK_u}\sum_{i\in \D_k^{(n)}}
	d_{k,i}^{(n)}\cdot \1_{p_{k,i}(\tilde{\Pi}_i^{(n)})}(a)
$
is a \emph{weighted sum} of mutually independent \emph{Bernoulli} random variables
$
		\1_{p_{k,i}(\tilde{\Pi}_i^{(n)})}(a),\quad \ i\in \D_{|\K_u}^{(n)}
		=\cup_{k\in \KK_u}\D_k^{(n)}.
$ \wu{Recall} that $p_{k,i}(\tilde{\Pi}_i^{(n)})$ 
is the random path sampled by agent $i$
using \wu{the} probability distribution $\tilde{\Pi}_i^{(n)}
=(\tilde{\Pi}_{i,p}^{(n)})_{p\in\P_k}$
for each $k\in\KK_u$ and $i\in\D_k^{(n)},$
and that $\1_{B}(b)$
is the indicator \WuSecondRevision{function} of \wu{the} membership relation ``$b\in B$''
for an arbitrary set $B$ and an arbitrary element $b.$

Fact~\ref{fact:Prob-Bounds-Weighted-Sum}a--d below \WuSecondRevision{show}
useful lower and upper probability bounds
for a weighted sum of \emph{arbitrary} Bernoulli random variables, 
\WuSecondRevision{and thus apply} to
\wu{the weighted sum} $\tilde{f}_{ran,a|\K_u}^{(n)}.$
\begin{fact}\label{fact:Prob-Bounds-Weighted-Sum}
	Consider $n$ mutually independent Bernoulli 
	random variables $X_1,\ldots,X_n$ with 
	success probabilities $q_1,\ldots,q_n\in [0,1],$
	respectively. Let $v_1,\ldots,v_n$
	be non-negative weights with sum $V_n:=\sum_{i=1}^n v_i$, and \WuSecondRevision{let}
	$Y_n=\sum_{i=1}^{n}v_i\cdot X_i$ \wu{be}
	the weighted sum of these $n$ random variables.
	If $v_i\le \upsilon$ for a constant $\upsilon>0,$ then 
	the following probability bounds hold.
	\begin{itemize}
		\item[a)] $\PP\big(Y_n\ge (1+\delta)\cdot E(Y_n)\big)
		\le e^{-\frac{(\delta+1)\cdot E(Y_n)}{v}\cdot \big(\ln(\delta+1)-\frac{\delta}{\delta+1}\big)}$
		for all $\delta>0.$
		\item[b)] $\PP\big(Y_n\!\le\! (1-\delta)\cdot E(Y_n)\big)
		\!\le\! e^{-\frac{V_n-(1-\delta)\cdot \E(Y_n)}{v}\cdot \big(\ln\frac{V_n-(1-\delta)\cdot \E(Y_n)}{V_n-\E(Y_n)}\!-\!\frac{\delta\cdot \E(Y_n)}{V_n\!-\!(1\!-\!\delta)\cdot \E(Y_n)}\big)}$
		for all $\delta\in (0,1).$
		\item[c)] If $\lim_{n\!\to\! \infty}\E(Y_n)=0$
		and $\varliminf_{n\to\infty}V_n>1,$ then
		there is an integer $N\in \N$ such that
		$\PP\big(Y_n\ge 1+\delta\big)
		\le e^{-\frac{\delta+1}{v}\cdot \big(\ln (\delta+1)-\frac{\delta}{\delta+1}\big)}$
		for all $\delta>0$ and all $n\ge N.$
		\item[d)] If $\lim_{n\!\to\! \infty}\frac{V_n}{\E(Y_n)}=1$
		and $\lim_{n\!\to\! \infty}V_n=\infty,$ then
		there is an integer $N\in \N$ s.t.
		$\PP\big(Y_n\!\le\! (1\!-\!\delta)\cdot (\E(Y_n)\!-\!c)\big)
			\!\le\! e^{-\frac{V_n\!-\!(1\!-\!\delta)\cdot (\E(Y_n)\!-\!c)}{v}\cdot \big(\ln \frac{V_n\!-\!(1\!-\!\delta)\cdot (\E(Y_n)\!-\!c)}{V_n\!-\!\E(Y_n)\!+\!c}\!-\!\frac{\delta\cdot \E(Y_n)\!-\!\delta\cdot c}{V_n\!-\!(1\!-\!\delta)\cdot (\E(Y_n)\!-\!c)}\big)}$
		for all $\delta\in (0,1),$ all
		$c\in (0,\ \E(Y_n)),$ and all $n\ge N.$
	\end{itemize}
\end{fact}
\textbf{Proof of Fact~\ref{fact:Prob-Bounds-Weighted-Sum}a:}
\wu{Our proof is} similar to that for the 
usual Chernoff bound in, e.g., \citep{Mitrinovic1970,Nowak2009}.
\wu{Using Markov's inequality and the fact that 
	$X_1,\ldots,X_n$ are mutually independent Bernoulli
	random variables with success probabilities 
	$q_1,\ldots,q_n$,} we obtain for an arbitrary $t>0$ and
an arbitrary $\delta>0$ that
\begin{equation}\label{eq:Prob-Trans}
\begin{split}
\PP&\big(Y_n\ge (1+\delta)\cdot E(Y_n)\big)
=\PP\big(e^{t\cdot Y_n}\ge e^{t\cdot (1+\delta)\cdot E(Y_n)}\big)
\le \frac{\prod_{i=1}^{n}\E(e^{t\cdot X_i\cdot v_i})
}{e^{t\cdot (1+\delta)\cdot E(Y_n)}}\\
&=\frac{\prod_{i=1}^{n}\big(q_i\cdot e^{t\cdot v_i}+(1-q_i)\big)}
{e^{t\cdot (1+\delta)\cdot E(Y_n)}}
=\frac{\prod_{i=1}^{n}\big(q_i\cdot v_i\cdot t\cdot 
	\frac{e^{t\cdot v_i}-1}{t\cdot v_i}+1\big)}
{e^{t\cdot (1+\delta)\cdot E(Y_n)}}.
\end{split}
\end{equation}
\wu{The} function $\frac{e^{x}-1}{x}$ is non-decreasing on 
$(0,\infty)$ and 
$1+x\le e^x$ holds for all 
$x\in [0,\infty).$ So we obtain by \eqref{eq:Prob-Trans}
that 
\begin{equation}\label{eq:Prob-Trans-UpperPre}
\begin{split}
\PP\big(Y_n\!\ge\! (1\!+\!\delta)\cdot E(Y_n)\big)
&\le\! \frac{\prod_{i=1}^{n}\big(q_i\cdot v_i\cdot t\cdot \frac{e^{t\cdot \upsilon}-1}{t\cdot \upsilon}\!+\!1\big)}
{e^{t\cdot (1+\delta)\cdot E(Y_n)}}\\
&\le\! \frac{e^{\sum_{i=1}^{n}q_i\cdot v_i\cdot \frac{e^{t\cdot \upsilon}-1}{\upsilon}}}{e^{t\cdot (1+\delta)\cdot E(Y_n)}}\!=\!e^{E(Y_n)\cdot \big(\frac{e^{t\cdot \upsilon}-1}{\upsilon}-t\cdot (1+\delta)\big)}
\end{split}
\end{equation}
for all $t>0.$
\eqref{eq:Prob-Trans-UpperPre} 
implies \wu{that}
$
\PP\big(Y_n\ge (1+\delta)\cdot E(Y_n)\big)
\le e^{-\frac{(\delta+1)\cdot E(Y_n)}{v}\cdot \big(\ln(\delta+1)-\frac{\delta}{\delta+1}\big)}
$
when we put $t=\frac{\ln(\delta+1)}{v}$ 
\wu{and} observe that $\ln(\delta+1)-\frac{\delta}{\delta+1}>0$
for all $\delta>0.$

\wu{\textbf{Proof of Fact~\ref{fact:Prob-Bounds-Weighted-Sum}b}}
Let $Z_n:=\sum_{i=1}^nv_i\cdot (1-X_i)=V_n-Y_n.$
Then $Z_n+Y_n=V_n$
and $\E(Z_n)+\E(Y_n)=V_n.$
Fact~\ref{fact:Prob-Bounds-Weighted-Sum}a) implies
for every $\delta\in (0,1)$ that 
\begin{displaymath}
	\begin{split}
	\PP&\big(Z_n\ge \E(Z_n)+\delta\cdot \E(Y_n)
	=\big(1+\frac{\delta\cdot \E(Y_n)}{\E(Z_n)}
	\big)\cdot \E(Z_n)\big)\\
	&\hspace{1cm}\le 
	e^{-\frac{\E(Z_n)+\delta\cdot \E(Y_n)}{v}\cdot \big(\ln\frac{\E(Z_n)+\delta\cdot \E(Y_n)}{\E(Z_n)}-\frac{\delta\cdot \E(Y_n)}{\E(Z_n)+\delta\cdot \E(Y_n)}\big)}.
	\end{split}
\end{displaymath}
Since \wu{the} random event
$Z_n\ge \E(Z_n)+\delta\cdot \E(Y_n)$
is equivalent to \wu{the} random event 
$Y_n\le (1-\delta)\cdot \E(Y_n),$
we obtain that
\begin{equation}\label{eq:Prob-Lower-Bound}
\begin{split}
\PP(Y_n\le (1-\delta)\cdot E(Y_n))
&\le 
e^{-\frac{\E(Z_n)+\delta\cdot \E(Y_n)}{v}\cdot \big(\ln\frac{\E(Z_n)+\delta\cdot \E(Y_n)}{\E(Z_n)}-\frac{\delta\cdot \E(Y_n)}{\E(Z_n)+\delta\cdot \E(Y_n)}\big)}\\
&=e^{-\frac{V_n-(1-\delta)\cdot \E(Y_n)}{v}\cdot \big(\ln\frac{V_n-(1-\delta)\cdot \E(Y_n)}{V_n-\E(Y_n)}-\frac{\delta\cdot \E(Y_n)}{V_n-(1-\delta)\cdot \E(Y_n)}\big)}
\end{split}
\end{equation}
\wump{for all $\delta\in (0,1).$}
\eqref{eq:Prob-Lower-Bound} proves Fact~\ref{fact:Prob-Bounds-Weighted-Sum}b.

\wu{\textbf{Proof of Fact~\ref{fact:Prob-Bounds-Weighted-Sum}c:}}
We \wu{say that} $n$ mutually independent Bernoulli random variables $X'_1,\ldots,X'_n$
\wu{with success probabilities
	$q'_1,\ldots,q'_n$ are \emph{stochastically larger than}}
$X_1,\ldots,X_n$ if 
$q'_i\ge q_i$ for each $i=1,\ldots,n.$
Clearly,  there are
$n$ mutually independent Bernoulli random variables $X'_1,\ldots,X'_n$ 
\wu{that are stochastically larger than}
$X_1,\ldots,X_n$ and satisfy $\E(Y'_n)=
\E(\sum_{i=1}^{n}v_i\cdot X'_i)=\sum_{i=1}^nv_i\cdot q'_i=1$
for large enough $n.$
This follows since $\E(Y_n)=\sum_{i=1}^{n}v_i\cdot q_i\to 0$
as $n\to \infty,$ $\varliminf_{n\to\infty}V_n=\varliminf_{n\to\infty}\sum_{i=1}v_i>1,$ and the continuous multi-variate
function $\alpha(x_1,\ldots,x_n):=\sum_{i=1}^{n}v_i\cdot (q_i+x_i)$
has $[\E(Y_n),\ V_n]$ \wu{as its range on the}
compact domain $\prod_{i=1}^{n} [0,1-q_i]$ for all $n\in\N.$ 

Fact~\ref{fact:Prob-Bounds-Weighted-Sum}c then follows
\wu{from Fact~\ref{fact:Prob-Bounds-Weighted-Sum}a},	if $\PP(Y_n\ge c)\le \PP(Y'_n=\sum_{i=1}^n
v_i\cdot X'_i\ge c)$
for an arbitrary constant $c\ge \E(Y_n),$ 
(since we can then obtain Fact~\ref{fact:Prob-Bounds-Weighted-Sum}c
by applying Fact~\ref{fact:Prob-Bounds-Weighted-Sum}a 
to $Y'_n$ with $c=1+\delta$ for large enough $n$).  

Consider now an arbitrary constant $c\ge \E(Y_n).$
We prove below \wu{that} $\PP(Y_n\ge c)\le \PP(Y'_n\ge c)$ only for \wu{the} particular case that 
$q'_1\ge q_1$ and $q'_i=q_i$ for
all $i=2,\ldots,n.$ One can
obtain a complete proof for the general case with
a simple induction over $\{2,\ldots,n\}.$

Note that\begin{equation*}
    \begin{split}
	\PP(Y'_n\ge c)&=\PP\big(\sum_{i=2}^nv_i\cdot X'_i\ge c-v_1\big)\cdot 
	\PP(X'_{1}=1)+\PP\big(\sum_{i=2}^nv_i\cdot X'_i\ge c\big)\cdot 
	\PP(X'_{1}=0)\\
	&=\PP\big(\sum_{i=2}^nv_i\cdot X'_i\ge c-v_1\big)\cdot 
	(q_1+q'_{1}-q_1)+\PP\big(\sum_{i=2}^nv_i\cdot X'_i\ge c\big)\cdot 
	(1-q_{1}+q_1-q'_1)\\
	&=\PP(Y_n\ge c)+(q'_1-q_1)\cdot \Big(\PP\big(\sum_{i=2}^nv_i\cdot X_i\ge c-v_1\big)
	-\PP\big(\sum_{i=2}^nv_i\cdot X_i\ge c\big)\Big)\\
	&=\PP(Y_n\ge c)+(q'_1-q_1)\cdot\PP\Big(c> \sum_{i=2}^nv_i\cdot X_i\ge c-v_{1}\Big)
	\ge \PP(Y_n\ge c).
	\end{split}
\end{equation*}This follows since the Bernoulli random variables $X_i$
and $X'_i$ can be \wu{identified} for
each $i=2,\ldots,n,$ \wu{as} they have the same success probability
$q_i.$ 

\wu{\textbf{Proof of Fact~\ref{fact:Prob-Bounds-Weighted-Sum}d:}} \wu{It} follows immediately from Fact~\ref{fact:Prob-Bounds-Weighted-Sum}b and
the fact that there are $n$ mutually independent Bernoulli
random variables $X'_1,\ldots,X'_n$ \wu{such that $X_1,\ldots,X_n$} are \wu{stochastically larger than $X'_1,\ldots,X'_n$}  and 
$\E(Y'_n)=\E(Y_n)-c$ for a constant
$c\in (0,\ \E(Y_n)).$ Note that such Bernoulli random variables exist since
$\lim_{n\!\to\! \infty}\frac{V_n}{\E(Y_n)}=1$
and $\lim_{n\!\to\! \infty}V_n=\infty.$

This completes \wu{the} proof of Fact~\ref{fact:Prob-Bounds-Weighted-Sum}.
\hfill$\square$

The two probability bounds 
in Fact~\ref{fact:Prob-Bounds-Weighted-Sum}a--b are similar to  \emph{Chernoff's bounds}
and \emph{Hoeffding's bounds}, see, e.g., 
\citep{Hoeffding1963,Mitrinovic1970,Nowak2009}.
However, a direct application of these known
bounds to $\tilde{f}_{ran,a|\K_u}^{(n)}$ involves either 
the number $|\D_{|\K_u}^{(n)}|$ of \wu{agents in} subgame $\G_{n|\K_{u}}$, or 
the \emph{minimum} individual demand $\min_{k\in\K_u,i\in\D^{(n)}_k}d_{k,i}^{(n)}$. Note that 
this minimum individual demand may vanish 
quickly as $n\to \infty$ and so the number $|\D_{|\K_u}^{(n)}|$ \wu{of agents}
need not be in $\Theta(T_{n|\K_u})$. \WuSecondRevision{Therefore, we include a proof tailored to our  needs.}

Note also that Fact~\ref{fact:Prob-Bounds-Weighted-Sum}a does not apply
when $\E_{\tilde{\Pi}^{(n)}}(\tilde{f}^{(n)}_{ran,a|\K_u})\in o(1),$
and Fact~\ref{fact:Prob-Bounds-Weighted-Sum}b \wu{does not} apply
when $\lim_{n\!\to\! \infty}\frac{\E_{\tilde{\Pi}^{(n)}}(\tilde{f}^{(n)}_{ran,a|\K_u})}{T_{n|\K_u}}= 
\tilde{f}_{exp,a}^{(\infty,u)}=1.$
We will \wu{instead} use Fact~\ref{fact:Prob-Bounds-Weighted-Sum}c--d,
respectively, in \wu{the} proof of Lemma~\ref{lemma:Subgame-U-Sto-Exp} \wu{in these two cases}.

With \WuSecondRevision{all these preparations,} we
are now ready to prove Lemma~\ref{lemma:Subgame-U-Sto-Exp}.

\wu{The two limits} in Lemma~\ref{lemma:Subgame-U-Sto-Exp}
are equal to $0$ when $\lambda>\beta,$
since both $h(\E_{\tilde{\Pi}^{(n)}}(\tilde{f}^{(n)}_{ran,a|\K_u}))$
and
$\E_{\tilde{\Pi}^{(n)}}(h(\tilde{f}^{(n)}_{ran,a|\K_u}))$ are in $o(g_n)$
when $\lambda>\beta.$

We assume, w.l.o.g., that 
$\beta\!\ge\! \lambda\!>\!0$. 
Moreover, we assume that 
$\lim_{n\to\infty}\!\E_{\tilde{\Pi}^{(n)}}(\tilde{f}^{(n)}_{ran,a|\K_u})$ $
\in [0,\infty]$ exists.
Otherwise, we take an arbitrary infinite  subsequence $(n_j)_{j\in\N}$ satisfying
this condition.  \WuSecondRevision{To simplify notation, we write 
$Y_n:=\tilde{f}^{(n)}_{ran,a|\K_u},$ $E_n:=\E_{\tilde{\Pi}^{(n)}}(\tilde{f}^{(n)}_{ran,a|\K_u}),$
$\PP_{\tilde{\Pi}^{(n)}}(\cdot)=\PP(\cdot),$ and 
$\E_{\tilde{\Pi}^{(n)}}(\cdot)=\E(\cdot).$}

\wu{We distinguish four cases.}

\textbf{Case I:} \WuSecondRevision{$E_n
\in \Theta(1),$ i.e., $\lim_{n\!\to\! \infty}E_n
\in (0,\infty)$. }
Let $\xi\!:=\!\frac{\lambda}{2\cdot \beta}\!\in\! (0,1)$.
We obtain by
Fact~\ref{fact:Prob-Bounds-Weighted-Sum}a
with $\delta\!:=\!T_{n|\K_u}^{\xi}$
that
\WuSecondRevision{
	$\PP[Y_n
	\ge (1+\delta)\cdot E_n]
	\le e^{-\frac{(1+\delta)\cdot E_n}{\upsilon}\cdot \left(\ln (\delta+1)-\frac{\delta}{\delta+1}\right)}
	= e^{-\omega(T_{n|\K_u}^{\xi}/\upsilon)}.$%
}
This in turn implies that 
\WuSecondRevision{$
\E(h(Y_n))
\le e^{-\omega\big(T_{n|\K_u}^{\xi}/\upsilon\big)}
\cdot h(T_{n|\K_u})+ h\big((1+T_{n|\K_u}^{\xi})\cdot E_n\big)
\in  o(g_n).
$}
So \WuSecondRevision{$\lim_{n\!\to\! \infty}\frac{h(E_n)}{g_n}\!=\!0\!
=\!\lim_{n\!\to\! \infty}\frac{\E(h(Y_n))}{g_n}.$}

\textbf{Case II:} \WuSecondRevision{$E_n
\in o(1),$ i.e., $\lim_{n\!\to\! \infty}E_n=0.$}
We obtain by Fact~\ref{fact:Prob-Bounds-Weighted-Sum}c
that 
\WuSecondRevision{$	\PP\big[Y_n
	\ge 1+T_{n|\K_u}^{\xi}\big]\le e^{-\omega\big(T_{n|\K_u}^{\xi}/\upsilon\big)}.$
Then, $\lim_{n\!\to\! \infty}\!\frac{h(E_n)}{g_n}\!=0\!
=\!\lim_{n\!\to\! \infty}\!\frac{\E\!(h(Y_n))}{g_n}$.}

\textbf{Case III:} \WuSecondRevision{$\tilde{f}_{exp,a}^{(\infty,u)}
=\lim_{n\!\to\! \infty}\frac{E_n}{T_{n|\K_u}}=1.$}
\wu{We obtain} by Fact~\ref{fact:Prob-Bounds-Weighted-Sum}d that 
\WuSecondRevision{\[
\begin{split}
\PP\big[Y_n\le \big(1-\delta\big)\cdot \big(E_n-c\big)\big]&\le
e^{-\frac{T_{n|\K_u}\!-\!(1\!-\!\delta)\cdot (E_n\!-\!c)}{v}\cdot \big(\ln \frac{T_{n|\K_u}\!-\!(1\!-\!\delta)\cdot (E_n)\!-\!c)}{T_{n|\K_u}\!-\!E_n\!+\!c}\!-\!\frac{\delta\cdot E_n\!-\!\delta\cdot c}{T_{n|\K_u}\!-\!(1\!-\!\delta)\cdot (E_n\!-\!c)}\big)}\\
 &=e^{-\Omega\big( \delta\cdot T_{n|\K_u}\big)},
\end{split}
\]}
where $\delta\in (0,1)$ is an arbitrary constant
and $c:=\sqrt{T_{n|\K_u}}.$
Therefore, 
\WuSecondRevision{\begin{displaymath}
\begin{split}
\frac{\E(h(Y_n))}{h(E_n)}
\ge\! (1\!-\!e^{-\Omega( \delta\cdot T_{n|\K_u})})
\cdot\frac{h((1\!-\!\delta)\cdot (E_n\!-\!\sqrt{T_{n|\K_u}}))}{h(E_n)}.
\end{split}
\end{displaymath}}
This implies \wu{that}
\WuSecondRevision{$
\varliminf_{n\to \infty}\frac{\E(h(Y_n))}{h(E_n)}\ge (1-\delta)^{\beta}
$} by letting $n\to \infty$ \wu{on} both sides of the above inequality.
So \WuSecondRevision{$\varliminf_{n\to \infty}\frac{\E(h(Y_n))}{h(E_n)}\ge 1$ due to the arbitrary choice of 
$\delta\in (0,1).$}
However, \wu{on the other hand,}
\WuSecondRevision{$
	\varlimsup_{n\to \infty}\!\frac{\E(h(Y_n))}{h(E_n)}\!=\! \varlimsup_{n\to \infty}\!\frac{\E(h(Y_n))}{h(T_{n|\K_u})}\cdot\lim_{n\!\to\!\infty}\!\frac{h(T_{n|\K_u})}{h(E_n)}\!\le\! 1.
$}
Hence, we have \WuSecondRevision{$\lim_{n\to \infty}\frac{\E(h(Y_n))}{h(E_n)}= 1$}
when $\tilde{f}^{(\infty,u)}_{exp,a}=1.$

\textbf{Case IV:} $\tilde{f}^{(\infty,u)}_{exp,a}<1$
and \WuSecondRevision{$E_n\in \omega(1),$}
i.e., \WuSecondRevision{$\lim_{n\!\to\! \infty}E_n=\infty$
and
$ T_{n|\K_u}-E_n\in \Theta(T_{n|\K_u}).
$}
Clearly, 
Fact~\ref{fact:Prob-Bounds-Weighted-Sum}a--b
\wu{apply} in this case. \wu{We further distinguish two subcases.}

\wu{\textbf{(Subcase IV-I: \WuSecondRevision{$h(E_n)\in o(g_n)$})}} \wu{Then} 
\WuSecondRevision{$
E_n
\in o(T_{n|\K_u}^{\lambda/\beta})
$.}
  \WuSecondRevision{We obtain further} by
Fact~\ref{fact:Prob-Bounds-Weighted-Sum}a that  \WuSecondRevision{$\E(h(Y_n))\in o(g_n).$}
\wu{This follows since} 
\WuSecondRevision{$
	\PP(Y_n\!>\!\delta\cdot T_{n|\K_u}^{\lambda/\beta})
	\le e^{-\Omega(\delta\cdot T_{n|\K_u}^{\lambda/\beta})}
$}
for all $\delta>0$
when \WuSecondRevision{$E_n\in o(T_{n|\K_u}^{\lambda/\beta}),$}
and so 
\WuSecondRevision{$	\varlimsup_{n\to\infty}\frac{\E(h(Y_n))}{g_n}$ $\le$ $\delta^{\beta}\cdot O(1)$}
for all $\delta>0.$ 

\wu{\textbf{(Subcase IV-II: \WuSecondRevision{$h(E_n)\in \Omega(g_n)$})}}
\wu{Then} 
\WuSecondRevision{$E_n\in\Omega(T_{n|\K_u}^{\lambda/\beta}).$}
Fact~\ref{fact:Prob-Bounds-Weighted-Sum}a \WuSecondRevision{yields that}
\WuSecondRevision{$
	\PP\big[Y_n
	\ge E_n+E_n^{2/3}\big]=e^{-\frac{E_n+E_n^{2/3}}{\upsilon}
	\cdot \big(
	\ln (1+E_n^{-1/3})-\frac{E_n^{-1/3}}{1+E_n^{-1/3}}
	\big)}
	\le e^{-\Omega(E_n^{1/3})}.
$}
\WuSecondRevision{Hence,}
\WuSecondRevision{\begin{displaymath}
	\begin{split}
	\varlimsup_{n\!\to\! \infty}\frac{\E(h(Y_n))}{h(E_n)}\le \varlimsup_{n\!\to\! \infty} e^{-\Omega(E_n^{1/3})}\cdot 
	\frac{h(T_{n|\K_u})}{h(E_n)}\!+\varlimsup_{n\!\to\! \infty} \frac{h(E_n\!+\!E_n^{2/3})}{h(E_n)}\!=\!1.
	\end{split}
\end{displaymath}
Moreover, 
$
	\varliminf_{n\!\to\! \infty}\frac{\E(h(Y_n))}{h(E_n)}\ge 1
$}
 follows from Fact~\ref{fact:Prob-Bounds-Weighted-Sum}b,
since 
\WuSecondRevision{
	$\PP[Y_n
	\le (1-\delta)\cdot E_n]\le 
	e^{-\Omega(T_{n|\K_u})}$}
\wu{for each $\delta\in (0,1),$}
\wu{when} $T_{n|\K_u}-E_n\in \Theta(T_{n|\K_u}).$

\WuSecondRevision{All the above together prove} Lemma~\ref{lemma:Subgame-U-Sto-Exp}.
\hfill$\square$

\end{document}